\documentclass[fleqn,usenatbib]{mnras}

\usepackage{newtxtext,newtxmath}

\usepackage[T1]{fontenc}

\DeclareRobustCommand{\VAN}[3]{#2}
\let\VANthebibliography\thebibliography
\def\thebibliography{\DeclareRobustCommand{\VAN}[3]{##3}\VANthebibliography}

\usepackage{graphicx}	
\usepackage{amsmath}	
\usepackage{amsfonts}
\usepackage{amscd}
\usepackage{natbib}
\usepackage[flushleft]{threeparttable}
\usepackage{orcidlink}
\usepackage{booktabs}
\usepackage{multirow}
\usepackage[section]{placeins}

\newcommand{\kms}{\,\mathrm{km} \, \mathrm{s}^{-1}}
\newcommand{\gpcm}{\, \mathrm{g} \, \mathrm{cm}^{-3}}
\newcommand{\pcm}{\, \mathrm{cm}^{-3}}
\newcommand{\Msolpc}{\, \mathrm{M}_{\sun} \, \mathrm{pc}^{-2}}

\newcommand{\ar}{a_{\mathrm{R}}}
\newcommand{\fdustgas}{f_{\mathrm{dg}}}
\newcommand{\kappaR}{\kappa_{\mathrm{R}}}
\newcommand{\kappaP}{\kappa_{\mathrm{P}}}
\newcommand{\alphavir}{\alpha_{\mathrm{vir}}}
\newcommand{\cs}{c_{\mathrm{s}}}
\newcommand{\Tr}{T_{\mathrm{r}}}
\newcommand{\kappaPLT}{\kappa_{\mathrm{PLT}}}
\newcommand{\kappaSem}{\kappa_{\mathrm{Sem}}}
\newcommand{\Mcloud}{M_{\mathrm{cloud}}}
\newcommand{\Rcloud}{R_{\mathrm{cloud}}}
\newcommand{\Sigmacloud}{\Sigma_{\mathrm{cloud}}}
\newcommand{\Msun}{\mathrm{M}_{\sun}}
\newcommand{\forcerad}{\langle \dot{p}_{\mathrm{rad}} \rangle_{4\pi}}
\newcommand{\forcegrav}{\langle \dot{p}_{\mathrm{grav}} \rangle_{4\pi}}
\newcommand{\ftrap}{f_{\mathrm{trap}}}
\newcommand{\fgrav}{f_{\mathrm{grav}}}

\defcitealias{Skinner_2015}{SO15}
\defcitealias{Tsang_2018}{TM18}
\defcitealias{Crocker_2018b}{C18}
\defcitealias{Krumholz_Thompson_2012}{KT12}

\title[IR Radiation Pressure in Star Cluster Formation]{Infrared Radiation Feedback Does Not Regulate Star Cluster Formation}

\author[S.~H.~Menon et al.]{
Shyam H.~Menon$^{\orcidlink{0000-0001-5944-291X}}$,$^{1}$\thanks{E-mail: shyam.menon@anu.edu.au (SHM)}
Christoph Federrath$^{\orcidlink{0000-0002-0706-2306}}$,$^{1,2}$\thanks{E-mail: christoph.federrath@anu.edu.au (CF)}
Mark R.~Krumholz$^{\orcidlink{0000-0003-3893-854X}}$$^{1,2}$
\\
$^{1}$Research School of Astronomy and Astrophysics, Australian National University, Canberra, ACT~2611, Australia\\
$^{2}$ARC Centre of Excellence for Astronomy in Three Dimensions (ASTRO-3D), Canberra, ACT~2611, Australia
}

\date{Accepted XXX. Received YYY; in original form ZZZ}

\pubyear{2022}

\begin{document}
\label{firstpage}
\pagerange{\pageref{firstpage}--\pageref{lastpage}}
\maketitle

\begin{abstract}
We present 3D radiation-hydrodynamical (RHD) simulations of star cluster formation and evolution in massive, self-gravitating clouds, whose dust columns are optically thick to infrared (IR) photons. We use \texttt{VETTAM} -- a recently developed, novel RHD algorithm, which uses the Variable Eddington Tensor (VET) closure -- to model the IR radiation transport through the cloud. We also use realistic temperature ($T$) dependent IR opacities ($\kappa$) in our simulations, improving upon earlier works in this area, which used either constant IR opacities or simplified power laws ($\kappa \propto T^2$). We investigate the impact of the radiation pressure of these IR photons on the star formation efficiency (SFE) of the cloud, and its potential to drive dusty winds. We find that IR radiation pressure is unable to regulate star formation or prevent accretion onto the star clusters, even for very high gas surface densities ($\Sigma > 10^5 \Msolpc$), contrary to recent semi-analytic predictions and simulation results using simplified treatments of the dust opacity. We find that the commonly adopted simplifications of $\kappa \propto T^2$ or constant $\kappa$ for the IR dust opacities leads to this discrepancy, as those approximations overestimate the radiation force. By contrast, with realistic opacities that take into account the micro-physics of the dust, we find that the impact of IR radiation pressure on star formation is very mild, even at significantly high dust-to-gas ratios ($\sim 3$ times solar), suggesting that it is unlikely to be an important feedback mechanism in controlling star formation in the ISM.  
\end{abstract}




\section{Introduction}
\label{sec:Introduction}

Star formation is an inefficient process, as evidenced by observed gas depletion times\footnote{The timescale required by a galaxy, or a part thereof, to consume 100\% of its molecular gas supply at its current star formation rate.}, which are two orders of magnitude above the dynamical time, both in galaxies \citep[e.g.,][]{Leroy_2017, Utomo_2018}, and in individual giant molecular clouds (GMCs) \citep[e.g.,][]{Krumholz_2007b, Evans_2014, Heyer_2016, Pokhrel_2020, Hu_2022}. Theoretical models explain this inefficiency through a combination of mechanisms that provide support against gravitational collapse, including turbulence, magnetic fields, stellar feedback and dynamical stabilisation \citep{Krumholz_2005, Ostriker_2010,Krumholz_2012c,Federrath_klessen_2012,Federrath_2013,Padoan_2014,Federrath2015,Burkhart_2018,Meidt_2018,Krumholz_Federrath_2019, Evans_2022}. Recent progress in both theory and observations have highlighted the pivotal role that feedback, especially due to massive (main-sequence) stars, plays in star/star-cluster formation \citep{Krumholz_2014,Krumholz_2019}, and the lifecycle of GMCs (see \citealt{Chevance_2020,Chevance_2022} for reviews). This massive-star feedback has been suggested to be largely responsible for limiting the integrated star formation efficiency ($\epsilon_*$) to low values in typical environments, where $\epsilon_*$ is given by
\begin{equation}
    \epsilon_* = \frac{M_{*}}{M_{\mathrm{gas}}},
    \label{eq:sfe}
\end{equation}
which quantifies the net efficiency of star formation over the lifetime of a GMC, i.e. the ratio of the final stellar mass $M_*$ and the available gas mass in the parent molecular cloud $M_{\mathrm{gas}}$. Feedback achieves this by i) disrupting GMCs in order $\sim$ unity dynamical timescales, through the momentum and energy carried by feedback processes \citep[e.g.,][]{Grudic_2018}, and ii) driving turbulent motions that could further provide support against collapse \citep[e.g.,][]{MacLow_Klessen_2004,Krumholz_2006,Elmegreen2009,Gritschneder_2009,Federrath_2010,Wibking_2018,Menon_2020,MenonEtAl2021,Garcia_2020}. 

Feedback from massive stars takes a number of forms: i) the ionisation of gas by the absorption of extreme ultraviolet (EUV) photons with energies that exceed their ionisation potential \citep[e.g.,][]{Dale_2012,Geen_2016,Gavagnin_2017,Kim_2018} ii) radiation pressure on dust grains by the absorption of UV and IR photons \citep[e.g.,][]{Krumholz_2009,Fall_2010,Murray_2010_clusters,Raskutti_2016,Thompson_Krumholz_2016} iii) winds driven by hot stars \citep[e.g.,][]{Castor_1975Bubbles,Weaver_1977,Lancaster_2021}, and supernovae \citep[SNe;][]{Rogers_2013,Calura_2015}. The first two of these are generally thought to be the most important. SNe, although important for driving turbulence in the diffuse ISM \citep[e.g.,][]{Padoan_2016} and controlling the dynamical state of gas at kpc scales \citep{Kim_Ostriker_2017}, are limited in their ability to control $\epsilon_*$ in GMCs due to the time delay of $\sim 3 \, \mathrm{Myr}$ between the onset of star formation and the first SNe -- a timescale within which either other ``early'' feedback mechanisms disrupt the cloud, or a large fraction of gas converts to stars before feedback has a chance to stop it \citep{Fall_2010,Geen_2016,Grudic_2022}. Indeed, observations confirm that most GMC disruption happens before SNe begin \citep[e.g.,][]{Hollyhead_2015, Grasha_2019, Chevance_2022b}. Stellar winds have also shown to be inefficient at regulating $\epsilon_*$ due to a leakage of hot gas through low-density channels, and/or energy losses due to turbulent mixing at the hot-cool interface \citep{Rogers_2013,Rosen_2014,Lancaster_2021}. Among the former two mechanisms, photoionisation has been understood to be the primary agent in regulating $\epsilon_*$ in environments with escape speeds lower than the sound speed of ionised gas ($\sim 10 \, \mathrm{km} \, \mathrm{s}^{-1}$), whereas radiation pressure becomes dominant for denser environments that are beyond this limit \citep{Krumholz_2009,Murray_2010_clusters,Kim_2018,Fukushima_2021}. Such environments have been observed to exist at the formation sites of massive clusters, and/or for younger/compact systems \citep{Lopez_2011,Lopez_2014,Barnes_2020, Olivier_2021}.  

An additional source of radiation pressure can arise in high (surface) density environments that are optically thick to the dust-reprocessed IR photons, thereby undergoing repeated cycles of absorption and emission, enhancing the imparted momentum over the stellar UV photon momentum \citep{Thompson_2005,Murray_2010_clusters}. This is the so-called \textit{multiple-scattering} regime, to differentiate it from the single-scattering regime, where the dust is optically thin to IR photons, and only the stellar photons can contribute. The enhancement in momentum imparted can be quantified by the trapping factor $\ftrap = \dot{p}_{\mathrm{IR}}/(L_*/c)$, where $\dot{p}_{\mathrm{IR}}$ is the momentum imparted per unit time integrated over all of a cloud by the IR photons, and $L_*/c$ is the momentum per unit time carried by the direct stellar radiation field, which is also the momentum per unit time deposited in the gas in the single-scattering regime. Environments in the multiple-scattering regime in the local universe are primarily found in extreme regions such as dwarf starbursts and ultra-luminous infrared galaxies (ULIRGs) like Arp 220. These environments potentially host the formation sites of super-star clusters \citep[SSC's; e.g.,][]{Mcgrady_2005,Zwart_2010,Turner_2015,LindaSmith_2020}, and represent a dense mode of star formation that might have existed more commonly at high redshift. Observationally probing the natal phase of star formation in these environments is difficult due to the highly embedded nature of young forming SSCs. However, recent efforts using the Atacama Large Millimeter/Submillimeter Array (ALMA) have managed to do so at relatively high resolution \citep[$\sim 2\,\mathrm{pc}$; e.g.,][]{Turner_2017,Leroy_2018,Villas_2020}, with future instruments such as the James Webb Space Telescope (JWST) expected to enable further study. It is therefore of interest to develop theoretical models and conduct numerical simulations to determine the effectiveness of feedback in such environments and its subsequent influence on the efficiency of star formation.

The dynamical competition between gravity and (IR) radiation pressure has previously been studied through semi-analytical models and numerical simulations. For instance, \citet{Murray_2010_clusters} built semi-analytic models, and found that IR radiation pressure is the dominant feedback mechanism in regulating $\epsilon_*$ in high-density environments. However, the findings of those studies were put into question by radiation-hydrodynamic (RHD) simulations of atmospheres in a gravitational field exposed to IR photon fluxes; these simulations found that semi-analytic models overestimate the amount of momentum that is imparted, as they cannot capture the complex interactions between radiation and matter in higher-dimensional, non-trivial (e.g., turbulent) density distributions \citep{Krumholz_Thompson_2012,Krumholz_Thompson_2013,Davis_2014,Rosdahl_2015,Tsang_2015}. However, these simulations were also highly idealised, with plane-parallel geometries, and static, fixed radiation fluxes and gravitational potentials. They also use an approximation for the IR opacities that we will show in this study, can significantly overestimate the effectiveness of the feedback. In the \textit{direct} context of star cluster formation in turbulent, self-gravitating GMCs, \citealt{Skinner_2015} (\citetalias{Skinner_2015} hereafter) performed the first systematic study with 3D RHD simulations. They found that the dust IR opacity ($\kappa$) is the most important parameter in determining $\epsilon_*$: values of $\kappa \ga 10 \, \mathrm{cm}^2 \, {g}^{-1}$ are required for having any regulatory effect on $\epsilon_*$, which likely apply in dust-enriched environments. \citet{Tsang_2018} reached similar conclusions using a more accurate RHD method, but only simulated one point in the parameter space explored in \citetalias{Skinner_2015}. 

Although the results from these simulations are suggestive and provide intuition, they do not fully resolve the question of how radiation feedback regulates $\epsilon_*$ in different environments. This is primarily because these studies probe a relatively narrow range in parameter space, specifically in surface density $\Sigma$, which is expected to be the most important environmental parameter when it comes to determining the value of $f_{\rm trap}$, and thus the effectiveness of IR radiation feedback \citep{Murray_2010_clusters}. Recent semi-analytic work by \citet[\citetalias{Crocker_2018b} hereafter]{Crocker_2018b} suggests the existence of a critical surface density $\Sigma_{\mathrm{crit}}$, such that clouds with $\Sigma>\Sigma_{\mathrm{crit}}$ may be unstable to the driving of winds by radiation pressure, effectively regulating $\epsilon_*$. \citetalias{Crocker_2018b}'s proposed surface density threshold is 
\begin{equation}
    \Sigma_{\mathrm{crit}} \simeq 1.3 \times 10^{5}\,\Msun \, \mathrm{pc}^{-2}\left(\frac{Z}{Z_{\sun}}\right)^{-1}\left(\frac{\Psi(Z)}{\Psi_{\sun}}\right)^{-1},
\end{equation}
where $Z$ is the gas metallicity, $\Psi$ is the light to mass ratio of the stellar population, and the subscript $\sun$ indicates values evaluated at Solar metallicity. The value of $\Sigma_{\rm crit}$ is comparable to the observed surface densities of the densest star clusters, but is larger than the highest $\Sigma$ explored in simulations thus far by about 2 orders of magnitude. Another important caveat of the study is the use of idealised, spatially and temporally constant opacities, while the true opacity is dependent on the (local) dust and radiation temperatures. The temperature dependence of the opacity has been argued by \citetalias{Crocker_2018b} to be instrumental in giving rise to a critical surface density. In addition, it is the temperature dependence of opacities that produces subtle interactions of radiation and matter that govern the evolution of such systems \citep{Krumholz_Thompson_2012}.

Another limitation of \citetalias{Skinner_2015} is associated with the Moment-1 ($M_1$) approximation for RHD that is used in their simulations, which is known to produce incorrect radiation field distributions with multiple/extended sources \citep[see, for instance, Fig.~13 in][]{Menon_2022}. In addition, the $M_1$ approach has been shown to underestimate the effective radiation forces, and to produce qualitatively different outcomes than are obtained with more accurate methods such as the Variable Eddington Tensor closure \citep[VET;][]{Davis_2014,Zhang_2017} or direct solutions of the radiative transfer equation \citep{Jiang_2021}. Although \citet{Tsang_2018} use an implicit Monte-Carlo RHD method in their study, which has been shown to be of comparable accuracy to VET \citep{Tsang_2015}, the fact that this study examined only a single point in parameter space makes it hard to draw conclusions about whether the $M_1$ results are robust more generally. 

In this study we address these limitations of prior work by studying the role of radiation feedback in regulating star formation in turbulent, self-gravitating clouds with more realistic IR dust opacities, dependent on temperature and density \citep[e.g.,][]{Semenov_2003}. We use the recently developed \texttt{VETTAM} \citep{Menon_2022} algorithm for our simulations, which uses a VET method for radiation transfer. We explore clouds with different $\Sigma$, both below and beyond $\Sigma_{\mathrm{crit}}$, to test the existence of a critical surface density, and if so, constrain its value. We also test environments that might be dust-enriched, to see what role this might play in controlling $\epsilon_*$. Our simulations help understand star cluster formation in extreme environments such as the SSC-forming clouds, which are present-day counterparts of the precursors of globular clusters \citep{Adamo_2020}. They also shed light on the intricacies governing the competition of IR radiation pressure and gravity, which is relevant to other systems, such as the driving of dusty winds in starburst galaxies \citep[e.g.,][]{Zhang_2018b, Zhang_2018} or AGN-driven outflows \citep{Bieri_2017, Costa_2018}.

The paper is organised as follows: In Section~\ref{sec:Methods} we describe the equations solved in our simulations, the numerical prescriptions we use, and the initial conditions of our clouds. 
In Section~\ref{sec:Results} we present the results of our simulation suite, exploring the effects of different treatments of the opacity and different environmental conditions.
In Section~\ref{sec:Discussion} we discuss the implications of our results, and in Section~\ref{sec:Conclusions} we conclude with a brief summary.

\section{Methods}
\label{sec:Methods}

\subsection{Equations solved}
\label{sec:Equations}
In this study, we run three-dimensional RHD simulations of gas clouds on Cartesian grids with gravity, and sink particles \citep{Federrath_2010_Sinks} to represent unresolved stellar clusters as point masses. We solve the non-relativistic, gray, RHD equations in the mixed-frame formulation \citep{Mihalas_1982}, retaining terms that are of leading order in all limiting regimes of RHD \citep[see, e.g., ][]{Krumholz_2007a}, given by 
\begin{gather}
	\frac{\partial \rho}{\partial t} + \nabla \cdot (\rho \textbf{v}) = 0 \label{eq:continuityeq} \\
	\frac{\partial (\rho\textbf{v})}{\partial t} + \nabla \cdot (\rho \textbf{v}\textbf{v}) = - \nabla P - \rho \nabla \Phi + \mathbfit{G} \label{eq:gasMom} \\
	\frac{\partial E_r}{\partial t} + \nabla \cdot \mathbfit{F} = -cG^0 + \dot{{\mathbfit{j}_*}} \label{eq:Erad} \\
	\frac{\partial \mathbfit{F}}{\partial t} + \nabla \cdot (c^2E_r\mathbfss{T}) = -c^2\mathbfit{G} \label{eq:Frad} \\
	P = \cs^2 \rho,
\end{gather}
where,
\begin{equation}
    \begin{aligned}
    G^0 =\;& \rho \kappaP \left(E_r - a_RT^4\right) + \rho \left(\kappaR - 2\kappaP \right) \frac{\mathbfit{v} \cdot \mathbfit{F}}{c^2} \\
    &+ \rho \left( \kappaP - \kappaR \right) \left[\frac{v^2}{c^2}E_r + \frac{\mathbfit{v}\mathbfit{v}}{c^2} :\mathbfss{P}_r\right],
\end{aligned}
\end{equation}
and
\begin{equation}
\label{eq:G}
\mathbfit{G} = \rho \kappaR \frac{\mathbfit{F}}{c} - \rho \kappaR E_r\frac{\mathbfit{v}}{c} \cdot (\mathbfss{I} + \mathbfss{T}),
\end{equation}
are the time-like and space-like parts of the specific radiation four-force density for a direction-independent flux spectrum \citep{Mihalas_2001} to leading order in all regimes. 
In the above equations $\rho$ is the mass density, $P$ the gas thermal pressure, $\mathbfit{v}$ the gas velocity, $\Phi$ the gravitational potential, $T$ the gas temperature, $\mathbfss{I}$ the identity matrix, and $c$ the speed of light in vacuum. In the radiation moment equations (Equations~\ref{eq:Erad} \& \ref{eq:Frad}), $E_r$ is the lab-frame radiation energy density, $\mathbfit{F}$ the lab-frame radiation momentum density, $\mathbfss{P}_r$ is the lab-frame radiation pressure tensor, and $\kappa_{\rm P}$ and $\kappa_{\rm R}$ are the Planck and Rosseland mean opacities. Note that we denote tensor contractions over a single index with dots (e.g., $\mathbfit{a} \cdot \mathbfit{b}$), tensor contractions over two indices by colons (e.g., \mathbfss{A}:\mathbfss{B}), and tensor products of vectors without an operator symbol (e.g., $\mathbfit{a}\mathbfit{b}$). 

The radiation closure relation is used to close the above system of equations, and is of the form
\begin{equation}
\label{eq:closure}
\mathbfss{P}_r = \mathbfss{T}E_r,
\end{equation}
where $\mathbfss{T}$ is the Eddington Tensor. We use an Eddington tensor directly calculated from angular quadratures of the frequency-averaged specific intensity $I_r(\hat{\mathbfit{n}}_k)$, using the relations
\begin{gather}
\label{eq:angularmoment0}
E_r=\int_{0}^{\infty} d \nu \int d \Omega\, I_r(\hat{\mathbfit{n}}_k, \nu)/c, \\
\mathbfss{P}_r=\int_{0}^{\infty} d \nu \int d \Omega\, \hat{\mathbfit{n}}_k \hat{\mathbfit{n}}_k\,I_r(\hat{\mathbfit{n}}_k, \nu)/c,
\label{eq:angularmoment2}
\end{gather}
where $I_r$ as a function of the spatial path length $s$ is calculated from a formal solution of the time-independent radiative transfer equation,
\begin{equation}
    \frac{\partial I_r}{\partial s} = \rho \kappa(S - I_r), \label{eq:RTeq}
\end{equation}
where $S$ is the source function, which, for the purposes of modelling the emission from dust grains, we set equal to the frequency-integrated Planck function, $B(T_{\rm d}) = ca_RT_{\rm d}^4/(4\pi)$, where $a_R$ is the radiation constant and $T_{\rm d}$ is the temperature of the dust grains that the radiation field most strongly interacts. The expression in Equation~\ref{eq:RTeq} neglects scattering, and assumes that the dust emits and absorbs radiation in the co-moving frame with the same grey opacity $\kappa = \int_{0}^{\infty} \kappa(v_{0}) d\nu_0$, where $\kappa(v_{0})$ is the material opacity at frequency $\nu_0$. In addition, we also ignore $\mathcal{O}(v/c)$ terms in this equation, which arise from the mixed-frame formulation, since we expect the contribution of these terms to the Eddington tensor to be relatively small. We also require a closure relation for the gas pressure $P$, the details of which we discuss in Section~\ref{sec:thermodynamics}.

\subsection{Numerical scheme}
\label{sec:NumMethod} 
We solve the equations outlined in the previous section using the Variable Eddington Tensor-closed Transport on Adaptive Meshes (\texttt{VETTAM}; \citealt{Menon_2022}) method, a state-of-the-art RHD algorithm integrated into a modified version of the \texttt{FLASH} magneto-hydrodynamics code \citep{Fryxell_2000,Dubey_2008}. \texttt{FLASH} solves the hydrodynamic equations on an Eulerian mesh, with Adaptive Mesh Refinement \citep{Berger_1989} using the \texttt{PARAMESH} library \citep{Macneice_2000}.  

\subsubsection{Hydrodynamics}
The hydrodynamic updates are performed using pre-existing infrastructure available in \texttt{FLASH}. We use an explicit Godunov method in the split, five-wave HLL5R (approximate) Riemann solver \citep{Waagan_2011}. This solver uses a piecewise linear reconstruction of the conserved variables (i.e., second-order accuracy) in such a way to ensure positivity, and has been shown to be of comparable accuracy to the Roe Riemann solver, but more stable \citep{Waagan_2011}.

\subsubsection{Gas and dust temperatures}
\label{sec:thermodynamics}
We assume an isothermal equation of state in our simulations, i.e. $P=\cs^2 \rho$, where $\cs$ is the thermal sound speed of the gas. We do not expect this approximation to impact our results significantly as the thermal pressure would be subdominant over the radiation pressure in our simulations, and presumably plays a minor role in affecting the dynamics of our clouds\footnote{It is possible that thermal pressure contributes to the dynamics at late times in some of our simulations, when radiation manages to push most of the gas in the cloud to the outskirts of the domain. However, at this point, most of the gas is already removed from the cloud by radiation pressure, and hence, thermal pressure does not play a significant role.}.

We now comment on our treatment of $T_{\rm d}$ in this study. Ideally, one would use a scheme where the gas, dust and radiation temperatures are treated separately and consistently, obtained by solving the coupled equations of their respective heating and cooling mechanisms \citep[e.g.,][]{Bate_2015}. Instead, we invoke the condition of radiative equilibrium, i.e. that all radiation absorbed by dust is instantly re-emitted locally, and $T_{\rm d} = \Tr$, which, in effect, means setting the radiation-dust energy exchange term $\rho \kappaP c (\ar T_{\rm d}^4 - E_r)$ to zero. This is a very reasonable approximation; we refer the reader to Appendix A of \citet{Krumholz_Thompson_2013} where this is justified.

The approximations mentioned above, in effect, implies that the gas temperature is completely decoupled from $T_{\rm d}$, and thus, $\Tr$. However, this assumption fails at the higher densities we probe ($n > 10^4-10^5 \, \pcm$), where dust and gas temperatures are expected to become coupled to each other by collisions, and this in turn couples the gas temperature to the radiation field (i.e., $T_{\rm d} = T_{\mathrm{g}} = \Tr$). In this limit, $T_{\mathrm{g}}$, and hence the associated thermal pressure $P$, would be affected by the distribution of $\Tr$, and our isothermal approximation is not correct. However, this matters little for the purposes of this study since, as noted above, thermal pressure is not a significant force in the systems we are simulating, regardless of the value of $T_{\rm g}$. The primary harm in using an incorrect gas pressure is that the fragmentation properties in our simulations are unreliable. However, as we discuss below, we lack the resolution to attempt to reproduce fragmentation into individual stars in any event. By contrast, properties like the star formation efficiency, and the radiation-gravity force balance -- which comprise the focus of this paper -- are unaffected. 

\subsubsection{Gravity and star formation}

We use a multi-grid algorithm implemented in \texttt{FLASH} \citep{Ricker_2008} to solve the Poisson equation to obtain the gas gravitational potential. To follow the evolution of gas at unresolved scales, we use sink particles, formed when the local gas properties satisfy a set of conditions \citep[see][]{Federrath_2010_Sinks}, which we briefly describe here. Sink particles may form when the local gas density exceeds the Jeans threshold density $\rho_{\mathrm{sink}}$, given in terms of the Jeans length $\lambda_{\mathrm{J}}$,
\begin{equation}
    \rho_{\mathrm{sink}} = \frac{\pi \cs^2}{G \lambda_{\mathrm{J}}^2}, \label{eq:rho_sink}
\end{equation}
where $\cs$ is the local sound speed, and $G$ the gravitational constant. The Jeans length is linked to the accretion radius of the sink particle by $r_{\mathrm{sink}} = \lambda_{\mathrm{J}}/2$. To avoid artificial fragmentation \citep{Truelove_1997}, the sink particle diameter is set to $2r_{\mathrm{sink}} = 5 \Delta x_{\mathrm{min}}$, where $\Delta x_{\mathrm{min}}$ is the cell size at the highest level of AMR. However, the condition that a local region exceeds the density threshold given by Eq.~(\ref{eq:rho_sink}) is only a necessary, but not sufficient condition for sink particle formation. In addition, the algorithm checks that all gas within the control volume with radius $r_{\mathrm{sink}}$ is gravitationally bound (taking into account thermal, kinetic, and magnetic energies), collapsing towards the centre of the control volume, and does not already contain a previously formed sink particle -- only then is a sink particle formed. This procedure is crucial to avoid spurious formation of sink particles \citep{Federrath_2010_Sinks}. Once created, a sink particle can accrete gas, if computational cells within $r_{\mathrm{sink}}$ exceed $\rho_{\mathrm{sink}}$, and satisfy the additional conditions above. During accretion, mass, momentum and angular momentum are conserved. We compute the gravitational interaction of sink particles with the gas by direct summation over all sink particles and grid cells \citep{FederrathBanerjeeSeifriedClarkKlessen2011}, and we use a second-order leapfrog integrator to advance the sink particle positions and velocities. 

\subsubsection{Radiation from sink particles}
We include the direct contribution of radiation energy from sink particles, which, given that our resolution is insufficient to capture individual stars, represent unsolved mini-star clusters. We determine sink particle luminosities by adopting a fixed light to mass ratio $\langle L_*/M_*\rangle = 1.7 \times 10^3 \, \mathrm{erg} \, \mathrm{s}^{-1} \, \mathrm{g}^{-1}$ appropriate for a young stellar population, so that a particle of mass $M_*$ has luminosity $L_* = \langle L_*/M_*\rangle M_*$. We inject this radiant energy into the simulation by adding a source term $\dot{{\mathbfit{j}_*}}$ in Equation~\ref{eq:Erad}. For simplicity, and to be consistent with earlier works, we model $\dot{{\mathbfit{j}_*}}$ as a Gaussian kernel of size $\sigma_*$, given by
\begin{equation}
    \label{eq:jstar}
    j_{*}(r)=\frac{L_{*}}{\left(2 \pi \sigma_{*}^{2}\right)^{3 / 2}} \exp \left(-\frac{r^{2}}{2 \sigma_{*}^{2}}\right), 
\end{equation}
where $r$ is the radial distance of a grid cell from the sink particle. We adopt a value of $\sigma_* = 2 \Delta x_{\mathrm{min}}$, where $\Delta x_{\mathrm{min}}$ is the minimum cell size in the domain. We demonstrate that our results are fairly insensitive to this choice by comparing runs with different $\sigma_*$ in Appendix~\ref{sec:AppendixSource}. We note that this term is meant to represent the UV radiation that has been reprocessed into the IR by dust on sub-grid scales that we do not resolve, and is somewhat idealised. A more accurate treatment would involve following the radiation field in both the UV and IR bands, and self-consistently capturing the reprocessing of UV to IR in the simulation domain. We intend to adopt this approach in future work.

\subsubsection{Radiation transport}
We operator-split the radiation moment equations from the hydrodynamic and gravity updates, and solve them using the \texttt{VETTAM} scheme described in \citet{Menon_2022}. We refer readers to that paper for a full description of the algorithm, and provide a brief summary here. \texttt{VETTAM} solves the radiation moment equations using a first-order Godunov method with a piece-wise constant reconstruction of the conserved variables, and an HLLE Riemann solver for the discretised divergence terms in the transport equation. It uses a first-order backward Euler temporal discretisation to enable time steps set by the hydrodynamic signal speed rather than the speed of light. \texttt{VETTAM} solves the moment equations with a non-local, self-consistent, radiation closure relation (Equation~\ref{eq:closure}) obtained from a solution of the time-independent radiative transfer equation (Equation~\ref{eq:RTeq}) using a hybrid characteristics ray-tracing scheme \citep{Buntemeyer_2016}. The closure relation is computed once at the start of the timestep, and kept fixed for the update in a given timestep. The discretisation, along with the estimate of $\mathbfss{T}$, leads to a coupled global system of equations, which we express in a matrix form. We use sparse matrix solvers based on Krylov subspace methods \citep{Saad_2003} to invert the matrices and obtain the (approximate) solution. \texttt{VETTAM} uses the open-source \texttt{PETSc} library \citep{PetscRef} for this purpose. 

We note that the solution of the time-independent transfer equation obtained with our ray-tracer provides the equilibrium solution for the given gas distribution. This would represent the true solution for situations where the timescale of radiation is short compared to dynamical timescales. While this condition is largely satisfied, it is possible that for some of our simulations with high optical depths, the timescale of radiation diffusion be comparable to dynamical timescales, and therefore the solution of the radiation intensity with the ray-tracer may not be entirely accurate. However, since this is only used for calculating $\mathbfss{T}$ -- the radiation is still transported with a time-dependent moment method that retains the right signal speeds -- we do not expect any significant impacts on our results.

\subsection{Dust IR opacities}
\label{sec:dustopacities}

\begin{figure}
    \centering
    \includegraphics[width = 0.5 \textwidth]{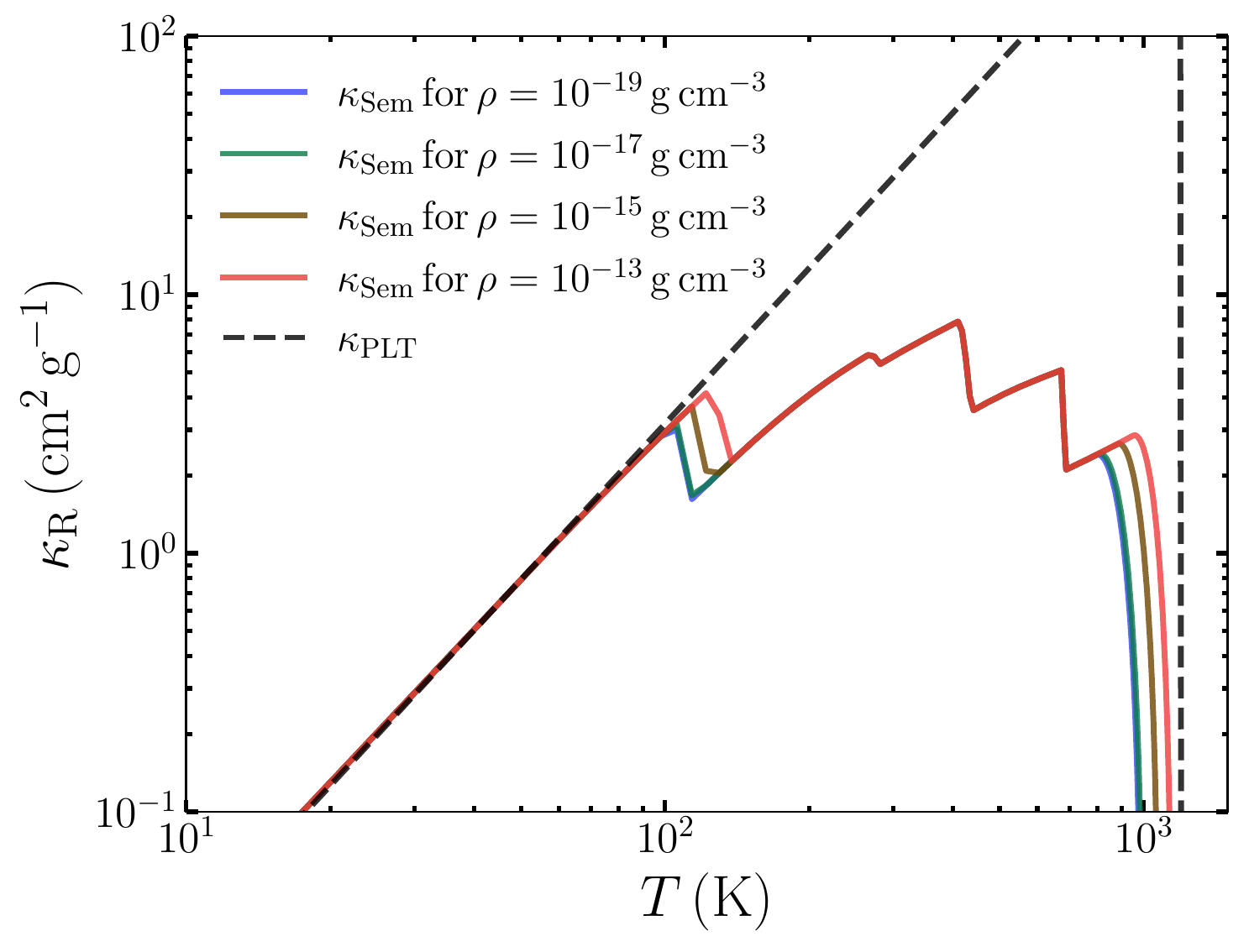}
    \caption{Plot showing $\kappaSem(\rho, \Tr)$, the IR Rosseland opacities obtained with the \citet{Semenov_2003} model as a function of radiation temperature ($\Tr$) for various gas densities $\rho$ (solid lines). Black dashed lines indicate $\kappaPLT$, the IR opacities obtained with the power-law approximation given by Equation~\ref{eq:kappaT2}. The sharp drop for this curve is associated with a floor of $\kappaR = 10^{-3}\, \mathrm{cm}^2 \, \mathrm{g}^{-1}$ we set to emulate dust sublimation for $T>T_\mathrm{sub}$, where $T_{\mathrm{sub}} = 1200 \, \mathrm{K}$ is the sublimation temperature of dust assumed.}
    \label{fig:SemenovOpacities}
\end{figure}

The IR dust opacity is central to the magnitude of the radiation force and thus the competition between radiation and gravity forces in this problem. We thus test three different (approximate) dust opacity laws in our study, which represent three different levels of accuracy. Since we neglect the gas-radiation energy exchange term, we take $\kappaP = 0$, and the only grey opacity in the problem is the Rosseland opacity $\kappaR$. Firstly, we test the simple assumptions of a spatially and temporally constant grey opacity with value $\kappa_{\mathrm{c}}$ as a starting point, and to serve as a tool of comparison with other works that adopt this approximation \citep{Skinner_2015,Tsang_2018}. However, this is not expected to be physically accurate, as the IR dust opacity is strongly dependent on the both the radiation colour temperature and the dust temperature -- which are nearly identical in an optically thick environment such as the one we are simulating (see Section~\ref{sec:Equations}). As discussed earlier in Section~\ref{sec:Introduction}, the temperature dependence of the IR opacities is important for force balance, an effect we intend to explore in this study. 

Our second approach is therefore to run simulations with a temperature- (and density-) dependent \citet{Semenov_2003} opacity law, based on the micro-physical, optical properties of dust grains. We refer to the $\kappaR$ returned by the Semenov opacity model as $\kappa_{\mathrm{Sem}} \equiv \kappa_{\mathrm{Sem}} (\rho, \Tr)$, based on a publicly available code\footnote{\url{https://www2.mpia-hd.mpg.de/~semenov/Opacities/opacities.html}} by \citet{Semenov_2003}. We use the model of ``normal'' silicates (NRM) with a relative iron content in silicates of Fe/(Fe+Mg) = 0.3, and assume that dust grains can be modelled as composite aggregates. However, we demonstrate in Appendix~\ref{sec:AppendixDustModel} that adopting different dust models has a negligible effect on our results. We show example curves of $\kappaSem$ versus $\Tr$ in Figure~\ref{fig:SemenovOpacities}. The discontinuous features in the curves of $\kappaSem$ arise due to different chemical species in the dust sublimating at those temperatures \citep{Semenov_2003}. The exact sublimation temperature of various species depends on the gas (or dust) density $\rho$, which leads to the subtle differences in the curves for different values of $\rho$ near the discontinuities.\footnote{There is a small subtlety here worth pointing out. The \citet{Semenov_2003} model is derived under the assumption of full local thermodynamic equilibrium between dust, gas, and radiation, so a single temperature characterises all three, but at lower densities we are in a regime where $T_{\rm r}$ and $T_{\rm d}$ are nearly equal, but $T_{\rm g}$ may be different. This raises the question of which temperature we should use when evaluating the \citeauthor{Semenov_2003} model opacity. We choose to use $T_{\rm r} = T_{\rm d}$, because this captures the two most important temperature-dependent effects: the change in quantum efficiency of grain absorption in response to the change in $\lambda/a$, where $\lambda$ is the radiation wavelength and $a$ is the grain size, and the sublimation of dust grains once their temperature rises too high. The main effect that our approximation misses is that changes in the rate of gaseous deposition onto grains (or spallation off grains) depends on the kinetic energy of gas molecules, and thus on $T_{\rm g}$. However, this is a second-order effect compared to quantum efficiency and grain sublimation, so, in the absence of a more general grid of dust models that considers the case $T_{\rm g}\neq T_{\rm d}$, using $T_{\rm r} = T_{\rm d}$ in the \citealt{Semenov_2003} model is the best alternative.}

Our third method of treating opacities is to explore simulations where we adopt a power-law dependence on the temperature for $\kappaR$, which we refer to as $\kappaPLT$ given by 
\begin{equation}
    \kappaPLT (\Tr) = \left\{
    \begin{array}{ll}
    10^{-1.5}\,\mathrm{cm^2\,g^{-1}} \left( \frac{\Tr}{10\,\mathrm{K}} \right)^2, & \Tr < 1200\,\mathrm{K} \\
    10^{-3} \,\mathrm{cm^2\,g^{-1}}, & \Tr \geq 1200\,\mathrm{K}
    \end{array}
    \right.
    .
    \label{eq:kappaT2}
\end{equation}
This opacity law has been adopted in various studies that investigated balance of radiation and gravity forces in idealised experiments \citep[e.g.,][]{Krumholz_2012,Krumholz_Thompson_2013,Davis_2014}, and is a central ingredient in the semi-analytical model of \citet{Crocker_2018b}; note that the break at 1200 K in this model simply represents the temperature at which dust sublimes and the opacity drops to tiny values. While this approximation is in accord with the model of \citet{Semenov_2003} at temperatures $\leq 150 \,\mathrm{K}$, at higher temperatures, the values of $\kappaR$ are greatly overestimated. We demonstrate this by comparing $\kappaPLT$ as a function of $\Tr$ with $\kappaSem$ in Figure~\ref{fig:SemenovOpacities}. Even at low values of $\Tr \sim 300 \, \mathrm{K}$, we see that $\kappaR$ is overestimated by a factor $\sim 2$, and this gets progressively worse at higher values of $\Tr$. 

\subsection{Initial and boundary conditions}
\label{sec:setup}
We now describe our initial and boundary conditions. We initialise our simulations as a uniform spherical cloud with mass ($\Mcloud$) and radius $\Rcloud$, which together define a cloud mass density $\rho_{\mathrm{cloud}} = \Mcloud/[(4/3) \pi \Rcloud^3]$ and a mass surface density $\Sigmacloud = \Mcloud/(\pi \Rcloud^2)$. We list the value of $\Mcloud$ and $\Rcloud$ for all our simulations in Table~\ref{tab:Simulations}.

In order to ensure pressure equilibrium, we place the cloud in a lower-density, hotter, ambient medium. The density and temperature contrasts are set such that $\rho_{\mathrm{cloud}} = \chi\,\rho_{\mathrm{ambient}}$ and $T_{\mathrm{cloud}} = \chi^{-1}\,T_{\mathrm{ambient}}$, with $\chi=10^2$. We further initialise a mass scalar to represent the cloud material $C_{\mathrm{cloud}}$, whose value lies between 0 and 1, representing purely ambient and cloud material, respectively. This allows us to set the correct thermal pressure for a mixture of cloud and ambient material at all times during the evolution. The thermal pressure $P$ is then given by 
\begin{equation}
    P = \cs^2 \rho \left[C_{\mathrm{cloud}} + \chi \left(1 - C_{\mathrm{cloud}}\right)\right].
\end{equation}
The side length of the cubic computational domain is set to $L = 4\,\Rcloud$, such that there is sufficient volume to track potentially expanding material (due to radiation feedback) before it leaves the domain through the boundaries.

We initialise the velocities in our clouds with supersonic turbulent fluctuations that follow a power spectrum $E(k) \propto k^{-2}$ \citep{FederrathDuvalKlessenSchmidtMacLow2010,Federrath2013,FederrathEtAl2021} for $k/(2\pi/L) \in \left[2,64 \right]$, using the publicly available code by \citet{FederrathEtAl2022ascl}. We scale the magnitude of the fluctuations such that we obtain a target initial velocity dispersion $\sigma_v$, such that we obtain a target virial parameter \citep[e.g.,][]{Federrath_klessen_2012},
\begin{equation}
    \alphavir = \frac{2E_{\mathrm{kin}}}{E_{\mathrm{grav}}} = \frac{5\Rcloud \sigma_v^2}{3G\Mcloud},
    \label{eq:alphavir}
\end{equation}
where $E_{\mathrm{kin}} = (1/2) \Mcloud \sigma_v^2$ and $E_{\mathrm{grav}} = (3/5) G \Mcloud^2/\Rcloud$. Most of our simulations use $\alphavir = 2$, but we also vary this parameter in some cases; again, we provide a summary of all our simulations in Table~\ref{tab:Simulations}.

We note that the initial turbulence imposed on the clouds, is not maintained by continuous driving, and is instead allowed to decay\footnote{We note, however, that the gravitational collapse and radiation could itself drive turbulent motions over the course of the simulation.}. This misses the turbulent energy that might cascade from larger scales (effectively from the surrounding ISM, in which the cloud is embedded, but not included in our simulations), and could therefore give gravity an unfavourable advantage, leading to a higher star formation rate than would be expected in reality \citep[see, e.g.,][]{Lane_2022}. However, since our goal is to perform a controlled experiment to study the effects of radiation feedback, this matters little for our purpose, since even if our star formation rate is too high overall, we can still determine the relative efficiency of simulations with differing treatments of IR radiation.

We set the sound speed $\cs$ such that the sonic Mach number $\mathcal{M} = \sigma_v/\cs = 11.5$, identical to the value of $\mathcal{M}$ used by \citet{Skinner_2015}. We note that the resulting values of $\cs$ for most of our simulations are much higher than would be expected for the cold, dense, star-forming phase of the ISM that we are simulating. Our choice is dictated largely by numerical convenience, since extremely large values of $\mathcal{M}$, as would be realistic for the clouds we are simulating, require extremely small hydrodynamic time steps to remain numerically stable. As noted above, we do not expect this choice to matter significantly because the thermal pressure is negligible compared to the radiation pressure, making its exact value dynamically unimportant. We do not include magnetic fields in any of our simulations. We initialise the radiation field such that the initial radiation temperature is $\Tr = T_{\mathrm{r},0} = 40 \, \mathrm{K}$ everywhere, and the radiation flux is zero.

We use diode boundary conditions for the hydrodynamics, i.e., gas is allowed to flow out of the domain, but not allowed to enter it. For the radiation moment equations, we adopt Marshak boundary conditions \citep{Marshak_1958}, such that the domain is bathed in a radiation field with temperature $T_{\mathrm{r},0} = 40 \, \mathrm{K}$, but also allows radiation generated inside the domain to escape freely. Consistent with this choice, for the purpose of solving the time-independent RT equation (Equation~\ref{eq:RTeq}) for the Eddington tensor $\mathbfss{T}$, we set the intensity of rays entering the domain to a value given by the frequency-integrated Planck function evaluated at temperature $T_{\mathrm{r},0}$. This ensures that the boundary conditions in the two independent forms of the transfer equation solved by our scheme are consistent with each other, which has been shown to be important for stability and accuracy \citep[e.g.,][]{Park_2012,Davis_2014}.

\subsection{Simulation suite}

\renewcommand{\arraystretch}{1.0}
\begin{table*}
\caption{Main simulation parameters.}
\centering
\label{tab:Simulations}
\begin{threeparttable}
\begin{tabular}{l c c c c c c c c c}
\toprule
\multicolumn{1}{l}{Model}& \multicolumn{1}{c}{$M_{\mathrm{cloud}}$}& \multicolumn{1}{c}{$R_{\mathrm{cloud}}$}& \multicolumn{1}{c}{$\Sigma_{\mathrm{cloud}}$}& \multicolumn{1}{c}{$n_{\mathrm{cloud}}$}& \multicolumn{1}{c}{$\alpha_{\mathrm{vir}}$}& \multicolumn{1}{c}{$\sigma_{v}$}& \multicolumn{1}{c}{$\mathcal{M}$}& \multicolumn{1}{c}{$f_{\mathrm{DG}}$}& \multicolumn{1}{c}{$\kappa_{\mathrm{R}}$}\\
& [$10^6 \, \Msun$]& [pc]& [$\Msolpc$]& [$\mathrm{cm}^{-3}$]& & [km/s]& & \\
\midrule
\multicolumn{9}{c}{\large \underline{$\kappa_{\mathrm{c}}$ Series: Section~\ref{sec:ConstantOpacity}}}\\
\texttt{S3K01A2F1} &$1.0$ &$10.0$ &$3.2$$ \times 10^{3}$ &$9.7$$ \times 10^{3}$ &2 &$23$ &11.5 &1 &$\kappa_{\mathrm{c}}=$$1$$ \, \mathrm{cm}^2 \, \mathrm{g}^{-1}$\\
\texttt{S3K05A2F1} &$1.0$ &$10.0$ &$3.2$$ \times 10^{3}$ &$9.7$$ \times 10^{3}$ &2 &$23$ &11.5 &1 &$\kappa_{\mathrm{c}}=$$5$$ \, \mathrm{cm}^2 \, \mathrm{g}^{-1}$\\
\texttt{S3K10A2F1} &$1.0$ &$10.0$ &$3.2$$ \times 10^{3}$ &$9.7$$ \times 10^{3}$ &2 &$23$ &11.5 &1 &$\kappa_{\mathrm{c}}=$$10$$ \, \mathrm{cm}^2 \, \mathrm{g}^{-1}$\\
\texttt{S3K20A2F1} &$1.0$ &$10.0$ &$3.2$$ \times 10^{3}$ &$9.7$$ \times 10^{3}$ &2 &$23$ &11.5 &1 &$\kappa_{\mathrm{c}}=$$20$$ \, \mathrm{cm}^2 \, \mathrm{g}^{-1}$\\
\texttt{S3K40A2F1} &$1.0$ &$10.0$ &$3.2$$ \times 10^{3}$ &$9.7$$ \times 10^{3}$ &2 &$23$ &11.5 &1 &$\kappa_{\mathrm{c}}=$$40$$ \, \mathrm{cm}^2 \, \mathrm{g}^{-1}$\\
\texttt{S3K80A2F1} &$1.0$ &$10.0$ &$3.2$$ \times 10^{3}$ &$9.7$$ \times 10^{3}$ &2 &$23$ &11.5 &1 &$\kappa_{\mathrm{c}}=$$80$$ \, \mathrm{cm}^2 \, \mathrm{g}^{-1}$\\
\midrule
\multicolumn{9}{c}{\large \underline{$\Sigmacloud$ Series: Section~\ref{sec:ResultsMW}}}\\
\texttt{S3KsemA2F1} &$1.0$ &$10.0$ &$3.2$$ \times 10^{3}$ &$9.7$$ \times 10^{3}$ &2 &$23$ &11.5 &1 &$\kappa_{\mathrm{Sem}}$\\
\texttt{S4KsemA2F1} &$1.0$ &$3.2$ &$3.2$$ \times 10^{4}$ &$3.1$$ \times 10^{5}$ &2 &$40$ &11.5 &1 &$\kappa_{\mathrm{Sem}}$\\
\texttt{S5KsemA2F1} &$1.0$ &$1.0$ &$3.2$$ \times 10^{5}$ &$9.7$$ \times 10^{6}$ &2 &$72$ &11.5 &1 &$\kappa_{\mathrm{Sem}}$\\
\texttt{S6KsemA2F1} &$1.0$ &$0.3$ &$3.2$$ \times 10^{6}$ &$3.1$$ \times 10^{8}$ &2 &$128$ &11.5 &1 &$\kappa_{\mathrm{Sem}}$\\
\midrule
\multicolumn{9}{c}{\large \underline{$\alphavir$ Series: Section~\ref{sec:alphavir}}}\\
\texttt{S5KsemA3F1} &$1.0$ &$1.0$ &$3.2$$ \times 10^{5}$ &$9.7$$ \times 10^{6}$ &3 &$88$ &11.5 &1 &$\kappa_{\mathrm{Sem}}$\\
\texttt{S5KsemA4F1} &$1.0$ &$1.0$ &$3.2$$ \times 10^{5}$ &$9.7$$ \times 10^{6}$ &4 &$102$ &11.5 &1 &$\kappa_{\mathrm{Sem}}$\\
\midrule
\multicolumn{9}{c}{\large \underline{$\fdustgas$ Series: Section~\ref{sec:ResultsDG}}}\\
\texttt{S4KsemA2F2} &$1.0$ &$3.2$ &$3.2$$ \times 10^{4}$ &$3.1$$ \times 10^{5}$ &2 &$40$ &11.5 &2 &$\kappa_{\mathrm{Sem}}$\\
\texttt{S4KsemA2F3} &$1.0$ &$3.2$ &$3.2$$ \times 10^{4}$ &$3.1$$ \times 10^{5}$ &2 &$40$ &11.5 &3 &$\kappa_{\mathrm{Sem}}$\\
\texttt{S5KsemA2F2} &$1.0$ &$1.0$ &$3.2$$ \times 10^{5}$ &$9.7$$ \times 10^{6}$ &2 &$72$ &11.5 &2 &$\kappa_{\mathrm{Sem}}$\\
\texttt{S5KsemA2F3} &$1.0$ &$1.0$ &$3.2$$ \times 10^{5}$ &$9.7$$ \times 10^{6}$ &2 &$72$ &11.5 &3 &$\kappa_{\mathrm{Sem}}$\\
\midrule
\multicolumn{9}{c}{\large \underline{$\kappaPLT$ Series: Section~\ref{sec:PLTOpacity}}}\\
\texttt{S3KplA2F1} &$1.0$ &$10.0$ &$3.2$$ \times 10^{3}$ &$9.7$$ \times 10^{3}$ &2 &$23$ &11.5 &1 &$\kappa_{\mathrm{PL}}$\\
\texttt{S4KplA2F1} &$1.0$ &$3.2$ &$3.2$$ \times 10^{4}$ &$3.1$$ \times 10^{5}$ &2 &$40$ &11.5 &1 &$\kappa_{\mathrm{PL}}$\\
\texttt{S5KplA2F1} &$1.0$ &$1.0$ &$3.2$$ \times 10^{5}$ &$9.7$$ \times 10^{6}$ &2 &$72$ &11.5 &1 &$\kappa_{\mathrm{PL}}$\\
\texttt{S6KplA2F1} &$1.0$ &$0.3$ &$3.2$$ \times 10^{6}$ &$3.1$$ \times 10^{8}$ &2 &$128$ &11.5 &1 &$\kappa_{\mathrm{PL}}$\\
\bottomrule
\end{tabular}
\begin{tablenotes}
\small
\item \textbf{Notes}: The table is divided into subparts studying the dependence of our results on various cloud conditions (see text) with a link to the section where each particular parameter series is presented. The fiducial run in our study is \texttt{S5KsemA2F1}. Columns in order indicate -- Model: model name, $M_{\mathrm{cloud}}$: mass of cloud, $R_{\mathrm{cloud}}$: radius of cloud, $\Sigma_{\mathrm{cloud}}$: mass surface density of the cloud given by $\Sigma_{\mathrm{cloud}} = M_{\mathrm{cloud}}/(\pi R_{\mathrm{cloud}}^2)$, $n_{\mathrm{cloud}}$: number density of the cloud given by $n_{\mathrm{cloud}} = 3M_{\mathrm{cloud}}/(4 \pi R_{\mathrm{cloud}}^3m_{\mathrm{H}})$ where $m_{\mathrm{H}}$ is the mass of atomic hydrogen, $\alpha_{\mathrm{vir}}$: virial parameter under the assumption of a spherical cloud, $\sigma_{v}$: turbulent velocity dispersion of the cloud, $\mathcal{M}$: initial turbulent Mach number of the cloud, $f_{\mathrm{DG}}$: dust-to-gas ratio enhancement factor over the solar value, $\kappa_{\mathrm{R}}$: Rosseland opacity used in the model; where $\kappa_{\mathrm{c}}$ denotes a constant opacity whose value is provided, $\kappa_{\mathrm{Sem}}$ the Semenov opacity model (see Sec.~\ref{sec:dustopacities}), and $\kappa_{\mathrm{PL}}$ is the power-law opacity model from Eq.~(\ref{eq:kappaT2}).
\end{tablenotes}
\end{threeparttable}
\end{table*}

We run a range of simulations across parameter space to explore the role of the IR radiation pressure mechanism in different environments. The entire suite of simulations is summarised in Table~\ref{tab:Simulations}. Our general naming convention for simulations is \texttt{SsKkAaFf}, where \texttt{s}, \texttt{k}, \texttt{a}, and \texttt{f} describe the surface density $\Sigma$, choice of opacity $\kappa$, virial parameter $\alpha$, and dust to gas ratio relative to the solar value $f_{\rm dg}$ of the simulation; we explain values these parameters can take on in more detail below. We broadly divide the simulation suite as follows.

\subsubsection{Idealised constant opacity runs}
First, we run simulations with idealised, constant opacities, i.e., $\kappaR = \kappa_{\mathrm{c}}$, testing various values of $\kappa_{\mathrm{c}}$. The intention of these idealised experiments is to compare with previous works \citep{Skinner_2015,Tsang_2018} and identify any differences. Since we use a more accurate method for radiation transport (VET) than earlier works, these simulations allow us to explore whether the RHD method has any effect on the results. In addition, these simulations also serve as a baseline to compare with our more physical, temperature-dependent opacity runs, and highlight how realistic opacities affect IR radiation feedback. For instance, although instabilities, such as the radiative Rayleigh-Taylor instability (RRTI), operate even for constant opacity laws \citep{Jiang_2013RRTI}, they are more pronounced for temperature-dependent opacity laws \citep{Jacquet_2011,Krumholz_2012}. For this series of runs, we perform 7~simulations with values of $\kappa_{\mathrm{c}} = 1,2,5,10,20,40$ \& $80$. We use $\Mcloud = 10^6 \, \Msun$ and $\Rcloud = 10\, \mathrm{pc}$ for all the constant opacity runs, identical to the fiducial simulation of \citet{Skinner_2015}. This corresponds to a value of $\rho_{\mathrm{cloud}} = 1.6 \times 10^{-20} \, \gpcm$, and $\Sigmacloud = 3.2 \times 10^3 \, \Msolpc$. We use $\alphavir=2$ for all these runs, which yields $\sigma_v = 23 \kms$ and $\cs = 2 \kms$ (see Section~\ref{sec:setup} for details). We discuss the results of these simulations in Section~\ref{sec:ConstantOpacity}.

\subsubsection{Milky Way-like dust opacity runs}
\label{sssec:MilkyWayDust}
Here we use more realistic, temperature-dependent IR opacities, which comprises the focus of our study. We explore the effects of such opacities across parameter space, with gas surface density $\Sigmacloud$ as the most important parameter. 
We test values of $\Sigmacloud = 3.2 \times [10^3,10^4,10^5,10^6]\, \Msolpc$ with a \citet{Semenov_2003} opacity law. The lowest $\Sigmacloud$ run has the same $\Sigma$ as the fiducial run of \citetalias{Skinner_2015}, while the latter two $\Sigma$ values exceed the critical value $\Sigma_{\mathrm{crit}}$ estimated in \citetalias{Crocker_2018b}; the former two are below the limit. Thus, this parameter exploration allows us to test the existence of -- and possibly constrain -- a value of $\Sigma_{\mathrm{crit}}$ above which IR radiation pressure might control $\epsilon_*$. We obtain our target values of $\Sigmacloud$ by keeping the mass of the clouds fixed to $\Mcloud = 10^6 \Msun$, and scaling $\Rcloud$ appropriately. We also tested alternative scalings to achieve a target $\Sigmacloud$ -- by i) keeping $\Rcloud$ fixed and modifying $\Mcloud$, or ii) keeping the ratio $\Mcloud/\Rcloud$ fixed -- and found that the evolution of the clouds between these scalings are indistinguishable. We retain $\mathcal{M} = 11.5$ and $\alpha_{\mathrm{vir}} = 2.0$ -- scaling $\sigma_v$ and $\cs$ appropriately, as outlined in Section~\ref{sec:setup}. We use a value of $\fdustgas = 1$ for these simulations, i.e., Milky Way-like dust conditions, and refer to this series as the $\Sigmacloud$ series (Table~\ref{tab:Simulations}). The results of these runs are summarised in Section~\ref{sec:ResultsMW}.

We also run two simulations with values of $\alphavir = 3$ and $4$, to test the effectiveness of feedback in (initially) unbound clouds. Such clouds may exist in starburst environments and extreme dynamical encounter regions, where large non-thermal motions may be driven by the high pressures and tidal forces in their environments. The names of these simulations are \texttt{S5KsemAaF1}, where the \texttt{S5} notation indicates that these simulations have surface densities of $3.2\times 10^5\,\mathrm{g\,cm^{-2}}$, and \texttt{a} can be either \texttt{3} or \texttt{4}, indicating the value of the virial parameter. The results of these runs are summarised in Section~\ref{sec:MW_alphavir}.

\subsubsection{Super-solar dust opacity runs}
Since the dust opacity $\kappaR$ is the primary parameter of importance in determining the competition between gravity and feedback, it is interesting to consider super-solar, dust-enriched clouds, where radiation could play a more important role. It has been suggested in earlier, idealised (constant $\kappaR$) simulations \citepalias{Skinner_2015} that dust-enriched environments may reach high enough $\kappaR$ values to limit star formation. It is important to test if this is the case with a more realistic $T$-dependent $\kappaR$, and if so, constrain the conditions where IR radiation feedback is able to significantly limit star formation. We emulate super-solar dust conditions with values of $\fdustgas>1$,
simply by scaling our default \citet{Semenov_2003} opacities by $\fdustgas$, i.e., for $\fdustgas = 2$ we simply double the Milky Way-like opacity.
We choose two values of $\Sigmacloud$ -- one below the $\Sigma_{\mathrm{crit}}$ proposed in \citetalias{Crocker_2018b}, i.e., $\Sigmacloud = 3.2 \times 10^4 \, \Msolpc$, and the other above, i.e., $\Sigmacloud = 3.2 \times 10^5 \, \Msolpc$. As we will see below, these cases also happen to be the most promising models in the $\Sigmacloud$-series in terms of the potential of radiation pressure to compete against gravity. We test values of $\fdustgas = 2$ and 3, chosen roughly to match the metallicities (relative to solar) found in the most metal-rich galaxies \citep[e.g.,][]{Curti_2020}. We refer to these simulations as the $\fdustgas$ series (Table~\ref{tab:Simulations}) and summarise the results of these experiments in Section~\ref{sec:ResultsDG}.

\subsubsection{Power-law approximation for $T$-dependent opacity runs}
\label{sssec:plt}
Finally, we repeat the $\Sigmacloud$ series with the $\kappaPLT$ power-law approximation for dust IR opacities. This approximation has been adopted in the semi-analytic model of \citetalias{Crocker_2018b}, which proposes the existence of a $\Sigma_{\mathrm{crit}}$. It has also been adopted by earlier studies that explored the competition of IR radiation pressure and gravity with numerical simulations \citep[e.g.,][]{Krumholz_Thompson_2012,Krumholz_Thompson_2013,Davis_2014,Rosdahl_2015,Tsang_2015, Zhang_2017}\footnote{However, \citet{Tsang_2015} deviate from a simple power-law at $T \sim 150 \, \mathrm{K}$, capping the opacity at $\kappaPLT(150)$ }. These experiments allow us to determine the effects of adopting this approximation by comparing with the $\kappaSem$ runs. We refer to these simulations as the $\kappaPLT$ series (Table~\ref{tab:Simulations}) and describe their outcome in Section~\ref{sec:PLTOpacity}.

\subsection{Resolution and numerical setup}
All our simulations use a uniform grid (UG) resolution of $N = 256^3$ grid cells; for our domain of size $L = 4 R_{\rm cloud}$, this corresponds to a resolution $R_{\rm cloud}/\Delta x = 64$. Although \texttt{VETTAM} supports AMR, we chose to use UGs here, motivated by a balance between accuracy and computational feasibility, considering the broad parameter space we intend to explore. Since we are primarily interested in the integrated star formation efficiencies set by the competition between gravity and feedback, and not in the detailed fragmentation properties or mass functions of the star clusters formed, we do not expect our grid resolution choice to be a major limitation. We verify this expectation in Appendix~\ref{sec:AppendixResolution}, where we describe convergence tests where we vary the dependence of our results on the UG resolution, and also compare to an AMR run with a maximum effective resolution of $1024^3$ grid cells. We find that our results are reasonably converged at our fiducial resolution. We use a CFL number of 0.4 in our simulations, and a relative tolerance of $10^{-8}$ for our implicit update of the radiation moment equations. The solution to the time-independent transfer equation (Equation~\ref{eq:RTeq}) is performed with 48~rays per cell with our ray-tracing scheme, which uniformly samples angles on the unit sphere (using the healpix algorithm); however, we have also verified that using 192~angles yields nearly identical results. We run all simulations to a time $t=8\,t_{\mathrm{ff}}$, where $t_{\mathrm{ff}}$ is the free-fall time of the cloud, at which point nearly all mass has been accreted onto sinks or expelled from the computational domain by radiation forces, depending on the particular simulation model (see Tab.~\ref{tab:Simulations}).

\section{Results}
\label{sec:Results}

\subsection{Idealised constant IR opacity clouds} \label{sec:ConstantOpacity}

\begin{figure}
    \centering
    \includegraphics[width = 0.5 \textwidth]{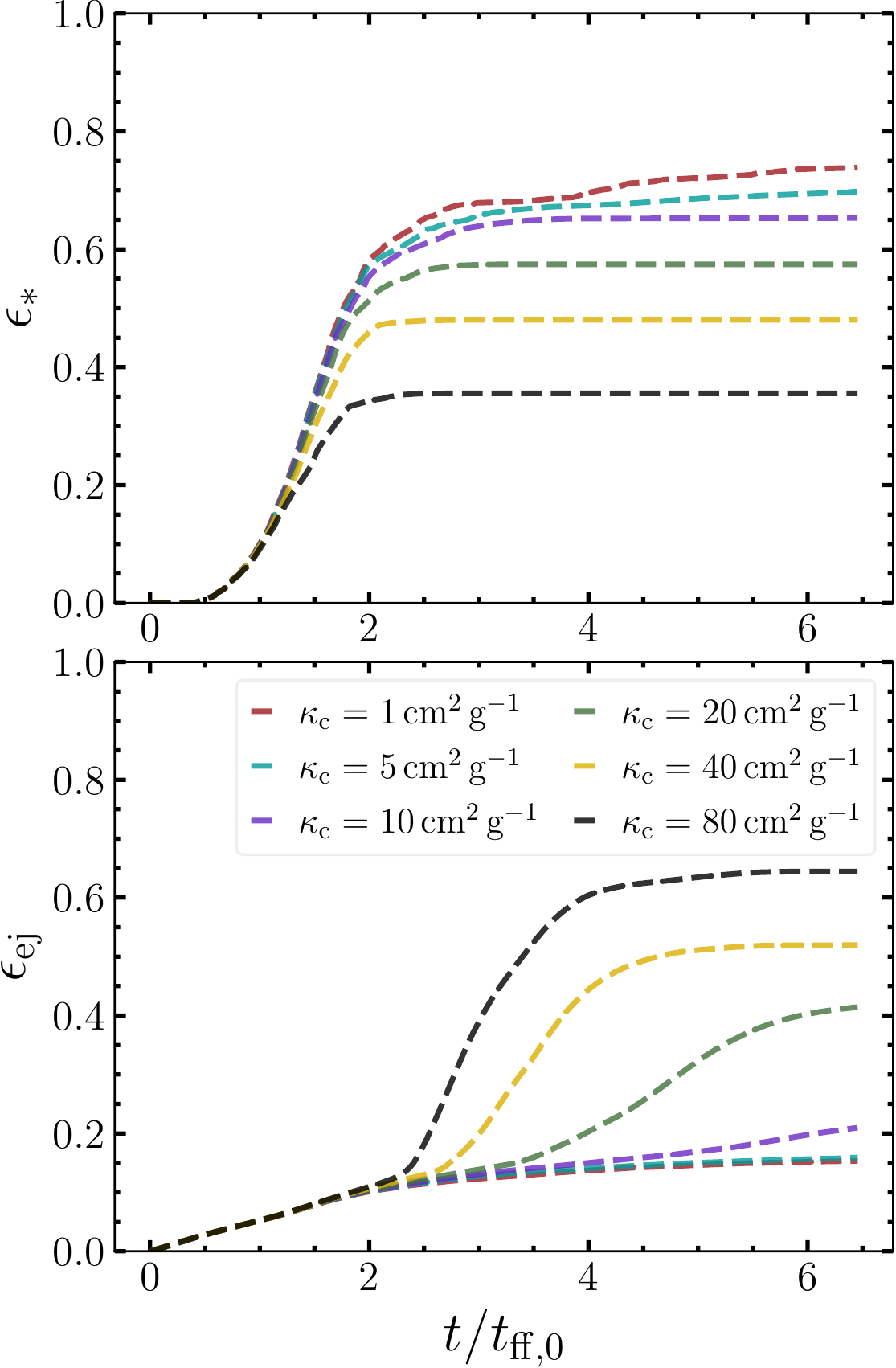}
    \caption{The star formation efficiency ($\epsilon_*$;top-panel) and fraction of mass ejected from the simulation volume ($\epsilon_{\mathrm{ej}}$;bottom-panel) for different values of $\kappa_\mathrm{c}$ ($\kappa_\mathrm{c}$ series; Table~\ref{tab:Simulations}).}
    \label{fig:ConstantKappa}
\end{figure}

\begin{figure}
    \centering
    \includegraphics[width = 0.5 \textwidth]{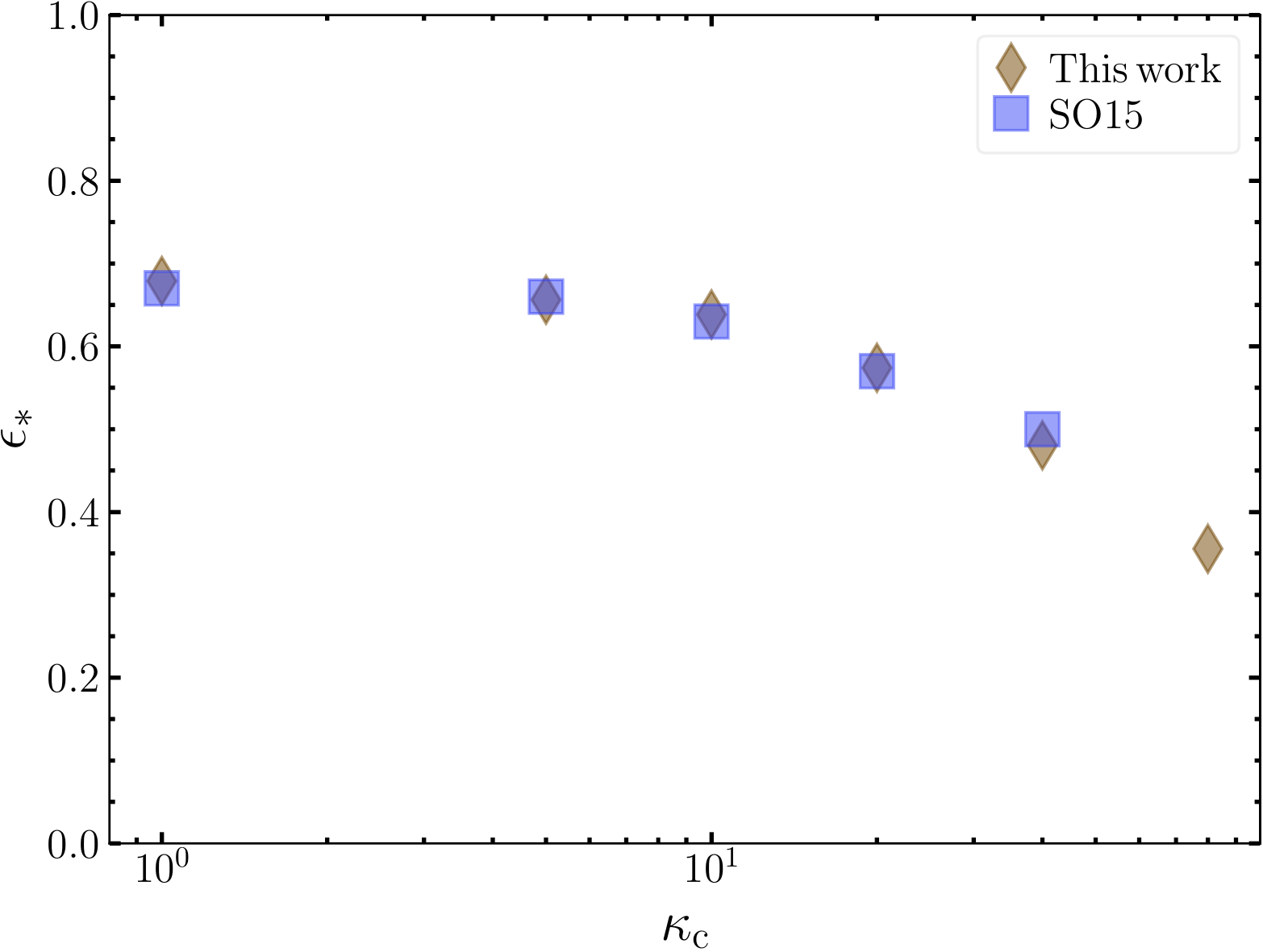}
    \caption{Comparison of $\epsilon_*$ values obtained in our study with those obtained in \citetalias{Skinner_2015}. We can see that the outcomes are nearly identical.}
    \label{fig:CompareSO15}
\end{figure}

Our first series of numerical experiments uses idealised constant opacities. We will not discuss these results in detail, since these runs simply repeat the earlier study of \citetalias{Skinner_2015}, and differ only in the numerical method used and in that we extend their study to higher opacities\footnote{\citetalias{Skinner_2015} were limited in the maximum opacity (and surface density) they could simulate by their explicit reduced-speed-of-light method for radiation, which becomes inaccurate in the dynamic diffusion regime, which is reached when the cloud optical depth is very large. The implicit Monte Carlo method used by \citet{Tsang_2015} suffers from similar limitations, and in addition becomes very computationally expensive when the optical depth is large. Our \texttt{VETTAM} method does not suffer from these limitations; see \citet{Menon_2022} for details.}. The purpose of this set of simulations is to explore whether the results depend on the numerical method, and to serve as a baseline for comparison to our runs with a more realistic treatment of opacity. For this reason, we simply present the star formation efficiencies ($\epsilon_*$; Equation~\ref{eq:sfe}) and mass-ejection fractions obtained in our simulations for different values of $\kappa_{\mathrm{c}}$, which can be compared to the earlier results. We show these quantities in Figure~\ref{fig:ConstantKappa}. We find that, qualitatively, our conclusions mirror those of \citetalias{Skinner_2015}, i.e., a constant dust opacity of $\kappa_\mathrm{c} \gtrsim 20 \, \mathrm{cm}^{2} \, \mathrm{g}^{-1}$ is required for the IR radiation feedback to make more than a small difference in the final $\epsilon_*$.

To provide a more direct quantitative comparison, we plot the $\epsilon_*$ we obtain at $t = 3 \, t_{\mathrm{ff}}$, as a function of $\kappa_{\mathrm{c}}$, along with the values reported in Table~2 of \citet{Skinner_2015} for the same quantity. We find that even quantitatively, the evolution of our clouds are very similar, suggesting that the $M_1$ radiation transport method used by \citetalias{Skinner_2015} does not yield results significantly different from our more accurate VET method for this particular setup. This is potentially because the limitations of the $M_1$ closure are more pronounced for regions close to the radiation sources, but do not play a significant role at larger distances, where the sources subtend a small solid angle \citep{Kim_2017}. Since we are in the multiple-scattering regime, it is possible that the exact nature of the radiation field in the close vicinity of the sources is not crucial in determining the dynamical outcome. In addition, the shortcomings of the $M_1$ approximation are most significant in the presence of multiple sources of radiation acting in an \textit{optically-thin} medium. In the higher $\kappa_{\mathrm{c}}$ runs where radiation is dynamically important, the optical depths are $>1$ even in lower-density channels, so the $M_1$ method performs well. On the other hand, the radiation field in the lower $\kappa_{\mathrm{c}}$ runs could be (significantly) affected by whether $M_1$ or VET is used; however, in these runs radiation forces are subdominant compared to gravity in any event, and thus inaccuracies in computing them do not change the basic outcome. Therefore, we caution that the similarity in outcomes with the $M_1$ and VET closure in these idealised simulations should not be generalised\footnote{For instance, in the single-scattering UV radiation pressure context, \citet{Kim_2017} -- who use an adaptive ray-tracing method (which is expected to be of similar accuracy to the VET) -- compare their results with \citet{Raskutti_2016}, and show that the radiation pressure is underestimated with the $M_1$ approximation, leading to a higher final $\epsilon_*$.}.

\subsection{Milky Way-like dust conditions} \label{sec:ResultsMW}

\subsubsection{Evolution of models}

\begin{figure*}
    \centering
    \includegraphics[width=0.9 \textwidth]{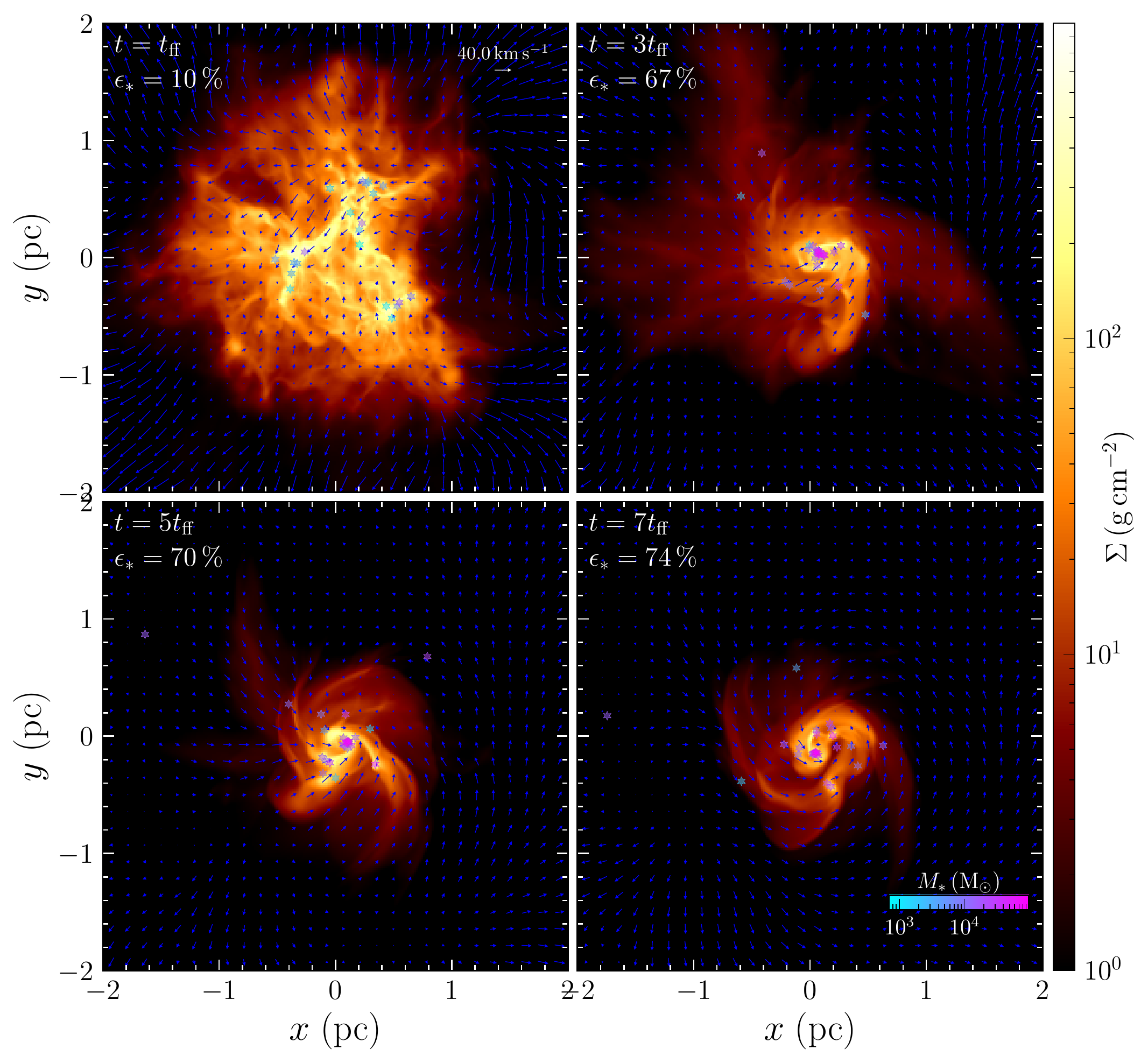}
    \caption{Surface density maps for model (\texttt{S5KsemA2F1}) at $t/t_{\mathrm{ff}}=1,3,5,7$ with the corresponding star formation efficiency ($\epsilon_*$) annotated. Star symbols indicate sink particles, coloured by their mass. Vectors (blue) indicate the mass-weighted projected velocity field, with arrow length indicating velocity magnitude. The scale for the velocity vectors is annotated in the top-left panel.}
    \label{fig:SigmaSemProjection}
\end{figure*}

\begin{figure*}
    \centering
    \includegraphics[width=0.9 \textwidth]{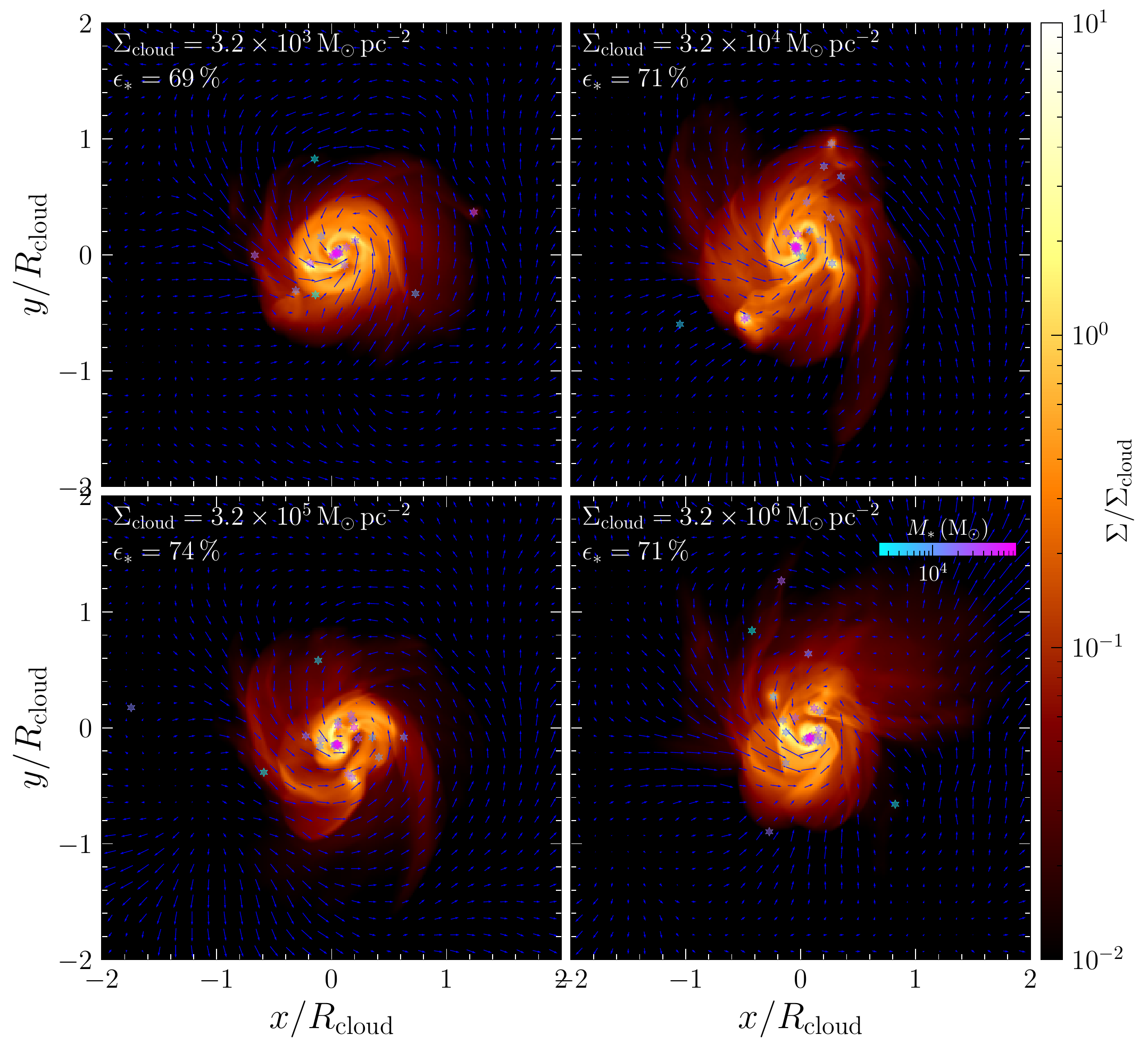}
    \caption{Same as Figure~\ref{fig:SigmaSemProjection}, but now the four panels show results for different values of $\Sigmacloud$ at $t = 7 t_{\mathrm{ff}}$, and with surface densities and positions scaled to $\Sigma_0$ and $R_{\rm cloud}$, respectively.}
    \label{fig:CompareSigmasProj}
\end{figure*}

Next we summarise the evolution of clouds with the fiducial initial conditions, i.e., $\fdustgas = 1$, $\alphavir = 2$, and $\Sigma = 3.2\times 10^3 - 3.2\times 10^6\, \Msolpc$ and our full treatment of temperature-dependent opacities ($\Sigmacloud$ series; Table~\ref{tab:Simulations}). As an example of the results for these models Figure~\ref{fig:SigmaSemProjection} shows snapshots of the surface density in run \texttt{S5KsemA2F1} at times $\left [1,3,5,7 \right] t_{\mathrm{ff}}$. We can see that the initial turbulent velocity fluctuations lead to a highly filamentary structure, where self-gravitating overdensities form star clusters. The self-gravity of the gas, combined with the gravity from the point sources, leads to global collapse of the cloud, increasing the total sink mass and hence $\epsilon_*$. We find that by $\sim 7\,t_{\mathrm{ff}}$, the star formation efficiency reaches $\epsilon_* \approx 75 \%$, with the remainder of the gas mass supported by systematic rotation seeded by the initial velocity fluctuations and then amplified by angular momentum conservation during the collapse. A central cluster is formed, with the highest mass sink particle sinking to the central regions. In addition, we see that the dynamical interaction between the star clusters leads to some of the lower-mass clusters being pushed out to larger radii. It is clear that overall, the radiation pressure is unable to prevent accretion, or to drive winds that are dynamically significant.

We find that the evolution of clouds for all other $\Sigmacloud$ cases are qualitatively similar to that shown in Figure~\ref{fig:SigmaSemProjection}. We compare the projected surface densities for runs \texttt{S3KsemA2F1}, \texttt{S4KsemA2F1}, \texttt{S5KsemA2F1}, and \texttt{S6KsemA2F1} at $t = 7 \, \mathrm{t}_{\mathrm{ff}}$ in Figure~\ref{fig:CompareSigmasProj}. We can see that, when we normalise positions to the initial cloud radius and surface densities to the initial cloud surface density, the final states of these cases are qualitatively identical.

\begin{figure}
    \centering
    \includegraphics[width =  0.48 \textwidth]{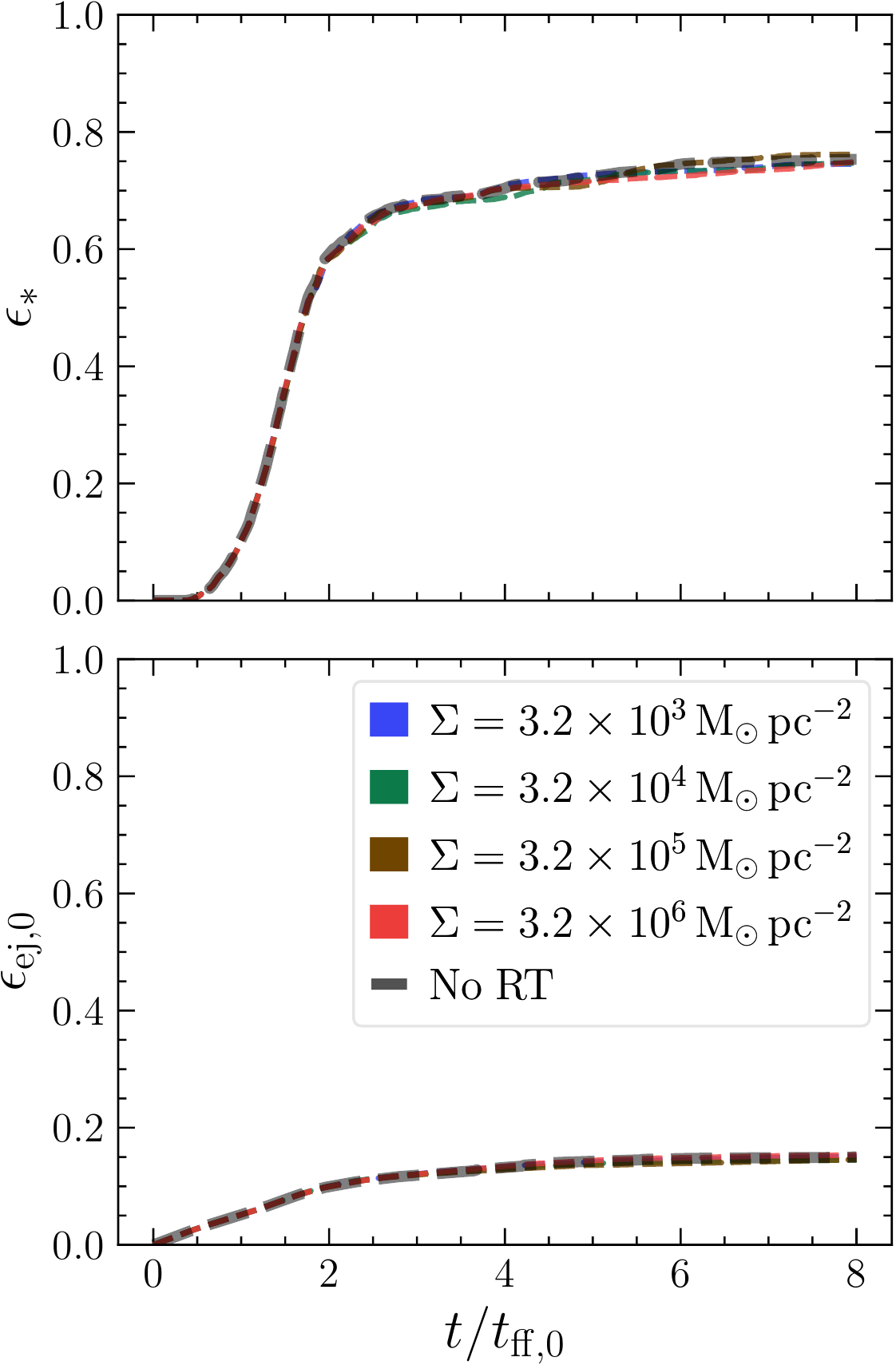}
    \caption{Time evolution of quantities compared for different values of $\Sigma$ (colours). Dark grey dashed lines indicate a control run without radiation feedback (No RT). \textbf{Top}: the integrated star formation efficiency ($\epsilon_* = M_*/\Mcloud$), \textbf{Bottom}: the fraction of mass ejected from the computational volume ($\epsilon_{\mathrm{ej}} = M_{\mathrm{ej}}/\Mcloud$ where $M_{\mathrm{ej}}$ is the ejected mass). }
    \label{fig:CompareSFEMW}
\end{figure}

To explore this more quantitatively, in Figure~\ref{fig:CompareSFEMW} we show the time evolution of the integrated star formation efficiency $\epsilon_*$, and the fraction of mass ejected from the simulation domain $\epsilon_{\mathrm{ej}}$, for the four values of $\Sigmacloud$ illustrated in Figure~\ref{fig:CompareSigmasProj}. We can see that the curves are nearly identical to one another, and to a control run without radiation. This clearly establishes that the radiation forces are unable to affect the dynamical outcomes of the clouds. In all cases $\epsilon_*$ rises sharply with time, starts to slow down by $t \sim 2\,t_{\mathrm{ff}}$, and by $t \sim 3\,t_{\mathrm{ff}}$ reaches $\epsilon_* \sim 80 \%$, with mass continuing to accrete at a very slow rate (as indicated by the positive slope) thereafter. Most of the remaining matter escapes from the simulation domain, $\epsilon_{\mathrm{ej}} \sim 20\%$, but this escape is due to the initial turbulence and not the IR radiation pressure, as evidenced by a saturation at later times and by the fact that $\epsilon_{\mathrm{ej}}$ is nearly identical in the control run and the radiation runs.

\subsubsection{Comparison of radiation and gravity forces}

\begin{figure}
    \centering
    \includegraphics[width=0.48\textwidth]{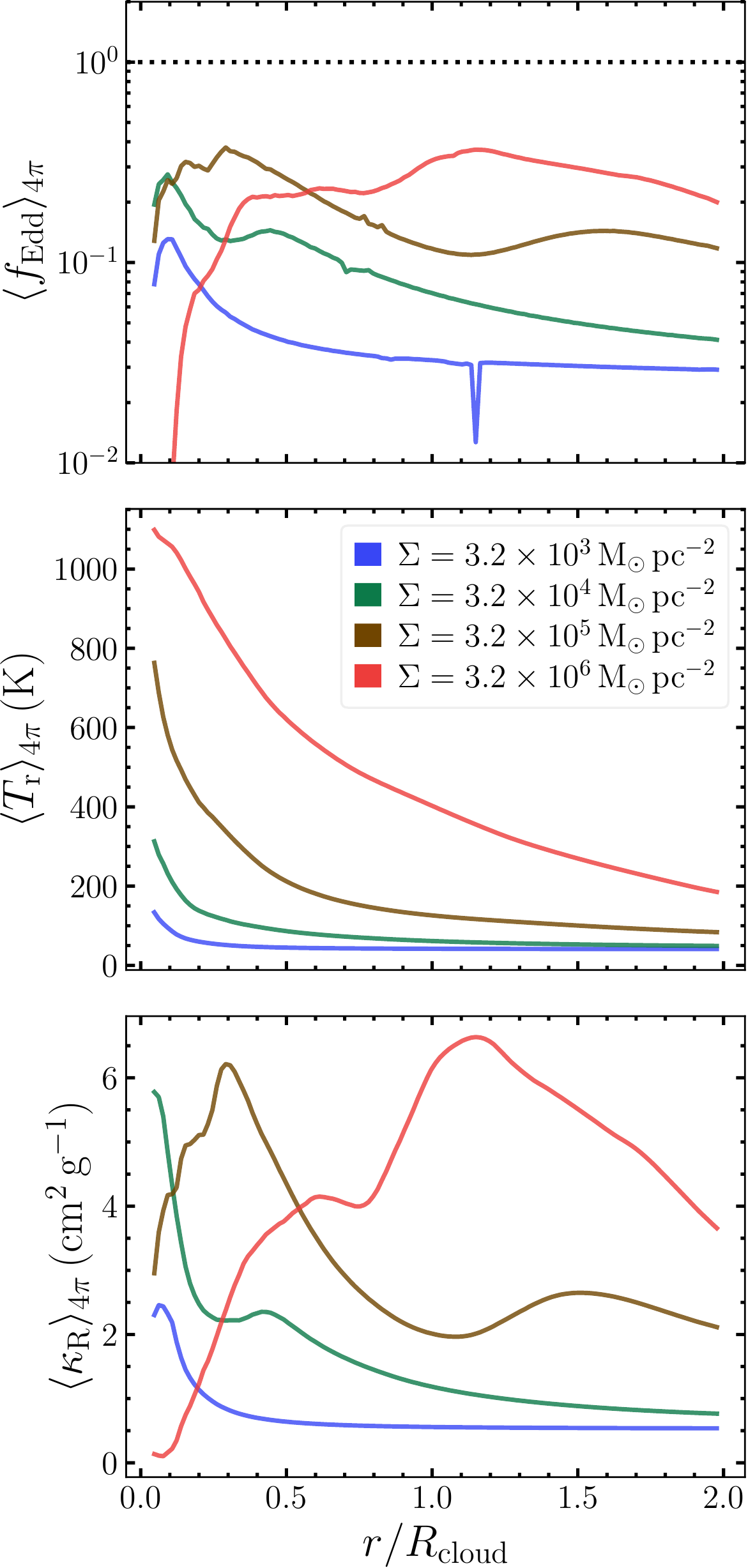}
    \caption{Balance of forces, and the physical properties that control it, compared for different $\Sigmacloud$. The top panel shows the volume-weighted Eddington ratio $f_{\mathrm{Edd}}$, i.e., the ratio between radiation and gravitational forces (Eq.~\ref{eq:fEdd}) with the dotted line indicating $\langle f_{\mathrm{Edd}} \rangle_{4\pi} = 1$. The middle panel shows the averaged radiation temperature $T_r$, and the bottom panel shows the averaged Rosseland opacity ($\kappaR$).}
    \label{fig:SigmaSemForces}
\end{figure}

To understand the balance of forces in the problem, and arrive at a physical explanation for the simulation outcomes, we compare the two dominant competing forces: radiation pressure and gravity. We examine the force balance as a function of radius using a method similar to that of \citet{Skinner_2015}. For a given snapshot in time, we define a spherical coordinate system centred on the instantaneous centre of mass of the sink particles, and assign every computational cell to one of $128$ radial bins relative to this point. We then compute the mean (specific) radial component of the radiation force, and the competing (specific) radial gravitational force (combination of gas self-gravity and sink particles), in each bin by averaging over all solid angles. We denote these averages by $\langle\dots\rangle_{4\pi}$. Formally, we define the mean radiation force as
\begin{equation}
    \forcerad = \left\langle \frac{\kappaR \mathbfit{F}_0}{c} \cdot \hat{\mathbfit{r}} \right\rangle_{4\pi},
    \label{eq:forcerad}
\end{equation}
where $\mathbfit{F}_0$ is the radiation flux in the co-moving frame of the fluid, which can be obtained from the lab-frame radiation flux to $\mathcal{O}(v/c)$ by the relation \citep[e.g.,][]{Castor_2004},
\begin{equation}
    \mathbfit{F}_0 = \mathbfit{F} - E_r \mathbfit{v} \cdot (\mathbfss{I} + \mathbfss{T}),
\end{equation}
and $\hat{\mathbfit{r}}$ is the unit vector in the radial direction in the coordinate system defined by the sink particles' centre of mass. To understand the radial variations of $\forcerad$ in further detail, we also separately compute the radially averaged radiation flux $F_{\mathrm{r}} = \langle \mathbfit{F}_0 \cdot \hat{\mathbfit{r}} \rangle_{4\pi}$, the radially averaged Rosseland opacity $\langle \kappaR \rangle_{4\pi}$, and the radially averaged radiation temperature $\langle \Tr \rangle_{4\pi}$. The corresponding (specific) gravitational force $\forcegrav$ is given by
\begin{equation}
    \forcegrav = \langle g_{\mathrm{gas}} \rangle_{4\pi} + \langle g_{*} \rangle_{4\pi},
\end{equation}
where $g_{\mathrm{gas}} = -\hat{\mathbfit{r}}\cdot \nabla \Phi_{\mathrm{gas}}$, $g_{*} = - \hat{\mathbfit{r}}\cdot \nabla \Phi_{*}$, and $\Phi_{\mathrm{gas}}$ and $\Phi_{*}$ are the gravitational potentials of the gas and sink particles, respectively \citep[we note that the sink particle potential is softened at radii of $r<2.5 \, \Delta x$ to prevent diverging values near the centre of the sink particle; see][]{Federrath_2010_Sinks}. We can use these relations to compute the ratio of these forces, i.e., the Eddington ratio $f_{\mathrm{Edd}}$ at each point, given by 
\begin{equation}
    f_{\mathrm{Edd}} = \frac{\dot{p}_{\mathrm{rad}}}{\dot{p}_{\mathrm{grav}}},
    \label{eq:fEdd}
\end{equation}
and average this quantity at each radial shell to obtain the volume-averaged Eddington ratio $\langle f_{\mathrm{Edd}}\rangle_{r}$.

In Figure~\ref{fig:SigmaSemForces} we show the radial variation of 
$\langle f_{\mathrm{Edd}}\rangle_{r}$ (top panel), $\langle\Tr\rangle_{4\pi}$ (middle panel), and $\langle \kappaR\rangle$ (bottom panel)
obtained for the different $\Sigmacloud$ models. We see from the top panel that $\langle f_{\mathrm{Edd}}\rangle_{r}<1$ for all the models, which explains why the results are nearly identical in all cases: the dynamics are dominated by $\forcegrav$, which is (once we normalise out the cloud size) more or less independent of $\Sigmacloud$. However, this does not yet explain what drives the differences in $\langle f_{\mathrm{Edd}}\rangle_{r}$ across the models, and thus, \textit{why} radiation is subdominant compared to gravity. There are clear differences in the values of $\langle f_{\mathrm{Edd}}\rangle_{r}$ for different $\Sigmacloud$, with higher $\Sigmacloud$ cases reaching, on average, higher values of $f_{\mathrm{Edd}}\rangle_{r}$, but also reaching their maxima at different locations. These differences cannot be driven by differences in the radiation flux, which on the grounds of energy balance must always be close to $L/(4\pi r^2)$, independent of $\Sigmacloud$; direct examination of the flux confirms this conclusion. Instead, as the middle and bottom panels of Figure~\ref{fig:SigmaSemForces} show, differences in the radiation force are primarily driven by how $\langle \Tr \rangle_{4\pi}$, and the corresponding $\langle \kappaR \rangle_{4\pi}$ (which depends on $T_r$) vary with radius. While $\langle \Tr \rangle_{4\pi}$ varies as roughly $r^{-0.5}$ in all cases, the normalisations of the different $\Sigmacloud$ runs are very different: higher $\Sigmacloud$ leads to higher $\langle \Tr \rangle_{4\pi}$, consistent with the expectation that there is enhanced trapping of photons at higher $\Sigmacloud$. This drives complex, non-monotonic variations in $\langle \kappaR \rangle_{4\pi}$, due to the non-trivial dependence of the $\kappaSem$ opacity law on $\Tr$ (see Figure~\ref{fig:SemenovOpacities}). The mean opacity varies smoothly and almost monotonically with radius for the models with lower values of $\Sigmacloud$, but for the cases with higher $\Sigmacloud$, $\langle \kappaR \rangle_{4\pi}$ has a distinct peak at which $\langle \kappaR\rangle_{4\pi} \sim 6 \, \mathrm{cm}^2 \, \mathrm{g}^{-1}$, and falls at radii on either side of this peak. This occurs due to the drop in $\kappaR$ at larger temperatures in the $\kappaSem$ opacity law. 

\subsubsection{Dependence on virial parameter}
\label{sec:MW_alphavir}

\label{sec:alphavir}
\begin{figure}
    \centering
    \includegraphics[width = 0.48 \textwidth]{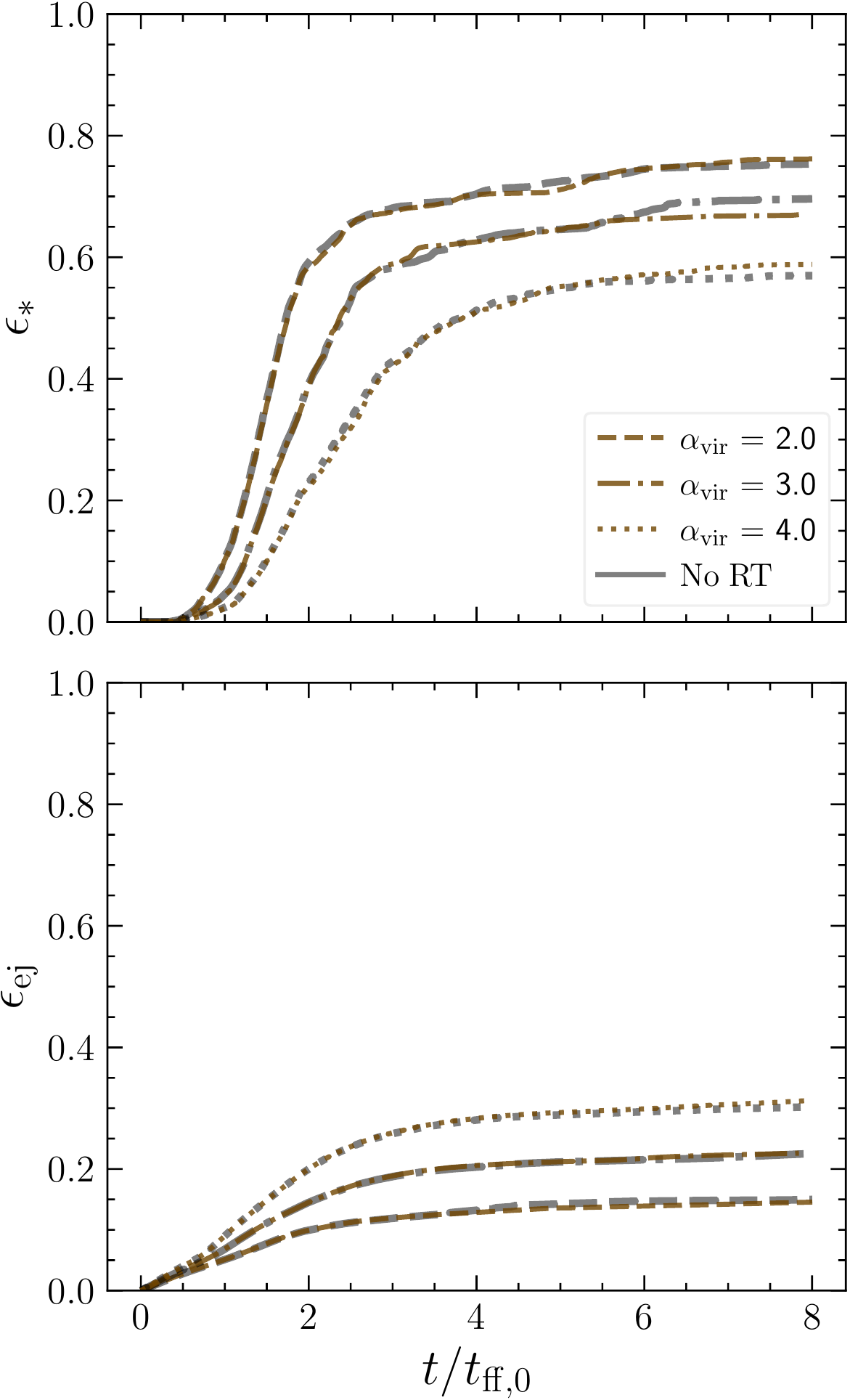}
    \caption{Same as Figure~\ref{fig:CompareSFEMW} for $\Sigma = 3.2 \times 10^5 \Msolpc$, but for different values of the initial virial parameter of the cloud ($\alphavir$). While the final SFE responds to changes in $\alphavir$, due to turbulence expanding the clouds, radiation feedback does not play any significant role in this process, as demonstrated by the three control runs (one for each $\alphavir$) without radiation (grey lines).}
    \label{fig:CompareSFEAlphaVir}
\end{figure}

\begin{figure}
    \centering
    \includegraphics[width= 0.5 \textwidth]{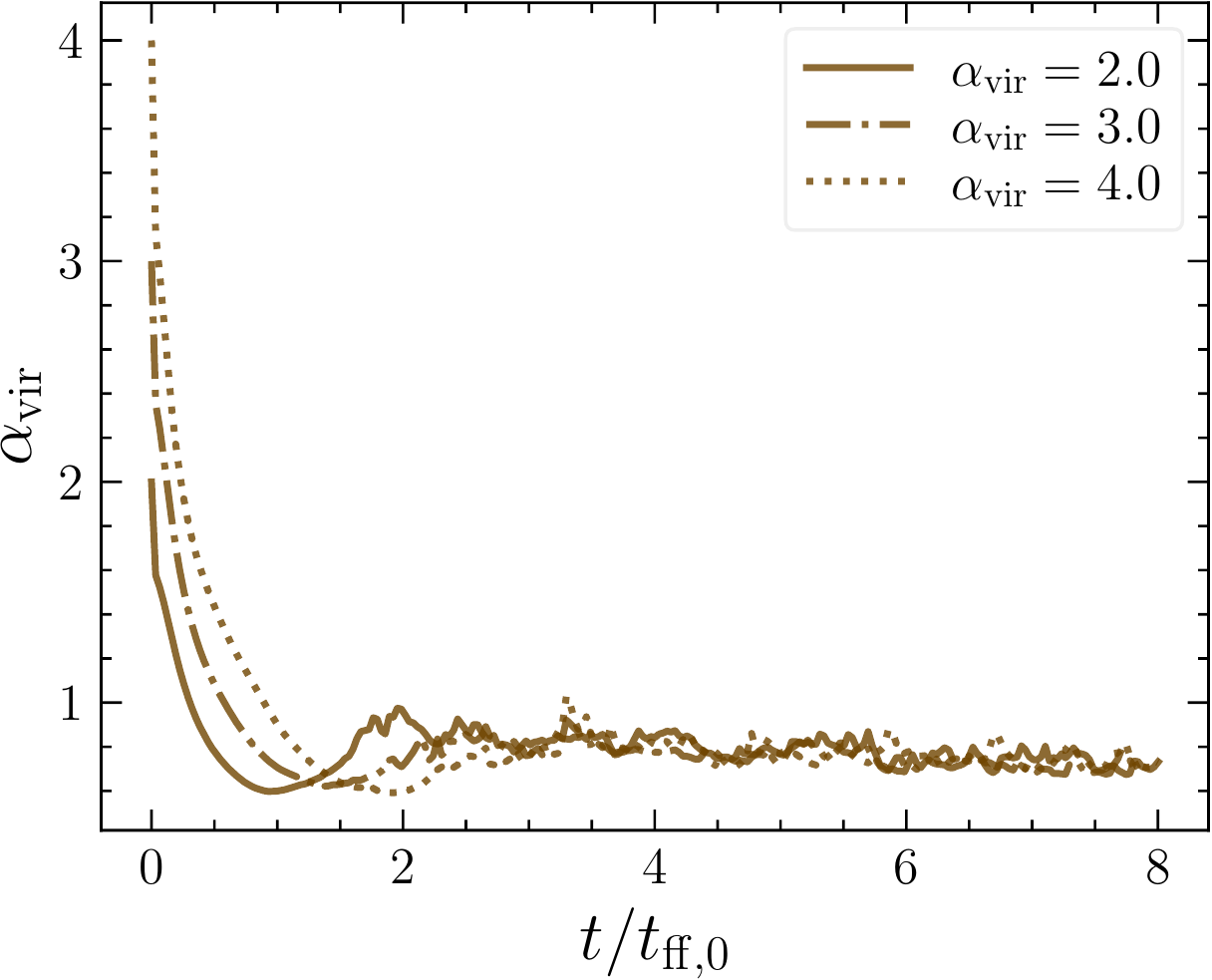}
    \caption{Time evolution of $\alphavir$ for the runs with different initial cloud virial parameters. We can see that even the initially super-virial clouds (i.e., $\alphavir = 3$ and 4) eventually reach sub-virial states, and saturate to $\alphavir \lesssim 1$.}
    \label{fig:CompareAlphaVir}
\end{figure}

Here, we compare the evolution of clouds with our fiducial setup at different values of the initial virial parameter $\alphavir$, testing values of $\alphavir = 3$ and 4, compared to the fiducial value of $2$. These runs allow us to explore whether radiation can drive winds from more weakly bound clouds. However, in Figure~\ref{fig:CompareSFEAlphaVir} we see that this is not the case. We find that $\epsilon_*$ is lower, and $\epsilon_{\mathrm{ej}}$ higher, in runs with higher $\alphavir$, with $\epsilon_* \sim 70\%$ and 60\% in the $\alphavir=3$ and 4, respectively. However, this difference is entirely driven by the fact that the dominance of turbulence over gravity leads to expansion and loss of gas from the clouds, as demonstrated by the fact that the control runs without radiation yield nearly indistinguishable curves in Figure~\ref{fig:CompareSFEAlphaVir}. Thus, even for clouds that are largely unbound, IR radiation feedback is unable to affect their dynamical evolution. This, along with the results of the previous sections, implies that the effects of IR radiation feedback are unambiguously negligible in regulating star formation for solar-neighbourhood dust conditions. We note, however, that the higher $\alphavir$ runs, although initially unbound, eventually become bound due to the decay of turbulence. This is demonstrated in a plot of $\alphavir$ (calculated using Equation~\ref{eq:alphavir}) with time in Figure~\ref{fig:CompareAlphaVir}. It is possible that IR radiation might be more effective in an environment where the turbulence is driven by a cascade from larger scales \citep[e.g.,][]{Federrath_2008}, rather than freely decaying.

\subsection{Super-solar dust conditions} \label{sec:ResultsDG}

\begin{figure}
    \centering
    \includegraphics[width = 0.48 \textwidth]{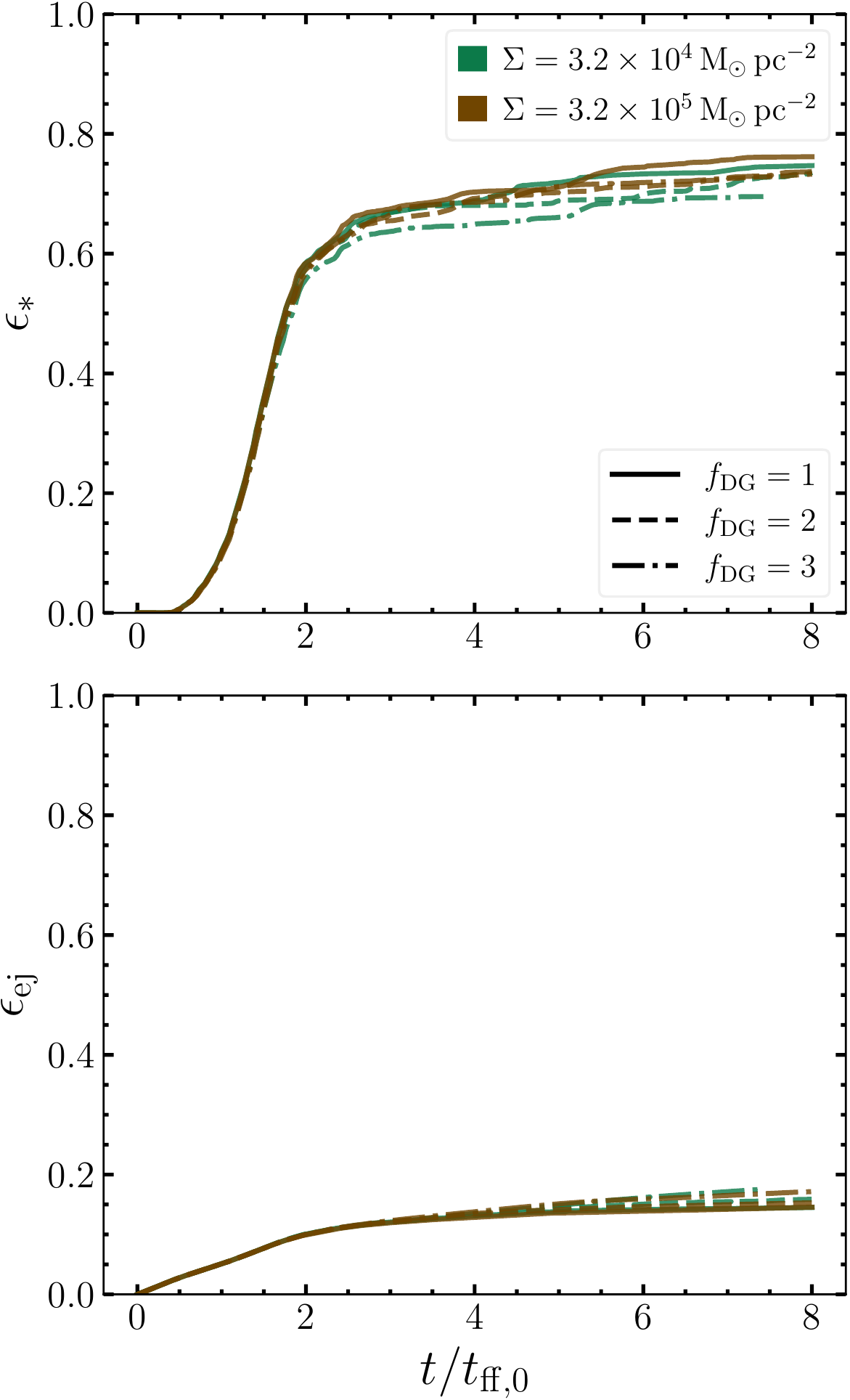}
    \caption{Same as Figure~\ref{fig:CompareSFEMW}, but for different values of the relative dust-to-gas ratio compared to solar conditions ($\fdustgas$).}
    \label{fig:CompareSFEDG}
\end{figure}

\begin{figure}
    \centering
    \includegraphics[width = 0.48 \textwidth]{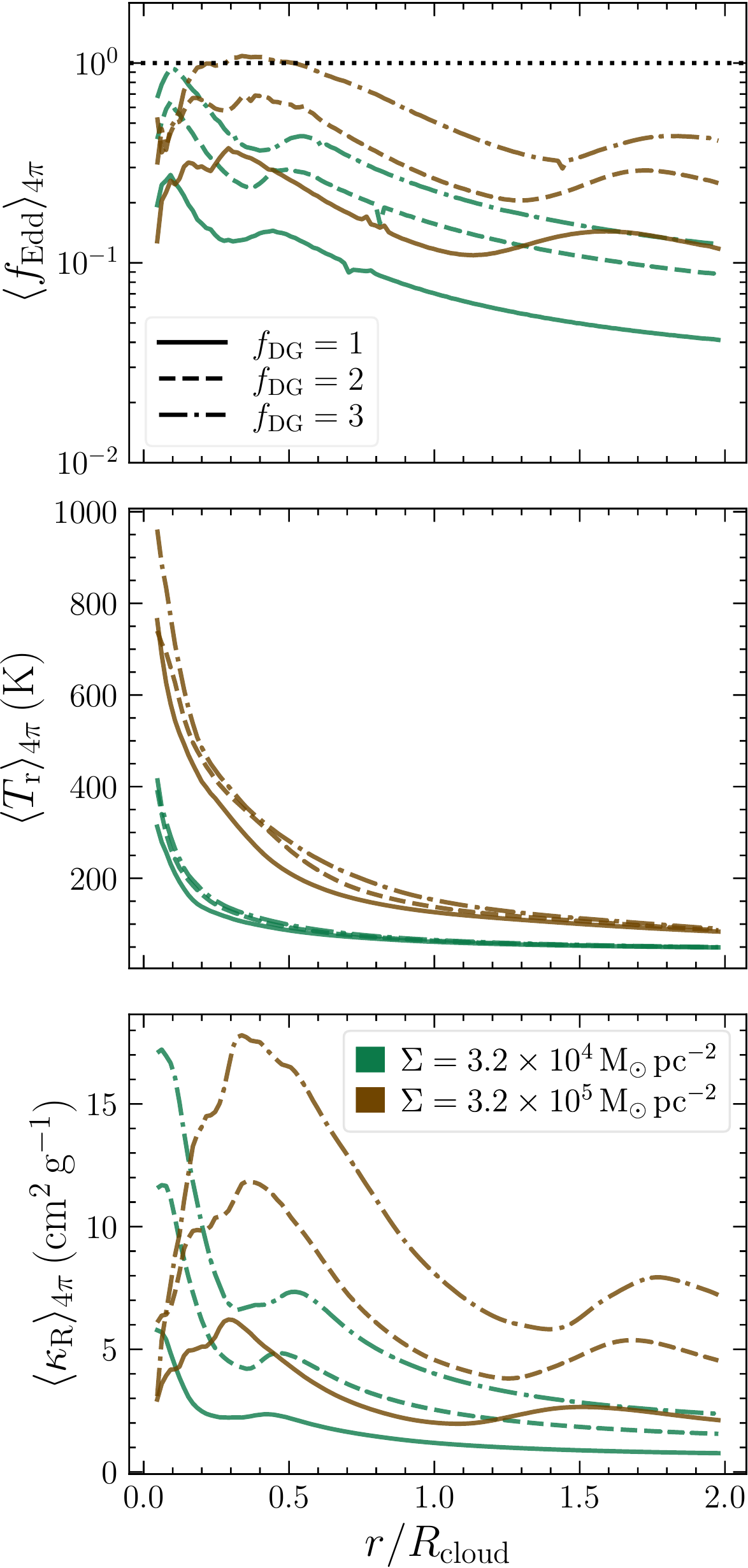}
    \caption{Same as Figure~\ref{fig:SigmaSemForces}, but for different $\fdustgas$.}
    \label{fig:CompareEddDG}
\end{figure}

\begin{figure}
    \centering
    \includegraphics[width = 0.45 \textwidth]{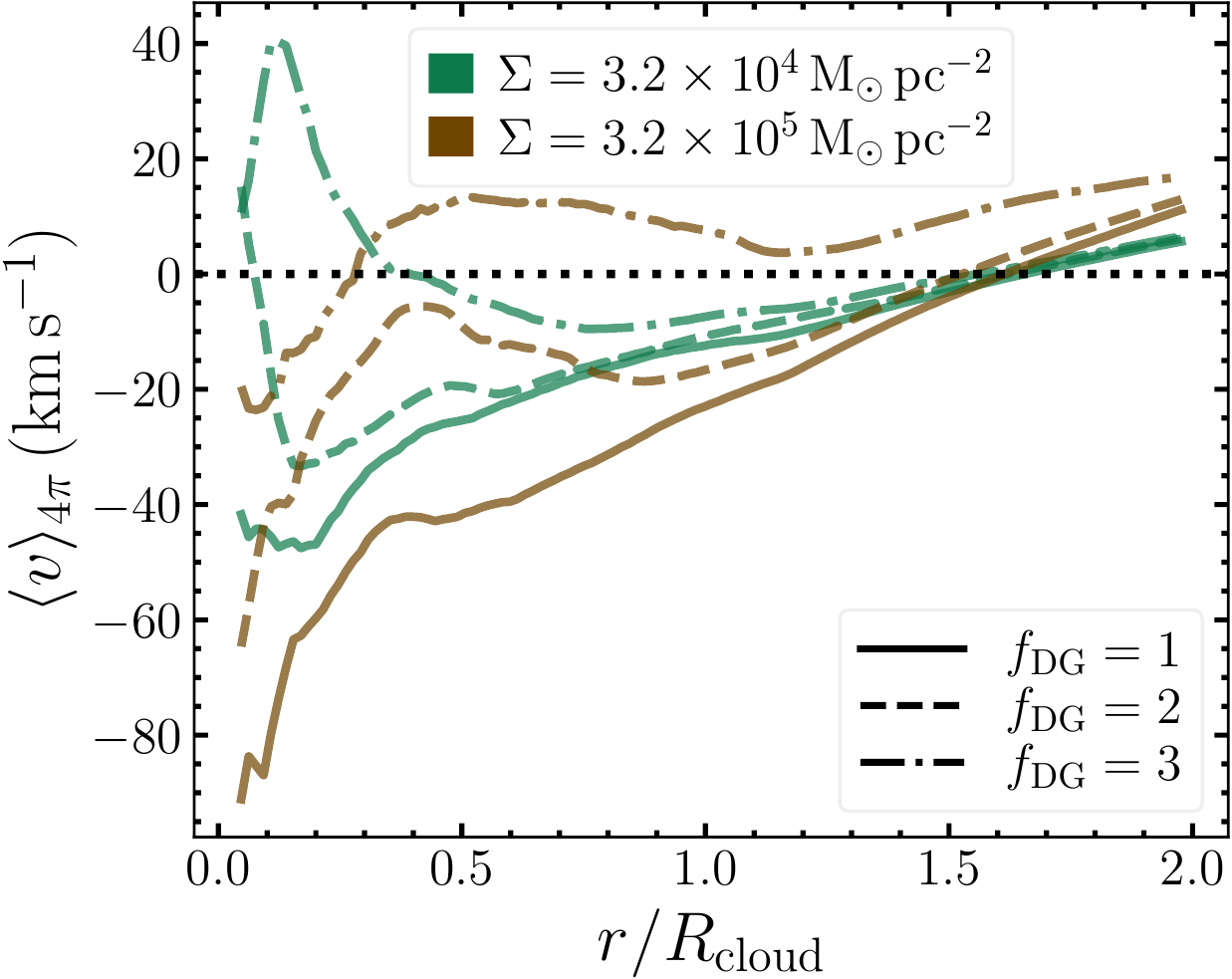}
    \caption{Volume-weighted radial profiles of the radial gas velocity ($v_r$) for the runs with different $\fdustgas$.}
    \label{fig:ComparevelDG}
\end{figure}

Next, we present the results of our runs with super-solar dust conditions, i.e., values of $\fdustgas>1$. In Figure~\ref{fig:CompareSFEDG} we compare the evolution of $\epsilon_*$ and $\epsilon_{\mathrm{ej}}$ over time for the different runs. The cases with $\fdustgas = 1$ and 2 reach final values of $\epsilon_*$ that are indistinguishable, suggesting that radiation is still largely subdominant in these cases. Even with a higher value ($\fdustgas = 3$) the effects on $\epsilon_*$ (and $\epsilon_{\mathrm{ej}}$) remain minor. Despite being small, these differences are worth investigating in order to gain physical insight. First consider the case $\Sigmacloud = 3.2 \times 10^4 \Msolpc$, $\fdustgas=3$, which deviates from the other cases at $t\sim 2t_{\mathrm{ff}}$, after which star formation slows down, eventually saturating at $\epsilon_* \sim 70\%$, slightly lower than the $\epsilon_* \sim 80\%$ attained by the corresponding $\fdustgas=1$ case. Differences in $\epsilon_{\mathrm{ej}}$ are substantially smaller than for $\epsilon_*$, suggesting that radiation pressure is sufficient to inhibit star formation, but not to eject gas completely for this run. Interestingly, the $\Sigmacloud = 3.2 \times 10^5 \Msolpc$ run, which showed the greatest promise for radiation in the $\fdustgas=1$ runs (see Figure~\ref{fig:SigmaSemForces}), shows weaker radiation effects at $\fdustgas=3$ case than the $\Sigmacloud = 3.2 \times 10^4 \Msolpc$ case.

To understand this behaviour, we again construct radial profiles of the volume-weighted Eddington ratio $\langle f_{\mathrm{Edd}} \rangle_{4\pi}$ as a function of radius $r$ for these runs, which we show in Figure~\ref{fig:CompareEddDG}. We see that higher $\fdustgas$ yields larger $\langle f_{\mathrm{Edd}} \rangle_{4\pi}$, as expected, but for $\fdustgas$ values of 1 (as shown earlier) and 2, the profiles are sub-Eddington over all $r$, explaining why radiation does not modify the evolution of these models at all. The $\fdustgas = 3$ cases, however, shows signs of trans- and even super-Eddington profiles at some radii, although they remain sub-Eddington over most of the cloud. In the $\Sigmacloud = 3.2 \times 10^4 \Msolpc$ model, this occurs very close to the sinks, indicating that radiation provides support against gravity in these regions. We confirm that gas is flowing away from the sinks locally at small $r$ in this simulation by plotting $\langle v_r \rangle_{4\pi}$ -- the radial profile of the volume-weighted average radial velocity of gas -- in Figure~\ref{fig:ComparevelDG}. However, outflow is restricted to the polar directions, and accretion continues to occur along dense, optically-thick filaments, very similar to the configuration found in simulations of the formation of individual massive stars \citep[e.g.,][]{Krumholz_2009b,Kuiper_2012,Rosen_2016}. Moreover, even the gas that is driven outwards at small radii does not escape, as is clear both from the reversal of $\langle v_r\rangle_{4\pi}$ at larger radii, and from the fact that $\epsilon_{\rm ej}$ does not increase in this run even though $\epsilon_*$ declines. This is because such gas eventually reaches larger radii, where it sees a cooler radiation field and becomes sub-Eddington. It is also noteworthy that, although the case $\Sigmacloud = 3.2 \times 10^5 \Msolpc$ also reaches $\langle f_{\mathrm{Edd}} \rangle_{4\pi} > 1$, for this run radiation is less effective at reducing $\epsilon_*$ than in the case of lower surface density. This suggests that it matters \textit{where} the gas is super-Eddington; closer to the sinks yields more regulation of star formation.

Overall, our models indicate that even for the most metal-rich observed systems, IR radiation feedback is unlikely to be significant. 

\subsection{Power-law opacities} \label{sec:PLTOpacity}
\begin{figure}
    \centering
    \includegraphics[width = 0.48 \textwidth]{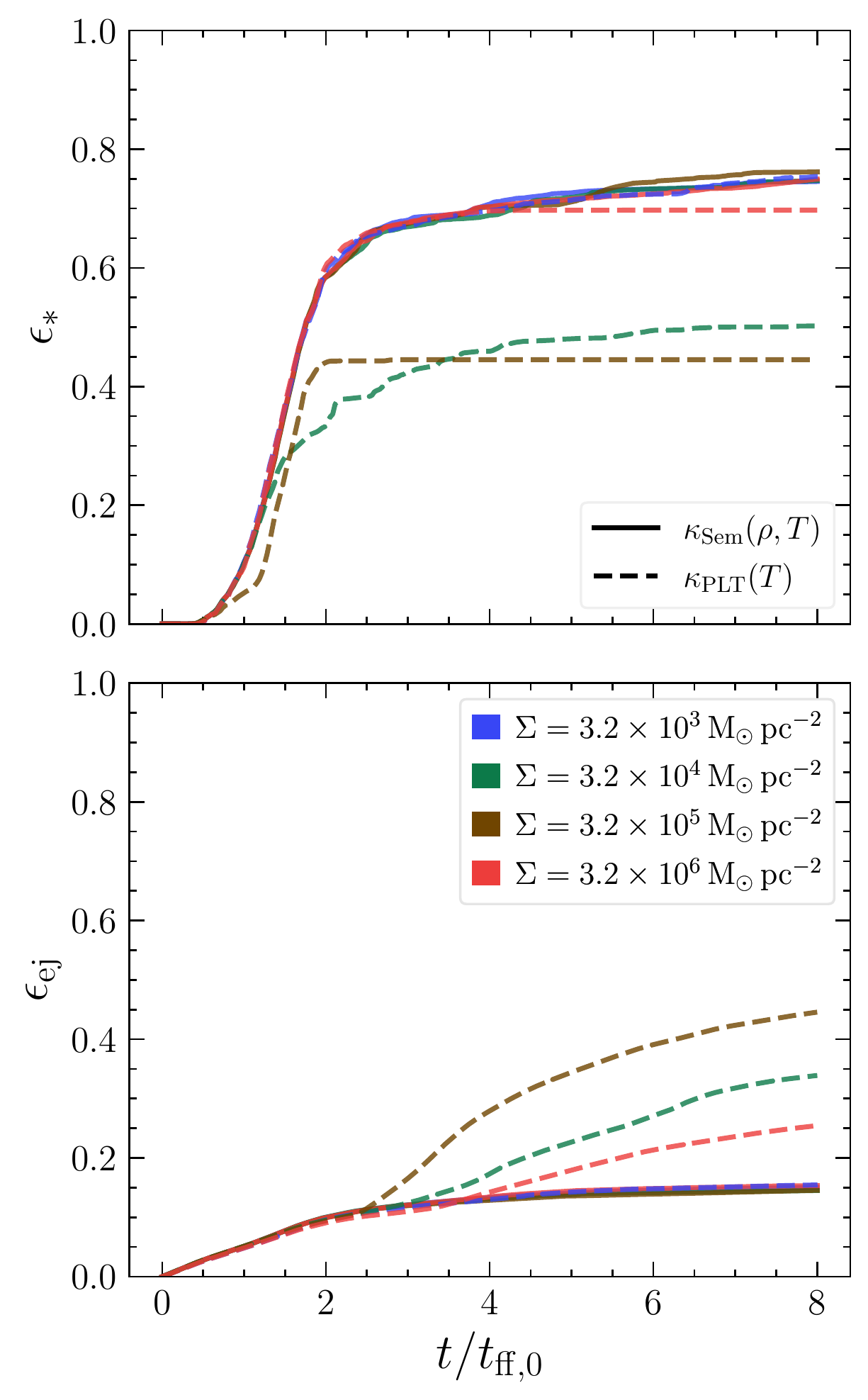}
    \caption{Same as Figure~\ref{fig:CompareSFEMW}, but for runs using the \citet{Semenov_2003} IR opacities ($\kappaSem$; solid lines), and the power-law approximation to the IR opacities ($\kappaPLT$; dashed lines) for the different values of $\Sigma$. }
    \label{fig:SFESemPLT}
\end{figure}

\begin{figure}
    \centering
    \includegraphics[width = 0.45 \textwidth]{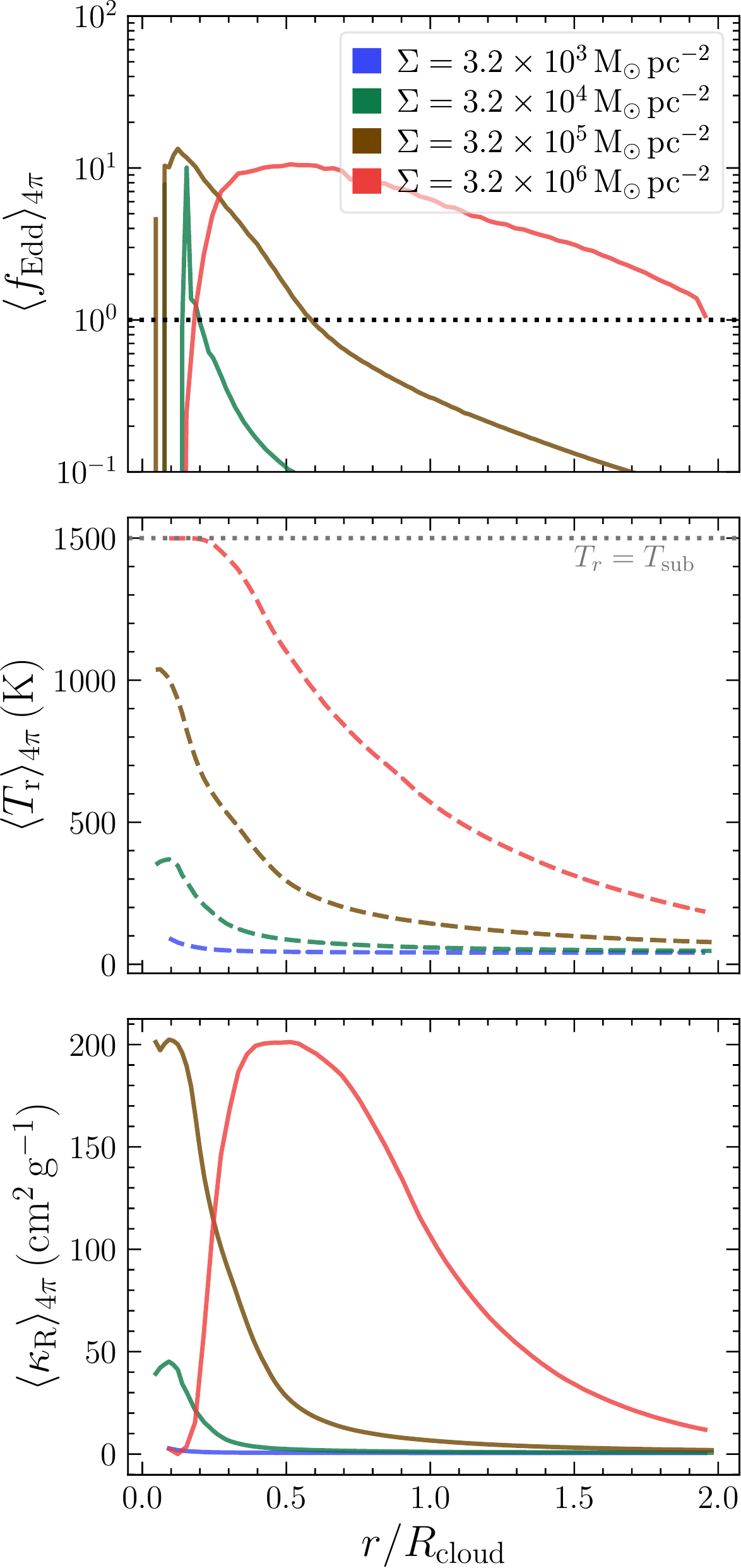}
    \caption{Same as Figure~\ref{fig:SigmaSemForces}, but for runs using the $\kappaPLT$ opacity law (Equation~\ref{eq:kappaT2}). Note that the dotted grey line in the middle panel indicates the dust sublimation temperature ($T_{\mathrm{sub}}$) beyond which $\kappaR = 10^{-3}$. We see that the $\kappaPLT$ approximation severely overestimates $\kappaR$ for the higher $\Sigmacloud$ runs compared to the $\kappaSem$ opacities (see Figure~\ref{fig:SigmaSemForces}), thereby producing super-Eddington states. This explains why the $\kappaPLT$ model significantly impacts $\epsilon_*$, whereas the $\kappaSem$ is unable to.}
    \label{fig:SigmaTSqrEdd}
\end{figure}

In our final set of experiments, we quantitatively compare the time evolution of clouds with the $\kappaSem$ and $\kappaPLT$ opacity laws; as discussed in Section \ref{sssec:plt}, the latter is a simplification commonly adopted in both simulations and semi-analytic models. In Figure~\ref{fig:SFESemPLT}, we plot $\epsilon_*$ and $\epsilon_{\mathrm{ej}}$ as a function of time for the different values of $\Sigmacloud$. As discussed in earlier sections, the cases with $\kappaSem$ opacities are qualitatively identical to one another and to a control run without radiation. On the other hand, the cases with the $\kappaPLT$ opacity law show a strong anti-correlation between $\epsilon_*$ and $\Sigmacloud$ -- with the exception of the $\Sigmacloud = 3.2\times 10^6 \, \Msolpc$ case, to which we shall return below. This exception aside, we find that for cases with higher $\Sigmacloud$, $\epsilon_*$ saturates at earlier times and reaches much lower final values. The bottom panel of Figure~\ref{fig:SFESemPLT}, which shows the time evolution of $\epsilon_{\mathrm{ej}}$, demonstrates that this is because these runs drive winds that escape the domain, which is not the case for the $\kappaSem$ runs. Interestingly, the transition to IR radiation ejecting $>50\%$ of the initial mass occurs at roughly $10^5 \, \Msolpc$, in good agreement with the critical value derived by \citetalias{Crocker_2018b} using the power-law opacity approximation.

To understand why the $\kappaPLT$ models are so different from the $\kappaSem$ ones, we look at the radial profiles of $\Tr, \kappaR$ and $f_{\mathrm{Edd}}$ for the $\kappaPLT$ models in Figure~\ref{fig:SigmaTSqrEdd}. We can compare this with Figure~\ref{fig:SigmaSemForces} for the $\kappaSem$ runs. We see that at higher $\Sigmacloud$ -- which corresponds to higher values of $\Tr$ near the clusters -- $\kappaPLT \gg \kappaSem$, rendering the system super-Eddington. The $\Sigmacloud = 3.2 \times 10^4 \, \Msolpc$ case is super-Eddington only in a small region close to $r=0$, while the $\Sigmacloud = 3.2 \times 10^5 \, \Msolpc$ case is super-Eddington over much of the cloud volume, at both small and large radii. Similar to the \texttt{S4KsemA2F3} run discussed in the previous section, this leads to increasing regulation of star formation, but not complete suppression, since accretion can still continue through opaque channels. The plot also explains why the $\Sigmacloud = 3.2 \times 10^6 \, \Msolpc$ is an exception to the trend of lower $\epsilon_*$ with higher $\Sigmacloud$: it is because a significant fraction of volume has $\Tr > 1200 \, \mathrm{K}$, which leads to very low values of $\kappaR$ due to dust sublimation for radii $r/R_\mathrm{cloud} \lesssim 0.3$; thus the deviation from the trend found elsewhere in the PLT models for this run lies in the fact that it is \textit{not} exactly a power law, but instead deviates from this behaviour at sufficiently high $\Tr$.

Thus we find that the power-law approximation significantly overestimates the effects of IR radiation pressure. We discuss the implications of this finding for earlier numerical results in Section~\ref{sec:PLTImplications}.

\section{Discussion}
\label{sec:Discussion}

\subsection{Why is IR radiation feedback dynamically unimportant?}

The most important conclusion obtained from the simulations presented in the last section is that IR radiation pressure is unable to compete with gravitational forces, and as a result, plays a negligible role in controlling the star formation efficiency of clouds. This holds across a wide range in $\Sigmacloud$ ($\sim 10^3$--$10^6 \, \Msolpc$), and even at dust-to-gas ratios (relative to solar) up to $\sim 3$, in disagreement with the qualitative expectations of semi-analytical studies \citep{Thompson_2005,Murray_2010_clusters,Crocker_2018b}. To understand why this is the case, we explore two possible mechanisms: 1) an anti-correlation between radiation flux and matter density, which is a critical mechanism in the formation of individual massive stars \citep{Krumholz_2009b, Rosen_2016} and has been suggested to be of importance for young stellar populations as well \citep[e.g.,][]{Jacquet_2011, Krumholz_2012,Davis_2014}, and 2) the fact that the IR dust opacities are never high enough for radiation forces to matter \citep[Fig.~\ref{fig:SemenovOpacities} and][]{Semenov_2003}. We explore the role of both effects in detail below.

\subsubsection{Radiation-matter anti-correlation}

\begin{figure*}
    \centering
    \includegraphics[width=0.9 \textwidth]{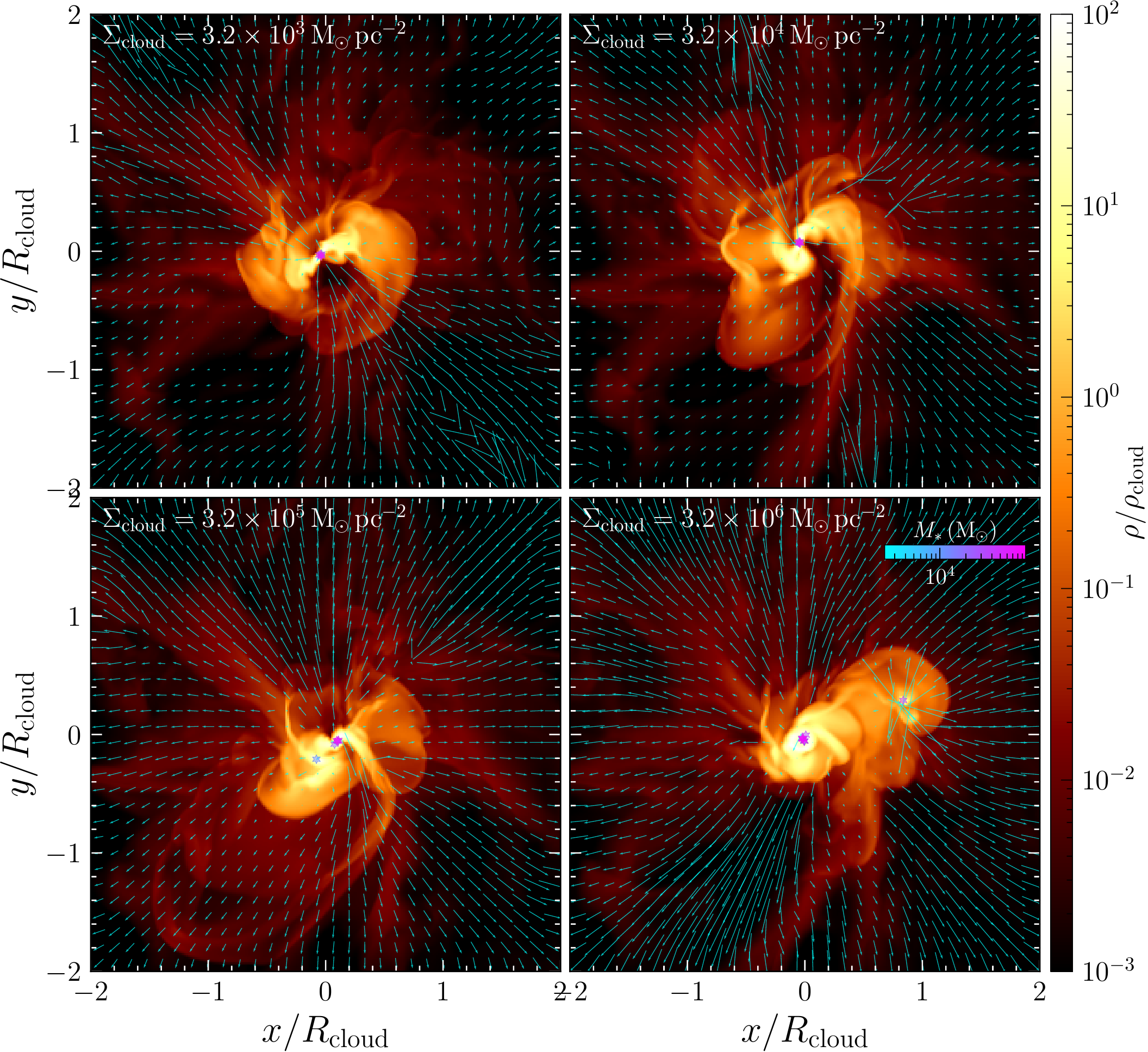}
    \caption{Comparison of slices of the gas density in the $x-y$ plane, for the runs with different $\Sigmacloud$ ($\kappaSem$ series; Table~\ref{tab:Simulations}) at $t =5 t_{\mathrm{ff}}$. The slice is centred on $x=y=0, \, z-z_{*,\mathrm{max}}$ where $z_{*,\mathrm{max}}$ is the $z-$coordinate of the sink with highest mass in each run. Star symbols indicate sink particles, coloured by their mass. Vectors (cyan) indicate the radiation flux field, scaled by $r^2$ to account for its geometrical dilution. We can visually identify that there is an anti-correlation between the radiation flux and matter density to some level in all four cases, with high values of flux in rarer regions and almost negligible in the dense filamentary structures near the stars.}
    \label{fig:SigmaSemDensSlice}
\end{figure*}

\begin{figure}
    \centering
    \includegraphics[width = 0.5 \textwidth]{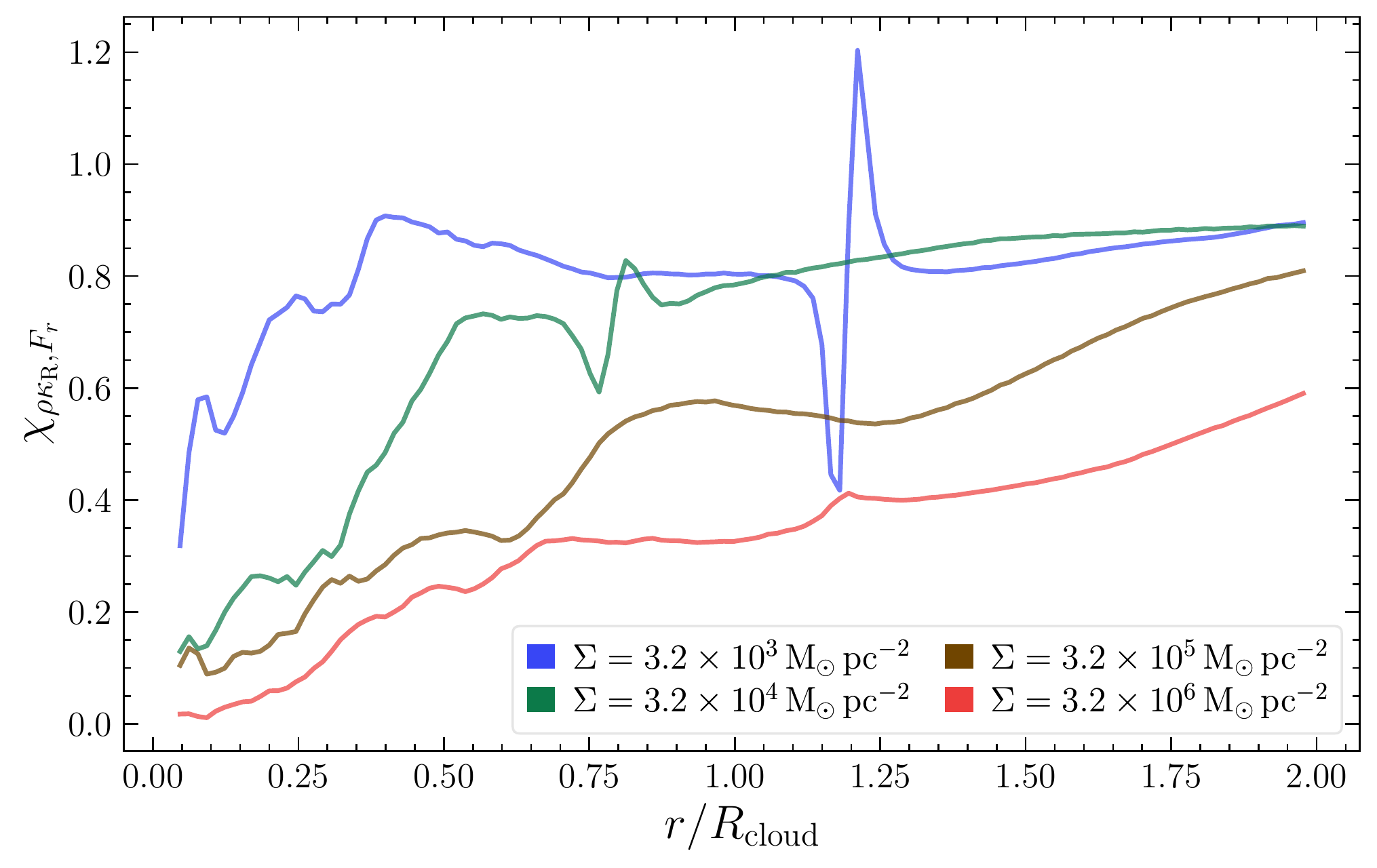}
    \caption{Correlation fraction $\chi_{\rho \kappaR,F_{\mathrm{r}}}$ (Equation~\ref{eq:corrFrac}) of the specific opacity $\rho \kappaR$ and the radial flux $F_{\mathrm{r}}$ for the runs from Fig.~\ref{fig:SigmaSemDensSlice}. We can see that there is significant radiation flux-matter anti-correlation, and increasingly so at higher $\Sigma$.}
    \label{fig:SigmaSemRadMatter}
\end{figure}

Figure~\ref{fig:SigmaSemDensSlice} provides visual evidence of radiation-matter anti-correlation: the flux has higher magnitude (indicated by longer vectors) in optically thinner channels, indicating that the radiation force is strongest where there is less matter and weakest where there is more matter. We quantify the extent of the anti-correlation by computing the correlation fraction of the specific opacity $\rho \kappaR$ and the flux $F_{\mathrm{r}}$, given by 
\begin{equation}
    \chi_{\rho\kappaR,F_{\mathrm{r}}} = \frac{\langle \rho \kappaR F_{\mathrm{r}} \rangle_{4\pi}}{\langle \rho \kappaR \rangle_{4\pi} \langle F_{\mathrm{r}} \rangle_{4\pi}}.
    \label{eq:corrFrac}
\end{equation}
A value of $\chi_{\rho\kappaR,F_{\mathrm{r}}} = 1$ indicates that the fluctuations of $\rho \kappaR$ and $F_{\mathrm{r}}$ are uncorrelated, whereas $\chi_{\rho\kappaR,F_{\mathrm{r}}} < 1$ indicates anti-correlations in the fluctuations about the mean. We compare the radial profiles of $\chi_{\rho\kappaR,F_{\mathrm{r}}}$ for the different $\Sigmacloud$ models in Figure~\ref{fig:SigmaSemRadMatter}. We find anti-correlation in all the model clouds, with the strongest anti-correlation at small radii (also visible in the slice plots of Fig.~\ref{fig:SigmaSemDensSlice}, which shows a disc-like structure close to the star through which the radiation flux is low), and at higher $\Sigmacloud$. These two trends suggest that the increase of $\kappaR$ found at higher $\Sigmacloud$ and smaller radii is balanced to an extent by an increased level of radiation-matter anti-correlation.

\subsubsection{Dust opacities}

\label{sec:ModelDiscuss}
\begin{figure}
    \centering
    \includegraphics[width=0.5\textwidth]{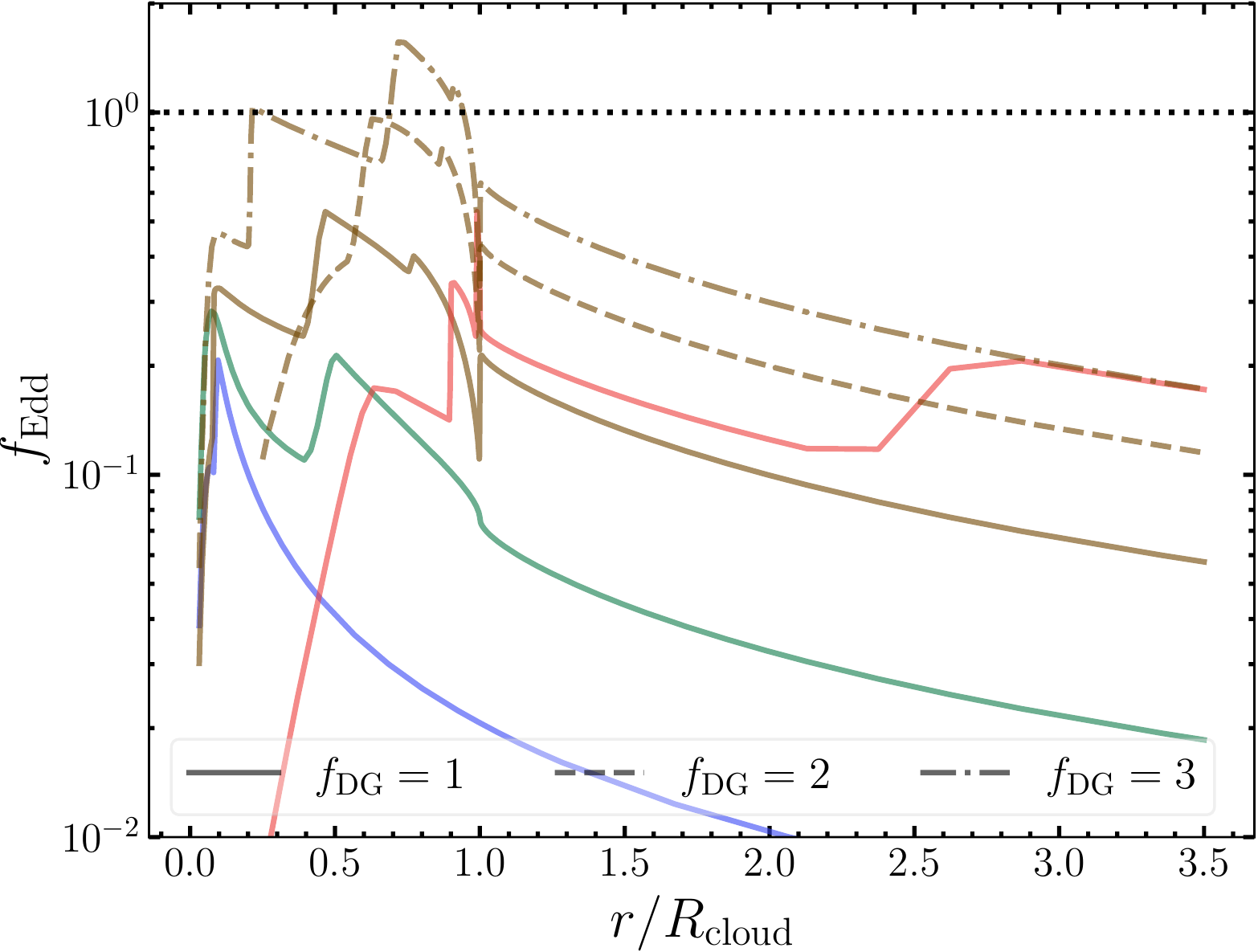}
    \caption{The Eddington ratio obtained with the idealised, steady-state model described in Section~\ref{sec:ModelDiscuss}. Colours indicate the four $\Sigmacloud$ values as in previous figures. Dashed and dash-dotted lines indicate $\fdustgas = 2$ and 3, respectively. The horizontal dotted line indicates $f_{\mathrm{Edd}}=1$, i.e., a balance between radiation and gravity forces. We see that $f_{\mathrm{Edd}}<1$ in all the cases with solar $\fdustgas$, indicating that even without matter-radiation anti-correlation, radiation forces are unlikely to compete with gravity. Even for $\fdustgas = 3$, only a small radial region has $f_{\mathrm{Edd}}\gtrsim1$.}
    \label{fig:fEddIdealised}
\end{figure}

To isolate the effects of realistic dust opacities from the radiation-matter anti-correlation, we ask whether radiation forces in a spherically symmetric cloud -- one where by construction there can be no radiation-matter anti-correlation, can be dynamically important if $\kappaR = \kappaSem$. For a spherical cloud in steady state, the radiation transport equations, Equations~(\ref{eq:Erad}) and~(\ref{eq:Frad}), reduce to a pair of ordinary differential equations that govern the run of radiation quantities through the cloud, and which can be solved given a density profile $\rho(r)$ and a radiation source term (given by Equation~\ref{eq:jstar}) for a source at $r=0$. We solve these equations for a cloud of mass $\Mcloud=10^6\,\mathrm{M}_{\sun}$, and a mass in stars of $M_* = \epsilon_*\Mcloud$. For our numerical solution we adopt $\epsilon_* = 0.7$, chosen to approximately match the steady-state conditions attained by our models at solar dust conditions (see Figure~\ref{fig:CompareSFEMW}), but the results are qualitatively similar for other choices. Given these fixed conditions, we can parameterise our models in terms of just two parameters: the mass surface density $\Sigma$, and a power-law index $\alpha$ such that $\rho(r) \propto r^{-\alpha}$. The complete description of the model, and our numerical approach to solving the resulting equations, are provided in Appendix~\ref{sec:AppendixModel}. We test values of $\Sigma$ that correspond to the four $\Sigmacloud$ values used in our various simulations ($\kappaSem$ series; Table~\ref{tab:Simulations}), and values of $\alpha = 0, 1$ and 2; the results prove to be almost completely insensitive to $\alpha$, and hence we do not show the variation with $\alpha$ here.

In Figure~\ref{fig:fEddIdealised}, we directly show the Eddington ratio $f_{\mathrm{Edd}}$ calculated from the steady-state distributions of radiation energy, flux and opacities in our models; we show profiles of these quantities in Figure~\ref{fig:fEddProfiles} of the Appendix. We show the profiles for a range of $\Sigmacloud$ at $\fdustgas=1$, and for $\fdustgas = 2$ and 3 for the fiducial $\Sigmacloud = 3.2 \times 10^5 \, \Msolpc$. We can see that irrespective of the value of $\Sigmacloud$, $f_{\mathrm{Edd}} < 1$ for $\fdustgas = 1$. In fact $f_{\mathrm{Edd}}$ is at most $\sim 0.5$ for the fiducial $\Sigmacloud$ case, reasonably close to the ranges found in the full 3D numerical simulations (see Figure~\ref{fig:SigmaSemForces}). At $\fdustgas \sim 3$, there is a small range of radii with $f_{\mathrm{Edd}}>1$, but most of the volume remains sub-Eddington. This strongly suggests that the inability of the radiation to dynamically alter the state of the clouds is driven by the dust opacity -- \textit{even with perfect radiation-matter coupling}, radiation forces are subdominant compared to gravity. 

\subsubsection{Relative importance of dust opacities and radiation-matter anti-correlation}

It is now interesting to ask which of these mechanisms is the more important contributor to keeping the radiation forces in our simulations sub-Eddington. The fact that we find $f_{\mathrm{Edd}}<1$ even in the spherically symmetric models suggests that the opacities alone are sufficient at $\fdustgas = 1$; although there is radiation-matter anti-correlation ($\chi_{\rho\kappaR,F_{\mathrm{r}}} < 1$), it does not change the qualitative results, as evidenced by the similar values of $f_{\mathrm{Edd}}$ found in our simulations (Figure~\ref{fig:SigmaSemForces}) and the spherically symmetric models (Figure~\ref{fig:fEddIdealised}).

However, radiation-matter anti-correlation becomes increasingly important when the opacities permit radiation forces to approach the Eddington limit; this can be seen in the larger differences between the spherically symmetric models and the simulations at higher $\fdustgas$, and in the constant opacity runs (Section~\ref{sec:ConstantOpacity}), and in the smaller values of $\chi_{\rho\kappaR,F_{\mathrm{r}}}$ found at higher $\Sigmacloud$. It is also apparent in the quantitative outcomes of the simulations: for constant $\kappa_{\rm c}$, a spherically symmetric cloud becomes super-Eddington once its star formation efficiency exceeds (\citealt{Fall_2010}, \citetalias{Skinner_2015})
\begin{equation}
    \epsilon_* = \frac{4\pi G c}{\kappa_{\rm c}\langle L_*/M_*\rangle}.
\end{equation}
For our highest opacity case, $\kappa_{\rm c} = 80$ cm$^2$ g$^{-1}$, this expression yields $\epsilon_* = 0.17$, whereas our simulation (\texttt{S3K80A2F1}) reaches $\epsilon_* \sim 0.4$; more generally, this expression predicts a relation $\epsilon_* \propto 1/\kappa_{\rm c}$, significantly steeper than the observed relationship (c.f.~Figure \ref{fig:CompareSO15}). The flattening at high $\kappa_{\rm c}$ is due to radiation-matter anti-correlation becoming increasingly effective lowering the mass-weighted Eddington ratio relative to the naive spherically-symmetric estimate as $\kappa_{\rm c}$ increases.

In summary, we find that at solar conditions, the IR opacities rule out the possibility of radiation pressure being dynamically important at all for objects whose light-to-mass ratios are in the range expected for fully-sampled stellar populations with a standard IMF (as opposed to single massive stars or top-heavy IMFs). However, even where radiation forces are enhanced -- such as metal-rich regions with higher $\fdustgas>1$, or around objects with higher $\langle L_*/M_* \rangle$ -- radiation-matter anti-correlation kicks in to reduce the effectiveness of feedback. We note, however, a subtle point: while the radiation-matter anti-correlation can (significantly) reduce the effectiveness of feedback in instances where the conditions permit super-Eddington states, it cannot reduce it sufficiently to render the system sub-Eddington altogether -- there is still some level of mass loss driven by radiation. This can, for instance, be seen in our runs with $\fdustgas = 3$.
Thus anti-correlation does not alter the \textit{conditions} where IR radiation becomes dynamically important, rather, it just reduces the \textit{effectiveness} of radiation forces once those conditions are satisfied. This echoes the picture outlined in the idealised experiments of IR radiation pressure in gravitationally confined columns of \citet{Davis_2014}, where the condition for radiation-driven instability is the same as that derived by \citet{Krumholz_Thompson_2012}, and once instability begins, anti-correlation acts to reduce the effectiveness of feedback, but steady winds are nonetheless still driven by radiation pressure. Our primary new finding here is that, for realistic rather than idealised opacities, stellar populations with normal IMFs embedded in a solar neighbourhood-like ISM will never cross the instability threshold in the first place.

\subsection{Implications for earlier studies of radiation-matter interaction using approximate opacities}

\label{sec:PLTImplications}
One of the interesting insights offered by our study is the importance of adopting realistic, temperature-dependent dust IR opacities \citep[e.g.,][]{Semenov_2003} in models of IR radiation feedback; commonly used alternatives such as an idealised constant opacity or a power-law ($\kappaPLT$) approximation to the \citet{Semenov_2003} opacities can lead to qualitatively different conclusions due to their tendency to overestimate the radiation forces. This finding has important implications for earlier studies in the literature that rely on approximate opacities. 

A set of studies have explored the role of IR radiation pressure in the context of driving (large-scale) dusty winds in starbursts and rapidly star-forming environments \citep[see,][for a review]{Zhang_2018}. This was first studied in models of dusty columns of matter confined in a gravitational field by \citet{Krumholz_Thompson_2012} (\citetalias{Krumholz_Thompson_2012} hereafter). They described the parameters of the models in terms of the dimensionless numbers $f_{\mathrm{E},*}$ and $\tau_{*}$: the (area-averaged) Eddington ratio of the system and the Rosseland mean optical depth of the slab of gas, respectively. They then derived a maximum Eddington ratio $f_{\mathrm{E},\mathrm{crit}}(\tau_*)$ for which such a slab can be in hydro-static equilibrium, and used simulations to explore the behaviour of slabs beyond this stability limit. Subsequent studies \citep{Davis_2014,Rosdahl_2015,Tsang_2015,Zhang_2017} confirmed the stability limit, but found that the \citetalias{Krumholz_Thompson_2012} simulations, which relied on the flux-limited diffusion approximation, underestimate the radiation forces, which leads to qualitatively different behaviour in models that lie in certain regions of parameter space \citep{Davis_2014,Zhang_2017}. However, some of these studies use the $\kappaPLT$ approximation for their dust opacities, which we show in Section~\ref{sec:PLTOpacity} to significantly overestimate the radiation forces and lead to qualitatively different conclusions than the more realistic dust opacities provided by \citet{Semenov_2003}. This is fundamentally due to the significantly higher $\kappaR$ with the $\kappaPLT$ power law, which even at modest temperatures of $\sim 200 \, \mathrm{K}$ can lead to $\sim 2$ times higher $\kappaR$, with the discrepancy getting progressively worse at higher temperatures (see Figure~\ref{fig:SemenovOpacities}).

A similar point can be made about another popular approach, which has been to adopt a constant (but high) opacity for IR photons; for example, \citet{Hopkins_2011, Hopkins_2012}, \citet{Bieri_2017}, \citet{Costa_2018}, and \citet{Hopkins2018} all adopt $\kappa_{\rm c} = 10\,$cm$^2\,$g$^{-1}$. A comparison to Figure~\ref{fig:SigmaSemForces} shows that, even for extremely high surface densities, the \textit{maximum} opacity in any radial bin is a factor of two smaller. In addition, the volume average is a factor of 3--4 smaller, since the opacity drops with radius, as the spectrum gets red-shifted (i.e., lower $\Tr$). This point has previously been made by \citet{Reissl_2018}, and our findings here confirm this.

This finding calls into question the results of the aforementioned models, primarily for cases with $\tau_*\gg 1$, i.e., models with more trapping of radiation, and hence higher (mid-plane) temperatures. For instance, figure~9 of \citet{Krumholz_Thompson_2013} shows that the mid-plane temperature in their run with $\tau_* = 10$, $f_{\mathrm{E},*} = 0.5 > f_{\mathrm{E},\mathrm{crit}}(\tau_*= 10)$ can be $\sim 400 \, \mathrm{K}$ using their dimensional scaling\footnote{The temperatures could be systematically higher in these simulations since they neglect gravity; however, even in the corresponding run with gravity in \citetalias{Krumholz_Thompson_2012}, the temperatures were $\sim 300 \, \mathrm{K}$.}. This corresponds to $\kappaR$ values that are a factor $\sim 5$ larger with the $\kappaPLT$ opacity than the corresponding $\kappaSem$ opacities at the same temperature. Indeed, based on our findings it is likely that a system with these parameters would not be unstable using the correct opacities. It is therefore important to revisit the conditions for driving a wind in the $\tau_*>1$ regime with the $\kappaSem$ opacities, and/or recompute the numerical simulations with them. This is especially relevant as the conditions on $f_{\mathrm{E},*}$ and $\tau_*$ for driving a wind are used to infer the plausibility of radiation pressure driving powerful dusty winds in ULIRGs and starburst galaxies \citep[e.g., sec.~4.2 in][]{Zhang_2017}, which observations suggest can reach $\tau_*\sim 30$ \citep{Munoz_2017}. It has been assumed that this is sufficient to trigger IR radiation-driven winds, but our results suggest a need to revisit this conclusion.

\subsection{Trapping of infrared radiation}
\label{sec:ftrap}

\begin{figure*}
    \centering
    \includegraphics[width = 0.88 \textwidth]{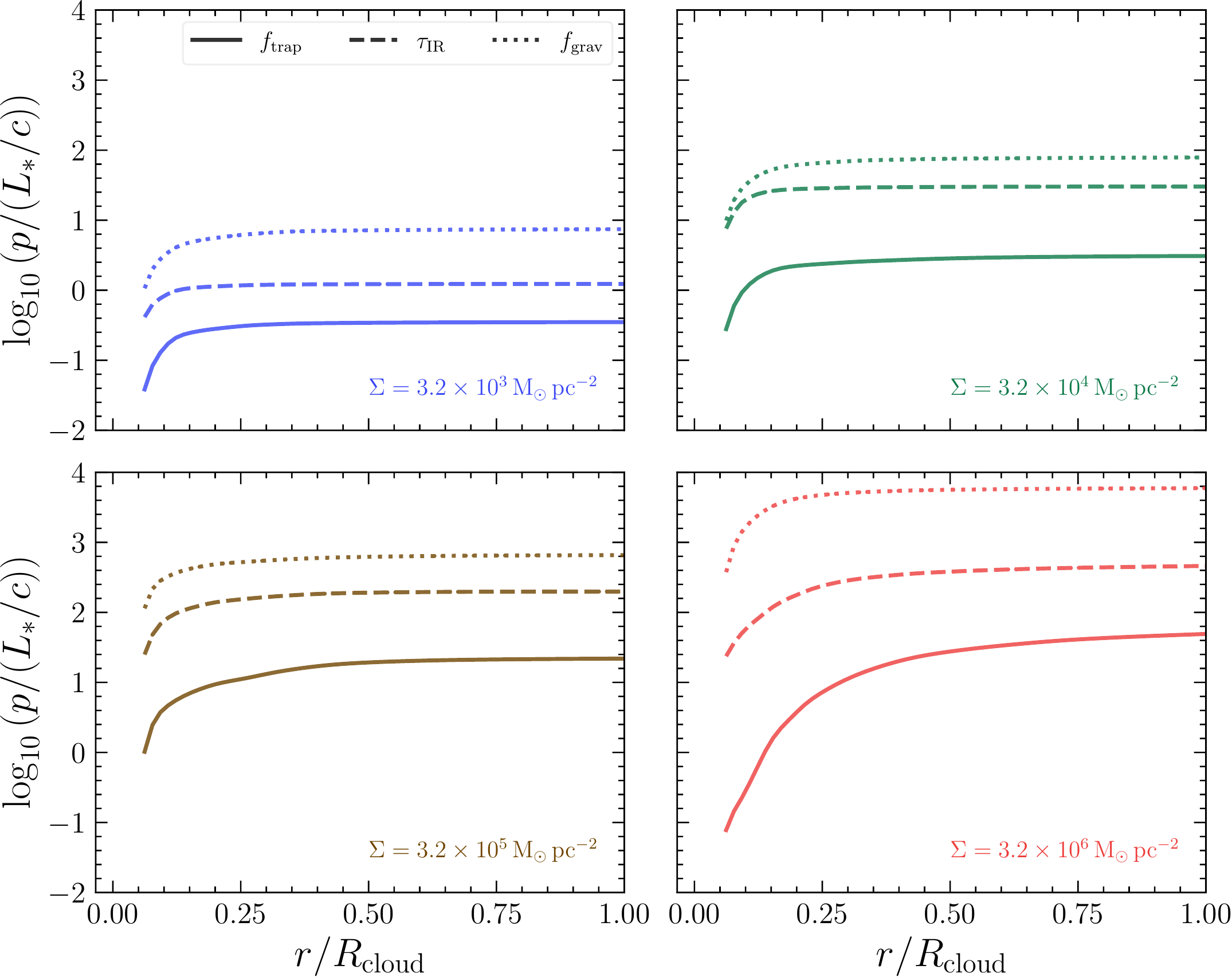}
    \caption{Comparison of the cumulative trapping factor ($\ftrap$; Equation~\ref{eq:fTrapsim}) in solid lines, the cumulative IR optical depth ($\tau_{\mathrm{IR}}$; Equation~\ref{eq:tauIR}) in dashed lines, and the (scaled) cumulative momentum imparted by gravity ($\fgrav$; Equation~\ref{eq:fGravsim}) shown as dotted lines, for different $\Sigmacloud$ shown in different panels. $\ftrap$ and $\tau_{\mathrm{IR}}$ respectively quantify the cumulative IR radiation momentum imparted to the gas in our simulations, and the cumulative momentum that would have been imparted if the cloud were spherically symmetric and had identical average radial profiles of the specific opacity $\rho \kappaR$; the difference between them arises primarily due to the radiation-matter anti-correlation. However, we also note that \textit{even} $\tau_{\mathrm{IR}}$ is lower than $\fgrav$.}
    \label{fig:ComparefTrapRadial}
\end{figure*}

\begin{figure}
    \centering
    \includegraphics[width = 0.48 \textwidth]{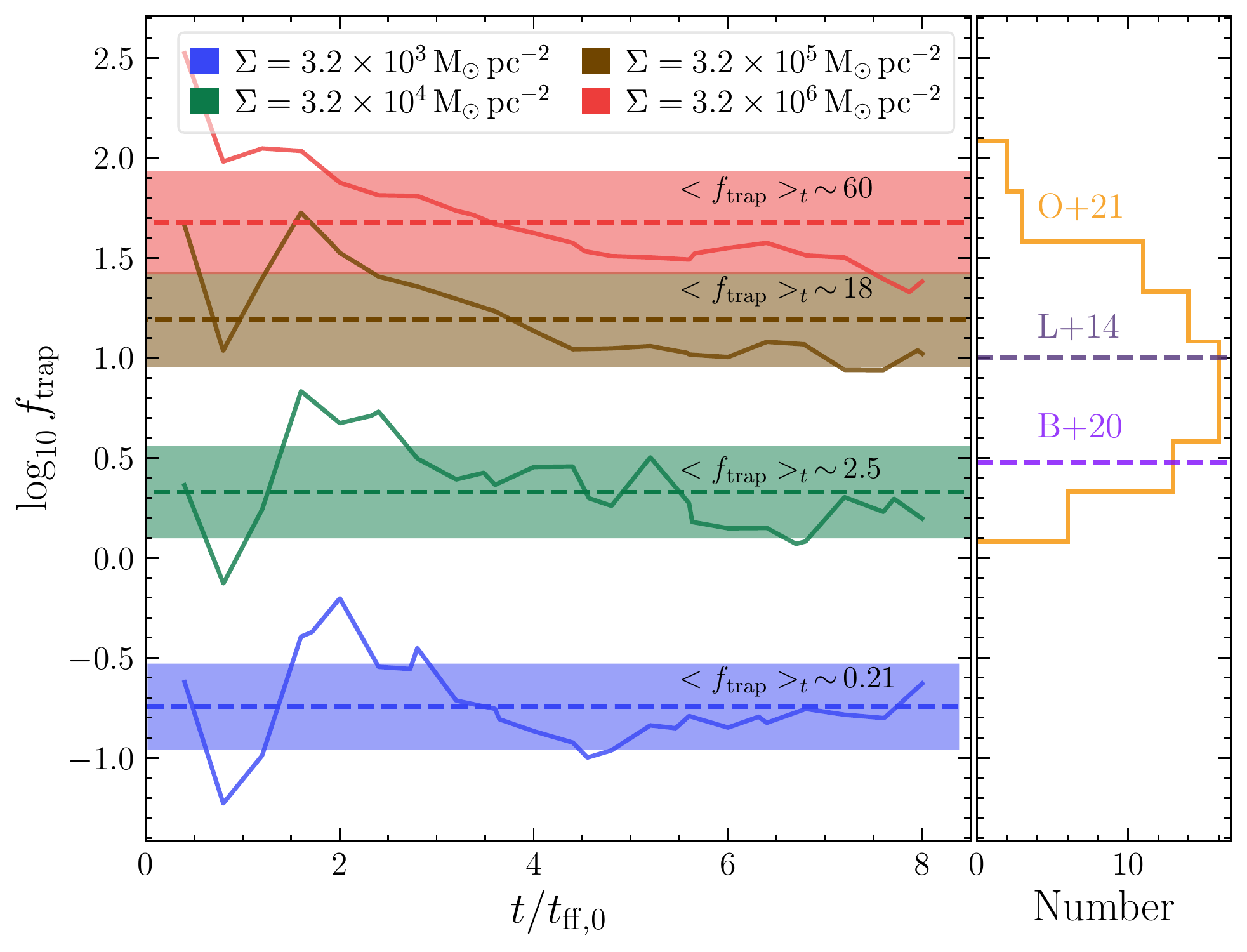}
    \caption{Comparison of the time evolution of the cumulative IR radiation trapping factor through the cloud ($f_{\mathrm{trap},\mathrm{cl}} \equiv \ftrap(r \to \Rcloud)$) for the four values of $\Sigmacloud$. Dashed lines indicate the time-averaged value of $f_{\mathrm{trap},\mathrm{cl}}$ for each case, which is also annotated in the figure, and the shaded bands indicate the $1\sigma$ variations. The panel on the right shows observational estimates of $\ftrap$: i) histogram of $\ftrap$ values derived for young, compact, $\ion{H}{ii}$ regions in the W49A star-forming region from \citet[][O+21]{Olivier_2021} ii) median value of $\ftrap$ inferred for $\ion{H}{ii}$ regions in the Central Molecular Zone (CMZ) from \citet[][B+20]{Barnes_2020}, and iii) the median value of $\ftrap$ for $\ion{H}{ii}$ regions in the Large and Small Magellanic clouds probed by \citet[][L+14]{Lopez_2014}.}
    \label{fig:ComparefTrap}
\end{figure}

Sub-grid models and observations of IR radiation feedback are frequently expressed in terms of the IR radiation trapping factor, $\ftrap$, defined as the ratio of the net momentum imparted to the gas by IR photons to the momentum flux carried by the direct stellar radiation field,
\begin{equation}
    \dot{p}_{\mathrm{rad},\mathrm{tot}} = (1 + \ftrap) L_*/c,
    \label{eq:fTrapdef}
\end{equation}
where $\dot{p}_{\mathrm{rad},\mathrm{tot}}$ is the total momentum per unit time transferred to the gas, and $L_*$ the luminosity of the star cluster. The factor of 1 in the above equation is the momentum imparted by the absorption of the direct (UV) photons from the star, and $\ftrap$ indicates the factor by which this is boosted due to the radiation pressure on dust by IR photons trapped in the opaque cloud. Spherically-symmetric semi-analytic models suggest values of $\ftrap \sim 4$--$5$ \citep[e.g.][]{Murray_2010} for gas conditions similar to massive clusters in the Milky Way, but \citetalias{Skinner_2015} find in their three-dimensional simulations that these models can overestimate $\ftrap$ by a factor of $\sim 5$ due to their inability to include radiation-gas anti-correlation. That said, \citetalias{Skinner_2015} use idealised constant opacities, which we have shown can yield misleading results, and they only explore clouds with relatively low values of $\Sigmacloud$. It is therefore of interest to investigate $\ftrap$ in our simulations, and their implications for both sub-grid models and for the interpretation of observations.

\subsubsection{The trapping factor}
We obtain $\ftrap$ from our simulations by directly computing the cumulative momentum imparted to gas by radiation pressure over the extent of the cloud, and taking the ratio of this to the momentum flux carried by the radiation field (i.e., $L_*/c$). The cumulative momentum imparted is given by integrating $\forcerad$ over spherical shells through the cloud, which gives
\begin{equation} \label{eq:fTrapsim}
    \ftrap(r) =  \frac{\int_{0}^{r} \forcerad 4\pi r^2 dr}{L_*/c}.
\end{equation}
Here the numerator is the rate of momentum deposition in gas out to radius $r$. For comparison, we also compute the cumulative optical depth through the cloud, given by
\begin{equation} \label{eq:tauIR}
    \tau_{\mathrm{IR}} (r) = \int_{0}^{r} \langle \rho \kappaR \rangle_{4\pi} dr,
\end{equation}
where $\langle \rho \kappaR \rangle_{4\pi}$ is the averaged specific opacity over all solid angles. This quantity represents the value of $\ftrap$ that would be expected \textit{if} our gas cloud were spherically-symmetric and had the same radial opacity distribution as our simulated clouds. Thus, if $\ftrap<\tau_{\mathrm{IR}}$, this indicates radiation-matter anti-correlation. 

We show the radial profiles of $\ftrap$ and $\tau_{\mathrm{IR}}$ in Figure~\ref{fig:ComparefTrapRadial}. We see that $\ftrap$ and $\tau_{\mathrm{IR}}$ both saturate at $r \sim 0.3\,R_\mathrm{cloud}$, suggesting that most of the momentum is imparted in the regions close to the star clusters. We find that $\tau_{\mathrm{IR}}$ is larger than $\ftrap$ by about a factor $\sim 10$, with larger differences at larger $\Sigmacloud$, pointing to coupling of the radiation to matter that is somewhat inefficient at low $\Sigmacloud$ and becomes increasingly inefficient as $\Sigmacloud$ increases, consistent with our findings above. However, in spite of the anti-correlation, $\ftrap$ is significantly higher than unity at higher $\Sigmacloud$.

\subsubsection{Implications for sub-grid models}

Our finding that $\ftrap > 1$ raises the following question: if high values of $\ftrap$ are possible, why do they not affect the dynamical state of the cloud, and what are the implications of this finding for semi-analytic models of IR radiation feedback where the strength of feedback is parameterised by $\ftrap$?
After all, semi-analytical models \citep{Thompson_2005,Murray_2010, Hopkins_2010}, and numerical simulations that adopt versions of them \citep{Hopkins_2011, Hopkins_2012}, find momentum-driven expansion of their model clouds even at modest $\ftrap$ values of $\sim 10$. Why do our simulations have $\ftrap > 10$ in some cases, yet IR feedback has no effect?

This question can be addressed by comparing the trapped IR momentum budget (i.e., $\ftrap L_*/c$) with the momentum per unit time provided by the gravitational force. We quantify this by calculating an effective gravity momentum, in analogy to Eq.~(\ref{eq:tauIR}),
\begin{equation} \label{eq:fGravsim}
    \fgrav(r) =  \frac{\int_{0}^{r} \forcegrav 4\pi r^2 dr}{L_*/c},
\end{equation}
where $\dot{p}_{\rm grav}$ is the change in gas radial momentum per unit time per unit volume due to gravitational forces. This quantity is directly comparable to $\ftrap(r)$, in that it describes the momentum per unit time imparted to gas at radius $<r$ by gravitational forces, again normalised to the momentum flux of the direct stellar radiation field. We over-plot $\fgrav$ in Figure~\ref{fig:ComparefTrapRadial}. We can see that $\fgrav$ exceeds $\ftrap$ at all $r$, explaining why even very high values of $\ftrap$ need not lead to winds driven by radiation pressure -- higher $\ftrap$ comes with even higher $\fgrav$. This is not surprising, and is just another way of stating the point that gravitational forces dominate over the radiation pressure for all $r$ in our clouds; in such a picture, it does not matter that $\ftrap$ is very large, because $\fgrav$ is always larger.

However, this finding has an important implication for sub-grid models applied to simulations that do not resolve the competition of radiation pressure and gravity: such simulations \textit{should not impart momentum with ad-hoc values of $\ftrap$} -- even ones that are calibrated to three-dimensional simulations that resolve the radiation feedback, such as ours. This is due to the fact that, since $\fgrav > \ftrap$, the gas that absorbs the trapped photon momentum gets accreted back to the sources of radiation, and hence cannot couple this momentum to the larger-scale ISM. 
Thus, it would be incorrect to compute an effective $\ftrap$ and impart this momentum in unresolved simulations without considering the corresponding enhancement of gravity due to the presence of unresolved bound structures, as embodied by $\fgrav$. To be consistent, a model should either set both $\ftrap$ and $\fgrav$ to zero (i.e., no sub-grid trapping), or include an explicit sub-grid estimate of both, with the condition that, if $\ftrap < \fgrav$, no momentum is imparted to the gas on resolved scales because all of the radiation momentum is advected back into the stars\footnote{See \citet{Krumholz_2018} for a sub-grid model that attempts this in the single-scattering limit.}. The reason that some previous models have found IR radiation feedback to be effective is that likely they were not consistent, because they set $\ftrap > 1$, but implicitly adopted $\fgrav = 0$.

\subsubsection{Comparison to observational estimates}
In Figure~\ref{fig:ComparefTrap} we plot the time evolution of $f_{\mathrm{trap},\mathrm{cl}}\equiv\ftrap(r \to \Rcloud)$, the cumulative momentum imparted through the entire cloud for our runs; we denote the time-average of this quantity throughout a simulation as $\langle f_{\mathrm{trap},\mathrm{cl}} \rangle_t$. We see that $f_{\mathrm{trap},\mathrm{cl}}$ increases for higher $\Sigmacloud$ -- consistent with the expectation of more trapping at higher surface densities -- with $\langle f_{\mathrm{trap},\mathrm{cl}} \rangle_t \sim 0.2$ for the lowest $\Sigmacloud$, and $\langle f_{\mathrm{trap},\mathrm{cl}} \rangle_t \sim 60$ for the highest $\Sigmacloud$ case. We also see significant time evolution in this quantity, especially at higher $\Sigmacloud$, displaying signs of a decline at later times. This variation is driven by the corresponding time variations in $\tau_{\mathrm{IR}}$, which decreases as the cloud is depleted by accretion through the cloud; clouds maintain a fairly (temporally) constant ratio of $f_{\mathrm{trap},\mathrm{cl}}$ to $\tau_{\rm IR}$. We also compare our (time-averaged) estimates of $\ftrap$ with observational estimates for it in Figure~\ref{fig:ComparefTrap}. We can see that the values of $\ftrap$ obtained in our simulations span similar ranges as observational estimates.

\subsection{Implications for star formation in dense/starburst environments}
\label{sec:SFEImplications}

The evolution of clouds in our simulations suggest that IR radiation pressure is unlikely to play a dynamically important role in regulating $\epsilon_*$ in extreme environments, which presumably are locations where SSCs would form. We find values of $\epsilon_*\sim 80\%$, in contrast to Milky Way-like GMCs, which reach integrated efficiencies of only $\epsilon_* \sim 5$--$20 \%$ \citep[and references therein]{Chevance_2022}. This is probably maintained in Milky Way GMCs through a combination of turbulence, magnetic fields and \textit{dynamically competitive} feedback mechanisms \citep[e.g.,][]{Federrath2015,Grudic_2018}. Our simulations show that IR radiation feedback cannot compete with gravity in this regime, and other feedback mechanisms are also expected to be ineffective in this regime of $\Sigma$ \citep[see the review by][]{Krumholz_2019}.
Turbulence driven by galactic-scale forces (e.g., tidal shear from a galactic collision -- common in galaxies forming SSCs) may be able to slow down star formation, and lead to lower integrated star formation efficiencies by dispersing unbound gas -- as evidenced by simulations with initially unbound clouds ($\alphavir$ series; Table~\ref{tab:Simulations}). However, even in these cases, we find $\epsilon_* \gtrsim 60 \%$.
Our findings therefore suggest that star formation in dense environments such as proto-SSCs and nuclear star clusters should achieve high efficiencies ($\sim 60$--$80 \%$), and are therefore very likely to be gravitationally bound at formation.

This is consistent with recent observational studies of SSC and proto-SSC environments in the local Universe. Efficiencies of $\epsilon_* \sim 50$--$80\%$ and high fractions of stars in bound clusters have been inferred for the dense star-forming regions in the starburst galaxies NGC~253 \citep{Leroy_2018,Villas_2020}, NGC~5253 \citep{Turner_2015}, NGC~4945 \citep{Emig_2020}, Arp~240 \citep{Hao_2020}, Antennae \citep{Hao_2022,Finn_2019}, and Henize~2-10 \citep{Costa_2021}. 
While high ratios of stellar to gas mass (which is ultimately what these observations measure) can be a sign of gas dispersal even when $\epsilon_*$ is low, line-widths of dense gas tracers, and an observed presence of quiescent dense gas within clusters \citep{Turner_2015,Villas_2020}, suggests this is unlikely to be the case. 

The main exception to this trend is presented in \citet{Levy_2021}, who analyse high-resolution ($\sim 0.5 \, \mathrm{pc}$) observations of SSCs in NGC~253 and find that a subset of their sample shows signs of (dense-gas) outflows indicated by P-Cygni profiles in the spectra of multiple tracers, with inferred mass fluxes large enough to substantially reduce $\epsilon_*$. They suggest that radiation pressure and/or stellar winds are the most likely candidates to drive this outflow, but our work shows that the former is unlikely. The low efficiency of wind-driven feedback found in recent models \citep{Lancaster_2021,Lancaster_2021b} suggests that winds are also unlikely to be the primary driver of these outflows. However, we note that the simulations in \citet{Lancaster_2021b} are limited to environments with $\Sigma \lesssim 3.2 \times 10^3 \, \Msolpc$; simulations of stellar wind feedback in more extreme environments -- such as NGC~253 -- have not been performed to our knowledge. It is also possible that the combined effects of stellar winds and radiation pressure are able to unbind gas, and drive outflows.  Therefore, it remains an open question as to what drives these observed outflows; future observations of proto-SSCs to compile a statistical sample, coupled with models/simulations that incorporate multiple feedback mechanisms, are needed to address this question.

\subsection{Caveats}
\label{sec:Caveats}
We end this discussion by pointing out that our numerical simulations are missing several physical mechanisms that might play a further role in regulating star formation, and hence stress that the reader interpret our results in light of these limitations. We briefly discuss these caveats and speculate the effects they might have on our results.

The primary caveat is that we do not model the direct stellar UV radiation field in our simulations, thereby omitting two other feedback mechanisms: photoionisation of hydrogen and the associated thermal pressure, and the single-scattering UV radiation pressure. The former would play a relatively minor role for the parameter space we are exploring: the escape speeds of even our lowest $\Sigmacloud$ run is $\sim 30 \, \kms$, significantly greater than the ionised gas sound speed ($\sim 10 \, \kms$), a limit at which earlier studies have shown that ionising feedback is dynamically subdominant to both radiation pressure and gravity \citep{Dale_2012,Kim_2018}. On the other hand, the UV radiation pressure on dust could be important as the dust opacity at these wavelengths can be $2-3$ orders of magnitude higher than in the IR \citep[e.g.,][]{Draine_2011}, and the available momentum would be imparted closer to the stars. Not directly including the UV systematically underestimates the forces of radiation in all our models by at most a value of $\sim L_*/c$. This might provide additional support against gravity that might be sufficient for radiation pressure to be dynamically relevant. However, in clouds with high $f_{\mathrm{trap}}$, including the direct UV force would have a very modest effect, as it would change $f_{\mathrm{Edd}}$ by about $\sim 1/f_{\mathrm{trap}}$ -- insufficient to render the cloud super-Eddington. However, this need not be the case for the lower $\Sigmacloud$ cases, where $f_{\mathrm{trap}}$ is of the order of a few; here the direct UV can be comparable, or even dominant over the total radiation force, and is crucial to obtain the correct total radiation force. Single-scattering radiation pressure also has the attractive property that even if the cloud is globally sub-Eddington to this force, it can eject gas in sight lines that have lower $\Sigma$ set by turbulence \citep{Thompson_Krumholz_2016, Wibking_2018}. In other words, there is always the possibility that \textit{some} fraction of the cloud be susceptible to being ejected by the direct UV radiation pressure. Combined with the IR radiation pressure, it is possible that the dynamical outcomes of some our model clouds be affected. 
We also note that, because we do not model the UV radiation field directly, we must rely on a spherically-symmetric source term for IR photons (Equation~\ref{eq:jstar}), which does not take into account the asymmetric density distribution of matter around the sources of radiation. This conceivably leads us to underestimate the effectiveness of IR radiation on very small scales, where direct radiation would deposit energy most efficiently in the densest structures. It is therefore of interest to model the radiation field in both the UV and IR bands, including the contribution of the direct UV radiation pressure along with the IR pressure, self-consistently injecting IR photons via the reprocessing of the absorbed UV photons by dust, and thereby accounting for non-symmetric density distributions.

We also do not include magnetic fields in our simulations. Magnetic fields have been shown to play a role in the effectiveness of feedback mechanisms -- by coupling forces from impulsive, localised mechanisms such as proto-stellar outflows/jets to a more extended region \citep[e.g.,][]{Nakamura_2007, Wang_2010, FederrathEtAl2014, Offner_2018}, or minimising energy losses due to mixing/conduction in energy-driven feedback mechanisms such as stellar winds or supernovae \citep[e.g.,][]{Markevitch_2007, Gentry_2019}. However, we argue that these effects are unlikely to be important for the IR radiation pressure force, which is a more volume-filling, momentum-driven mechanism, for which magnetic fields do not directly alter the force that can be imparted. However, magnetic fields can provide additional support against gravitational collapse and therefore possibly render higher fractions of gas unbound \citep[e.g.,][]{Krumholz_Federrath_2019}, in analogy to the effects we find by raising $\alphavir$, but with the advantage that magnetic support does not decay the way turbulent support does. We cannot entirely rule out the possibility that with this additional support, the radiation forces might become dynamically important. We also do not include explicit, continuous, turbulence driving, which emulates the turbulent cascade from larger scales in the ISM -- this too might act as an additional source of support. Therefore, the reader should interpret our high values of $\epsilon_*$ with caution, as there are additional mechanisms that might work together in a non-trivial way to potentially enhance the role of radiation feedback.

\section{Conclusions}
\label{sec:Conclusions}
We conduct 3D radiation hydrodynamic simulations of star cluster formation and evolution in massive, dusty, self-gravitating clouds in the multiple-scattering limit, with an aim to determine the star formation efficiency ($\epsilon_*$) of clouds set by the competition between gravity and IR radiation pressure. We use the recently developed \texttt{VETTAM} algorithm \citep{Menon_2022}, which uses the state-of-the-art Variable Eddington Tensor closure, and, unlike earlier radiative transfer methods, allows us to sample a broad range of cloud surface densities ($\Sigmacloud \sim 10^3$--$10^6 \, \Msolpc$). Thus, our simulations cover the range of conditions found in young massive clusters and super star clusters, in both the Milky Way and denser starburst environments. We also compare the results using models for dust opacity at increasing levels of realism -- constant opacities, a commonly used power-law approximation valid at low temperatures, and opacities taken from a more realistic, complex dust model \citep{Semenov_2003} -- and using a range of dust-to-gas ratios. We draw the following conclusions from our simulations:

\begin{enumerate}
    \item For realistic dust opacities, infrared radiation pressure is highly unlikely to regulate star formation, even at the highest cloud surface densities. We find that the star formation efficiency $\epsilon_* \sim 80 \%$ in our simulations, and radiation pressure does not show \textit{any} evidence of driving winds or inhibiting accretion. This occurs because the clouds are sub-Eddington at all radii.
    \item This conclusion continues to hold even for clouds that are initially unbound, or that are dust-enriched up to twice the solar-neighbourhood dust-to-gas ratio. Only for dust-to-gas ratios of $\sim 3$ times solar does IR radiation feedback play \textit{some} role, and even then it produces only a $\sim 10\%$ reduction in $\epsilon_*$.
    \item Radiation forces are relatively weak due to a combination of modest IR dust opacities, and radiation-matter anti-correlation that renders feedback inefficient. We find that with realistic IR opacities, the former is primary, as it is sufficient by itself to render clouds wholly sub-Eddington; the latter just aggravates the ineffectiveness of feedback, and only becomes significant for super-solar dust to gas ratios ($\gtrsim 3\times$ solar), or for radiation sources with significantly higher light-to-mass ratios than stellar populations with a standard IMF (e.g., individual massive stars).
    \item These results depend critically on using realistic temperature-dependent dust opacities. Both of the approximations we test -- using constant IR opacities sampled at the peak of the \citet{Semenov_2003} opacity curve ($\sim 10 \, \mathrm{cm}^2 \, \mathrm{g}^{-1}$) or using a pure power-law scaling for the opacity as a function of temperature -- overestimate the opacity and thus the effects of radiation. The constant-opacity approximation fails because only very small localised regions of the cloud are at these opacities, and the power-law approximation fails because it leads to opacities that are far too large in the warmer parts of clouds (at temperatures $\gtrsim 100\,\mathrm{K}$). The fact that these approximations yield qualitatively different conclusions than our simulations with more realistic opacities calls into question a number of results in the literature that are based on them (in particular the role of radiation in limiting $\epsilon_*$ and in driving large-scale galactic winds).
    \item Despite the ineffectiveness of IR radiation feedback, IR radiation can nonetheless be effectively trapped and impart significantly more momentum to dusty gas than would be the case in the single-scattering limit; trapping factors in our simulations span the range $\ftrap \sim 0.2$--$60$, depending on the cloud surface density, and those simulations with surface densities comparable to those seen around the Milky Way H~\textsc{ii} regions produce trapping factors in reasonable agreement with observationally-inferred values. However, this momentum does \textit{not} couple to the larger-scale ISM, because the gas that receives this momentum is eventually accreted, i.e., we show that under the conditions where $\ftrap>1$, gravity dominates radiation. Sub-grid models that include IR radiation feedback with $\ftrap > 1$, but do not include the effects of gravity at unresolved scales and its effects on confining the gas, are therefore inconsistent.
\end{enumerate}

Since IR radiation pressure is thought to be the only significant feedback mechanism at very high surface density, our results imply that regions such as (proto-) super star clusters (SSCs) in starburst environments \citep[e.g.,][]{Leroy_2018}, or other extreme stellar systems in our Universe such as nuclear star clusters, should form stars very efficiently and with a large fraction of the stars confined to bound structures. Only if another physical feedback mechanism that is unaccounted for in our simulations or a non-trivial combination of physical effects (such as magnetic fields, turbulence and jet/outflow feedback) becomes important, would this picture change. There is significant scope to study star formation in such extreme environments with more sophisticated numerical simulations, and statistically significant sample sizes of observed SSC-forming regions. 

\section*{Acknowledgements}

We thank Todd Thompson and the anonymous referee for constructive comments on the manuscript. S.~H.~M would like to thank Rolf Kuiper for assistance with the numerical setup of the simulation, Benjamin D.~Wibking for assistance with our idealised semi-analytic models, and Grace M.~Olivier for sharing their data on trapping factors. S.~H.~M also acknowledges useful discussions with M.~Aaron Skinner, Eve C.~Ostriker, Shane Davis, Ahmad Ali, and Dmitry A.~Semenov, which greatly assisted the progress of this study. C.~F.~acknowledges funding provided by the Australian Research Council through Future Fellowship FT180100495, and the Australia-Germany Joint Research Cooperation Scheme (UA-DAAD). M.~R.~K.~acknowledges funding from the Australian Research Council through its \textit{Discovery Projects} and \textit{Future Fellowship} funding schemes, awards DP190101258 and FT180100375. We further acknowledge high-performance computing resources provided by the Leibniz Rechenzentrum and the Gauss Centre for Supercomputing (grants~pr32lo, pn73fi, and GCS Large-scale project~22542), and the Australian National Computational Infrastructure (grants~ek9 and~jh2) in the framework of the National Computational Merit Allocation Scheme and the ANU Merit Allocation Scheme. Large parts of this paper were written at the RSAA student writing retreat at Batemans Bay, and S.~H.~M would like to acknowledge the organisers for providing this opportunity.

\textit{Software}: \texttt{PETSc} \citep{PetscConf,PetscRef}, \texttt{NumPy} \citep{numpy}, \texttt{SciPy} \citep{scipy}, \texttt{Matplotlib} \citep{matplotlib}, \texttt{yt} \citep{yt}. This research has made use of NASA's Astrophysics Data System (ADS) Bibliographic Services.

\section*{Data Availability}
Outputs of our simulations would be shared on reasonable request to the corresponding author.



\bibliographystyle{mnras}
\bibliography{RHDCluster,federrath} 




\appendix

\section{Numerical convergence studies}
In this section, we test the dependencies of our results on certain numerical choices that have been made in our simulations. Since the fiducial simulations are all sub-Eddington, the values of $\epsilon_*$ attained are virtually identical, irrespective of these numerical choices. Therefore, we found it more informative to compare the Eddington ratio $f_{\mathrm{Edd}}$ with different numerical choices. 

\subsection{Dependence on source size}
\label{sec:AppendixSource}

\begin{figure}
    \centering
    \includegraphics[width = 0.48\textwidth]{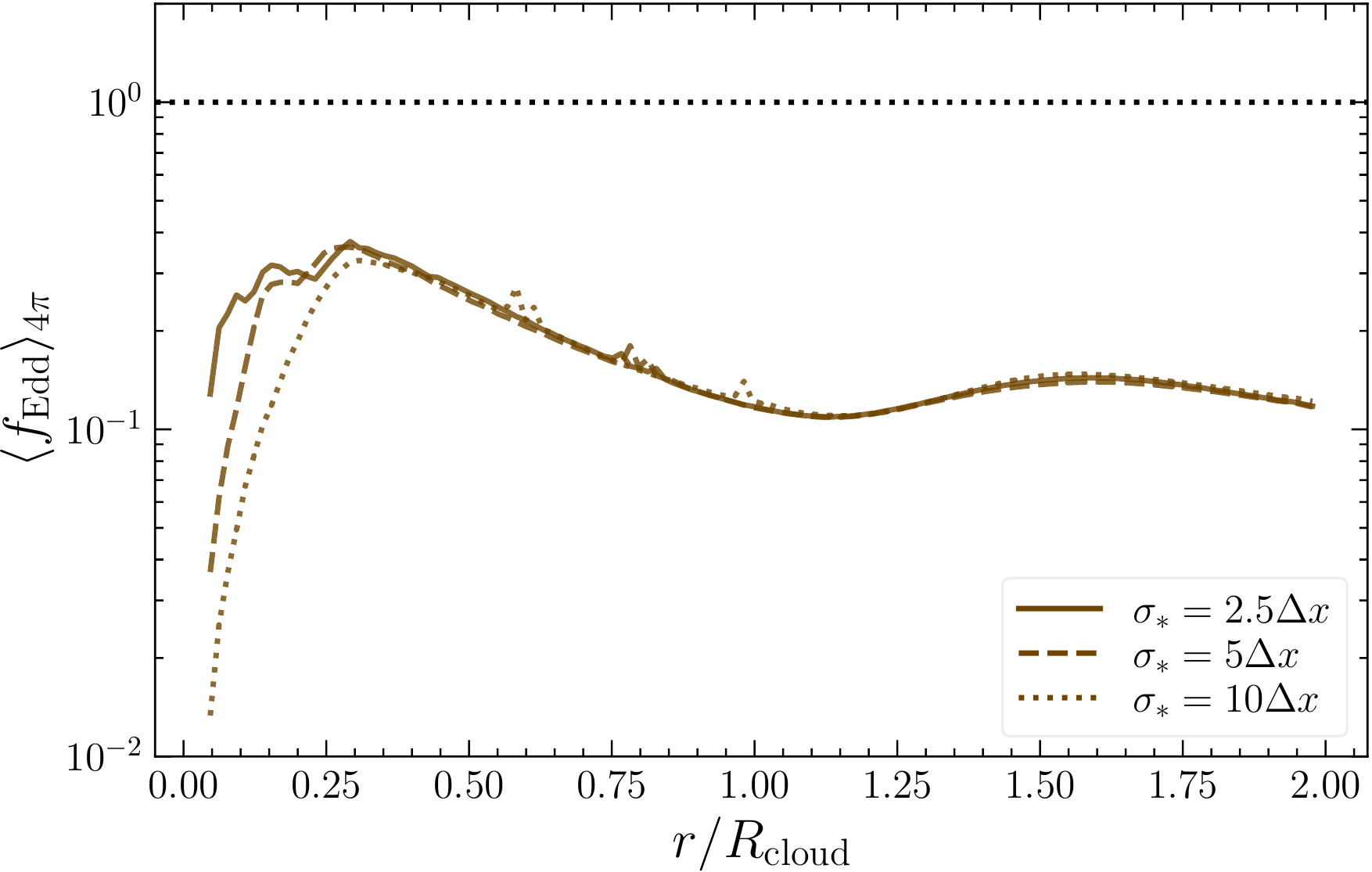}
    \caption{Comparison of the obtained Eddington ratios for the run \texttt{S5KSemA2F1} at $t = 3 \, t_{\mathrm{ff}}$, with different choices for the source size ($\sigma_*$). The choice adopted in our study is $\sigma_* = 2.5 \Delta x$, where $\Delta x$ is the cell width.}
    \label{fig:SourceSizeDependence}
\end{figure}

Here we test the dependence of our results on the adopted choice for the size of the source $\sigma_*$ in Equation~(\ref{eq:jstar}). Our fiducial choice is $\sigma_* = 2.5 \Delta x$, where $\Delta x$ is the size of a computational cell on the highest level of AMR where the sink particle is located; for our fiducial resolution, this corresponds to $\sigma_* = 0.039\, R_{\rm cloud}$. We check the dependence of our results on this choice by repeating run \texttt{S5KSemA2F1} with alternate choices of $\sigma_* = 5 \Delta x = 0.078 \, R_{\rm cloud}$ and $\sigma_* = 10 \Delta x = 0.16\, R_{\rm cloud}$. We plot the Eddington ratios we obtain at $t=3t_{\mathrm{ff}}$ for this test in Figure~\ref{fig:SourceSizeDependence}. We see that all the curves are nearly identical at all radii substantially larger than $\sigma_*$; at $r \lesssim \sigma_*$, the Eddington ratio $f_{\mathrm{Edd}}$ decreases because $j_*$ is smoothed in this region, and hence radiation forces are smaller. Based on this test we conclude that the effects of smoothing are minimal for our fiducial choice of $\sigma_*$, because Figure~\ref{fig:SourceSizeDependence} demonstrates that smoothing has negligible effects outside the smoothed volume, and the volume that is affected for $\sigma_* = 2.5\Delta x$ is only $\sim 10^{-5}$ of the total cloud volume.

\subsection{Dependence on the micro-physical dust model}
\label{sec:AppendixDustModel}

\begin{figure}
    \centering
    \includegraphics[width = 0.48\textwidth]{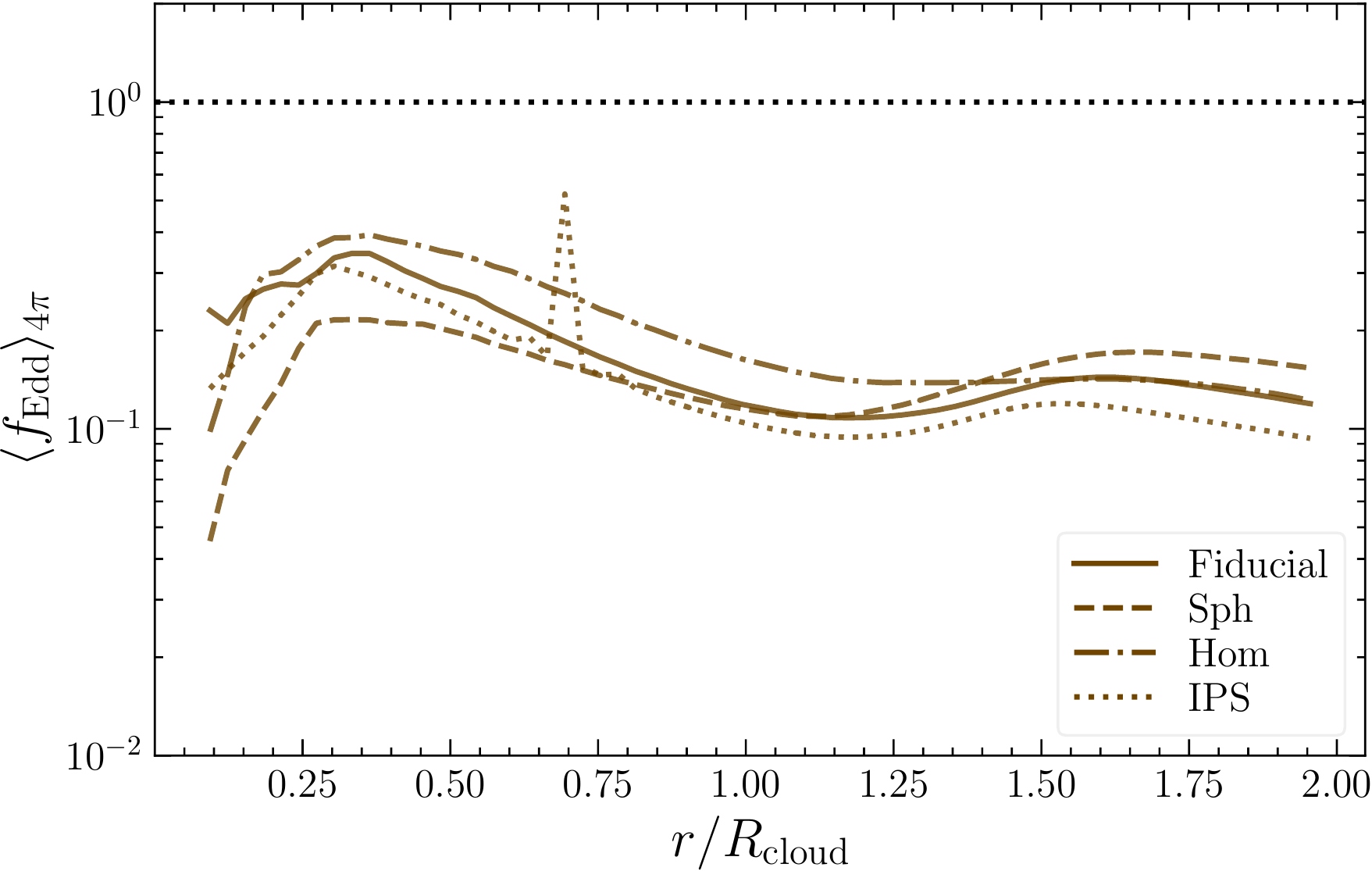}
    \caption{Eddington ratios at $t = 3 \, t_{\mathrm{ff}}$ in run \texttt{S5KSemA2F1} repeated using different micro-physical properties. The opacities in the fiducial run assume that dust grains are aggregate particles made of composite materials with normal iron content in the silicates. The alternative models shown use spherical (Sph) rather then aggregate particles, homogeneous (Hom) rather than composite dust, and iron-poor silicates (IPS) rather than normal iron content. See text for further details.}
    \label{fig:DustModelDependence}
\end{figure}

Since we find that realistic dust opacities (i.e., $\kappaSem$) are crucial to our results, it is important to check for any dependencies on the particulars of the dust model. \citet{Semenov_2003} present a range of possible models, based on differing assumptions regarding the micro-physical nature of dust grains -- specifically their topology, shape and chemical composition. Our default choices are grains with an ``aggregate'' shape (i.e., they are a cluster of small spherical sub-grains sticking together), composite'' composition (i.e., they incorporate a mixture of materials), and ``normal'' (NRM) iron content in the dust silicates; alternative assumptions for each of these properties lead to changes in the opacity \citep[see fig.~1 in][]{Semenov_2003}.

To test the sensitivity of our results to these choices, we again repeat run \texttt{S5KSemA2F1} using some of \citeauthor{Semenov_2003}'s alternative dust models. We test the following choices: ``spherical'' particles instead of a cluster of sub-grains, ``homogeneous'' rather than heterogeneous composition, and ``Iron-poor silicates'' (IPS) instead of normal (NRM) iron content; for each of these variations, we leave the remaining properties unchanged, e.g., our spherical grain models have composite composition and normal iron content. We plot the Eddington ratios $f_{\mathrm{Edd}}$ obtained from these experiments in Figure~\ref{fig:DustModelDependence}. We see that alternative dust models lead to mild changes in the Eddington ratio, but the results are qualitatively quite similar (the same to within a factor of $\lesssim2$), and most importantly, all are sub-Eddington. This suggests that our major results are robust against plausible changes in the assumed micro-physical dust properties.

\subsection{Dependence on resolution}
\label{sec:AppendixResolution}

\begin{figure}
    \centering
    \includegraphics[width = 0.48\textwidth]{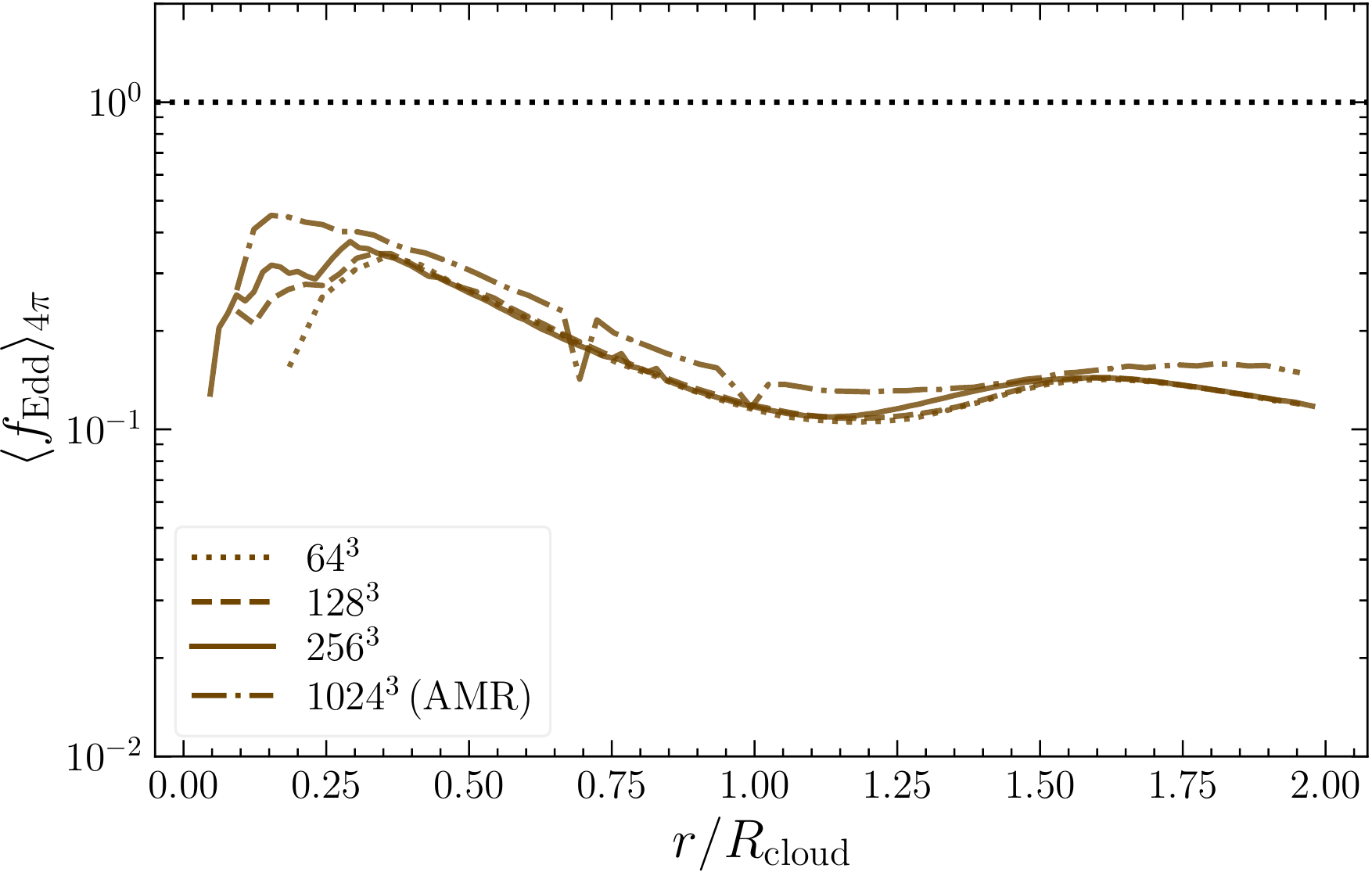}
    \caption{Comparison of the obtained Eddington ratios for the fiducial run at $t = 3 \, t_{\mathrm{ff}}$, with different grid resolutions. Solid lines indicate the resolution used in exploring parameter space in our study. The AMR run had an effective maximum resolution of $1024^3$ cells. We see that the runs are reasonably converged with resolution.}
    \label{fig:ResolutionStudy}
\end{figure}

We test for numerical convergence of our results by comparing runs with different grid resolutions. We again repeat simulation \texttt{S5KSemA2F1} with uniform-grid resolutions of $64^3$ and $128^3$ to compare with our fiducial choice of $256^3$, and we further compare to an AMR run with a base-grid resolution of $128^3$, and 3~levels of refinement, thereby leading to a maximum effective resolution of $1024^3$. For the AMR run we refine based on the Jeans length, requiring that it be resolved by at least 8~cells \citep{Truelove_1997,FederrathSurSchleicherBanerjeeKlessen2011}. We compare the radial profiles of Eddington ratios in these runs in Figure~\ref{fig:ResolutionStudy}. We see that at higher resolution, the radial location of maximum $f_{\mathrm{Edd}}$ shifts inwards with increasing resolution. However, the maximum value of $f_{\mathrm{Edd}}$ shows changes of at most $\sim 20 \%$, and over the bulk of the cloud, $f_{\mathrm{Edd}}$ is fairly independent of resolution. Therefore, we conclude that our qualitative results are insensitive to resolution, and thereby are reasonably converged.

\section{Idealised spherically-symmetric model cloud}
\label{sec:AppendixModel}

\begin{figure}
    \centering
    \includegraphics[width=0.5\textwidth]{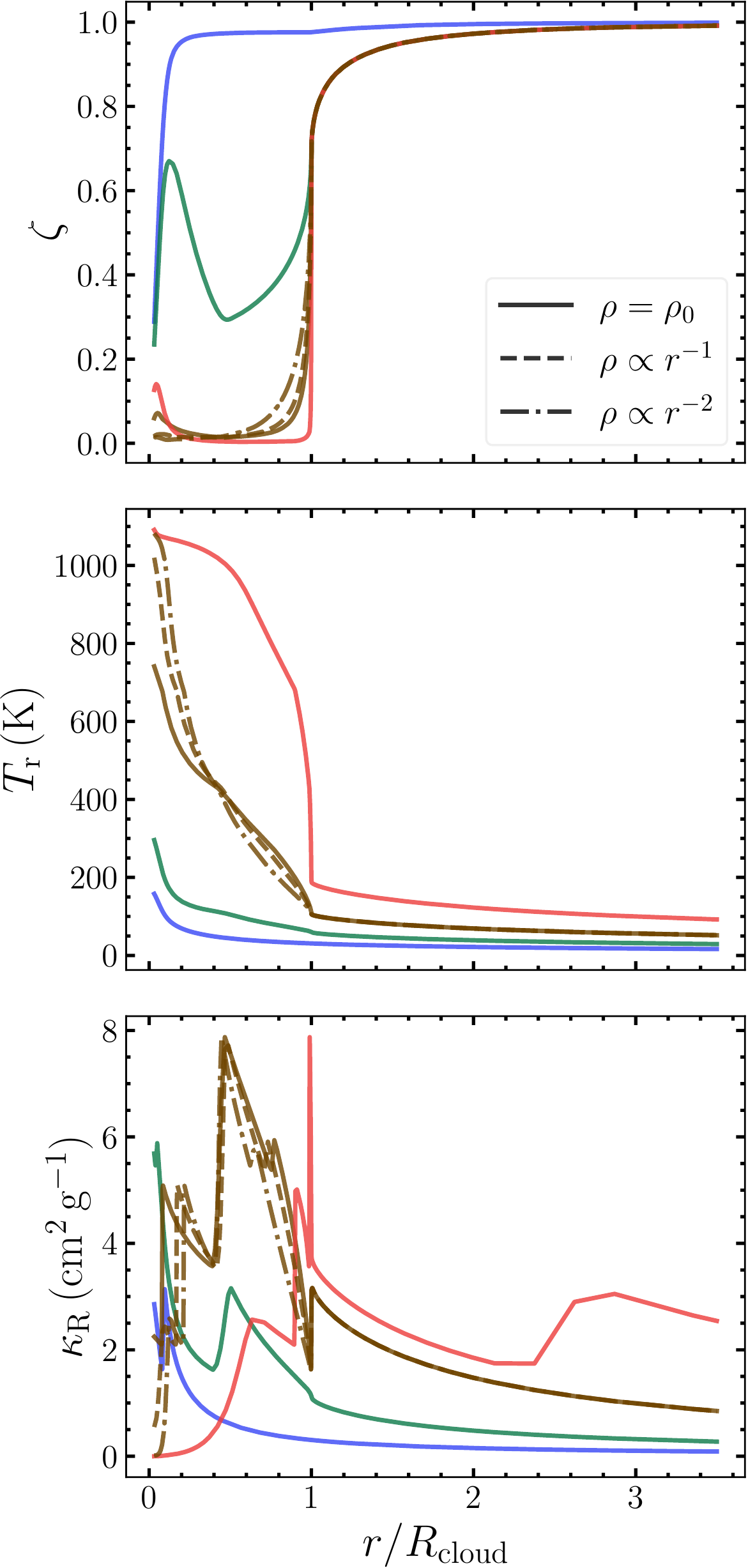}
    \caption{The reduced flux ($\zeta$; top), radiation temperature ($\Tr$; middle), and Rosseland opacity ($\kappaR$; bottom) profiles obtained with the steady-state semi-analytic models described in Section~\ref{sec:AppendixModel}. Colours indicate the four values of $\Sigmacloud$ probed in our study; solid, dashed and dash-dotted lines indicate values of $\alpha$ in the density profile.}
    \label{fig:fEddProfiles}
\end{figure}

Here we fully describe the semi-analytical model that we use  to obtain the steady-state radial profiles of matter and radiation quantities in spherical symmetry. Our approach follows the one described in section~4.3.6 of \citet{Skinner_2013}, modified to accommodate non-constant opacities. We model a spherically-symmetric cloud of gas and stars with a total mass $M = 10^6 \, \Msun$; the stars are located at the origin ($r=0$), and we inject (IR) photons with a radial distribution that follows Equation~(\ref{eq:jstar}). Their luminosity $L_* = \langle L_*/M_*\rangle M_*$, where $M_* = \epsilon_* M$ is the stellar mass, $\epsilon_*$ is the mass fraction in stars, and $\langle L_*/M_*\rangle = 1.7 \times 10^3 \, \mathrm{erg} \, \mathrm{s}^{-1} \, \mathrm{g}^{-1}$. We set the size of the source $\sigma_* = 0.03\,R_{\mathrm{g}}$, where $R_{\mathrm{g}}$ is the maximum radius of the cloud, but, as discussed in Appendix~\ref{sec:AppendixSource}, this choice makes little difference at radii $r > \sigma_*$.
We parameterise our model in terms of the gas mass surface density $\Sigma = M_{\mathrm{g}}/(\pi R_{\mathrm{g}}^2)$, and an index $\alpha$ such that $\rho \propto r^{-\alpha}$ describing the radial profile of gas density. For a given $\Sigma$ and $\alpha$, it is straightforward to obtain the normalisation of the density profile required to reach the desired mass $M$. We set $\rho(r) = 10^{-10}\rho_0$ for $r>R_{\mathrm{g}}$, where $\rho_0 = M_{\mathrm{g}}/(4/3 \pi R_{\mathrm{g}}^3)$, to emulate an optically-thin background outside the cloud. 

Given the parameters, the radial radiation flux $F$ can then be obtained by solving the steady-state equation of $E_r$ (Equation~\ref{eq:Erad}),
\begin{equation}
    \nabla \cdot F = j_*(r).
\end{equation}
This gives
\begin{equation}
F_{*}(r)=\frac{L_{*}}{4 \pi r^{2}}\left[\operatorname{erf}\left(\frac{r}{\sqrt{2} \sigma_{*}}\right)-\frac{2 r}{\sqrt{2 \pi \sigma_{*}^{2}}} \exp \left(-\frac{r^{2}}{2 \sigma_{*}^{2}}\right)\right].
\end{equation}
The equation for the radiation momentum (Equation~\ref{eq:Frad}) in steady state satisfies, in spherical coordinates, 
\begin{equation}
    \nabla \cdot \mathbfss{P} = -\rho \kappaR \frac{\mathbfit{F}}{c},
\end{equation}
where $\mathbfit{F}$ is the vector flux. Solving this equation for the radiation energy density $E_r$ requires a closure relation. We use the analytical $M_1$ closure in this model, since it is both purely local (removing the need for any iteration) and exact for a single source. This closure expresses $P_{rr}$, the radial component of \mathbfss{P}, in terms of $E_r$ as $P_{rr} = f_{rr}E_r$, where $f_{rr}$ is the radial-diagonal component of the Eddington tensor. Under the $M_1$ approximation, this is related to the reduced flux $\zeta = F/(cE_r)$ as
\begin{equation}
    \zeta = \frac{5-2\sqrt{4-3f_{rr}^2}}{3}.
\end{equation}
The non-radial components of $\mathbfss{P}$ that arise when applying the divergence operator can be written in terms of $P_{\mathrm{rr}}$ and $E_r$ by using the fact that the trace of $\mathbfss{P}$ is equal to $E_r$. Subsequent substitutions allow us to write an ODE for $\zeta$ given by
\begin{equation}
\begin{aligned}
\partial_{r} \zeta=& \frac{3 \zeta \sqrt{4-3 \zeta^{2}}}{5 \sqrt{4-3 \zeta^{2}}-8}\left[\frac{\partial_{r} \ln F}{3}\left(5-2 \sqrt{4-3 \zeta^{2}}\right)\right.\\
&\left.+\frac{2}{r}\left(2-\sqrt{4-3 \zeta^{2}}\right)+\rho \kappaR \zeta\right].
\end{aligned}
\label{eq:ModelODE}
\end{equation}
This ODE passes through a critical point when $\zeta_{\mathrm{crit}} = 2\sqrt{3}/5$, at a radial location $r_{\mathrm{crit}}$ that can be obtained numerically by solving
\begin{equation}
3 \partial_{r} \ln F+\frac{4}{r}+2 \sqrt{3} \rho \kappaR=0.
\end{equation}
Our numerical procedure is to solve for the location of the critical point, and then integrate outwards and inwards in radius.

Thus far our approach parallels that of \citet{Skinner_2013}. Where we diverge is that they treat the opacity $\kappaR$ as constant, whereas for us the opacity is dependent on the radiation temperature $T_r = (E_r/\ar)^{0.25}$, which itself can be rewritten in terms of $\zeta$ (and $F$). This renders Equation~(\ref{eq:ModelODE}) an ODE in one variable ($\zeta$), which can be integrated numerically to obtain $\zeta(r)$, and subsequently $\Tr(r)$, and $\kappaR(r)$. We show the resulting profiles for some examples of $\Sigmacloud$ and $\alpha$ in Figure~\ref{fig:fEddProfiles}. 

Once we have obtained the profiles, it is straightforward to compute the radiative force as $f_{\mathrm{rad}} = \rho \kappaR F/c$. The corresponding forces of gravity on the gas are given by the sum of their interaction with the source ($g_{\mathrm{*}}$) and self-gravitational interactions ($g_{\mathrm{self}}$) as $f_{\mathrm{grav}} = \rho (g_{\mathrm{sink}} + g_{\mathrm{self}})$. The former is simply $g_* = GM_*/r^2$, and $g_{\mathrm{self}} = G \int_{0}^{r}dM(r^{\prime})/r^2$, where $\int_{0}^{r} dM(r^{\prime}) = \int_{0}^{r} \rho(r^{\prime}) 4 \pi {r^{\prime}}^2 dr^{\prime}$ is the cumulative gas mass out to radius $r$. These can then be combined to construct the Eddington ratio $f_{\mathrm{Edd}} = f_{\mathrm{rad}}/f_{\mathrm{grav}}$. The numerical implementation for this model is available at the following github repository: \url{https://github.com/shm-1996/ModelCloud}.



\bsp	
\label{lastpage}
\end{document}